# AI-based Diagnosis of Acute Aortic Syndrome from Non-contrast CT

Yujian Hu[1,20], Yilang Xiang[1,20], Yan-Jie Zhou[2,3,4,20], Yangyan He[1,20], Dehai Lang[5,20], Shifeng Yang[6,20], Xiaolong Du[7], Chunlan Den[8], Youyao Xu[9], Gaofeng Wang[10], Zhengyao Ding[11], Jingyong Huang[12], Wenjun Zhao[13], Xuejun Wu[14], Donglin Li[1], Qianqian Zhu[1], Zhenjiang Li[1], Chenyang Qiu[1], Ziheng Wu[1], Yunjun He[1], Chen Tian[1], Yihui Qiu[11], Zuodong Lin[5], Xiaolong Zhang[13], Lin Hu[15], Yuan He[1], Zhenpeng Yuan[1], Xiaoxiang Zhou[1], Rong Fan[1], Ruihan Chen[1], Wenchao Guo[2,3], Jing Xu[2,3], Jianpeng Zhang[2,3,4], Tony C. W. MOK[2,3], Zi Li[2,3], Mannudeep K. Kalra[16], Le Lu[17], Wenbo Xiao[8,21], Xiaoqiang Li[7,21], Yun Bian[18,21], Chengwei Shao[18,21], Guofu Wang[10,21], Wei Lu[9,21], Zhengxing Huang[4,21], Minfeng Xu[2,3,21], Hongkun Zhang[1,19,21]

1. Department of Vascular Surgery, The First Affiliated Hospital of Zhejiang University School of Medicine, Hangzhou, China
2. DAMO Academy, Alibaba Group, Hangzhou, China
3. Hupan Laboratory, Hangzhou, China
4. College of Computer Science and Technology, Zhejiang University, Hangzhou, China
5. Department of Vascular Surgery, Ningbo No.2 Hospital, Ningbo, China
6. Department of Radiology, Shandong Provincial Hospital Affiliated to Shandong First Medical University, Jinan, China
7. Department of Vascular Surgery, Nanjing Drum Tower Hospital, Nanjing, China
8. Department of Radiology, The First Affiliated Hospital of Zhejiang University School of Medicine, Hangzhou, China
9. Department of Vascular Surgery, The Quzhou Affiliated Hospital of Wenzhou Medical University, Quzhou People’s Hospital, Quzhou, China
10. Department of Vascular Surgery, Shaoxing Central Hospital, Shaoxing, China
11. Polytechnic Institute of Zhejiang University, Hangzhou, China
12. Department of Vascular Surgery, The First Affiliated Hospital of Wenzhou Medical University, Wenzhou, China
13. Department of Vascular Surgery, Taizhou Hospital of Zhejiang Province, Taizhou, China
14. Department of Vascular Surgery, Shandong Provincial Hospital Affiliated to Shandong First Medical University, Jinan, China
15. Department of Emergency, The Second People's Hospital of Chun'an County, Hangzhou, China
16. Division of Thoracic Imaging, Department of Radiology, Massachusetts General Hospital and Harvard Medical School, Boston, Massachusetts, USA
17. DAMO Academy, Alibaba Group, New York, NY, USA
18. Department of Radiology, Shanghai Changhai Hospital, Shanghai, China
19. Key Laboratory of Clinical Evaluation Technology for Medical Device of Zhejiang Province, The First Affiliated Hospital of Zhejiang University School of Medicine, Hangzhou, China
20. These authors contributed equally: Yujian Hu, Yilang Xiang, Yan-Jie Zhou, Yangyan He, Dehai Lang, Shifeng Yang
21. These authors jointly supervised this work: Wenbo Xiao, Xiaoqiang Li, Yun Bian, Chengwei Shao, Guofu Wang, Wei Lu, Zhengxing Huang, Minfeng Xu, Hongkun Zhang

Email: **Wenbo Xiao:** xiaowenbo@zju.edu.cn, **Xiaoqiang Li:** flytsg@126.com, **Yun Bian:** bianyun2012@foxmail.com, **Chengwei Shao:** chengweishaoch@163.com, **Guofu Wang:** 13645758030@163.com, **Wei Lu:** Luwei-77@163.com, **Zhengxing Huang:** zhengxinghuang@zju.edu.cn, **Minfeng Xu:** eric.xmf@alibaba-inc.com, **Hongkun Zhang:** 1198050@zju.edu.cn

## Summary

The accurate and timely diagnosis of acute aortic syndromes (AAS) in patients presenting with acute chest pain remains a clinical challenge. Aortic CT angiography (CTA) is the imaging protocol of choice in patients with suspected AAS. However, due to economic and workflow constraints in China, the majority of suspected patients initially undergo non-contrast CT as the initial imaging testing, and CTA is reserved for those at higher risk. Although non-contrast CT can reveal specific signs indicative of AAS, its diagnostic efficacy when used alone has not been well characterized. In this work, we present an artificial intelligence-based warning system, iAorta, using non-contrast CT for AAS identification in China, which demonstrates remarkably high accuracy and provides clinicians with interpretable warnings. iAorta was evaluated through a comprehensive step-wise study. In the multi-center retrospective study (n = 20,750), iAorta achieved a mean area under the receiver operating curve (AUC) of 0.958 (95% CI 0.950-0.967). In the large-scale real-world study (n = 137,525), iAorta demonstrated consistently high performance across various non-contrast CT protocols, achieving a sensitivity of 0.913-0.942 and a specificity of 0.991-0.993. In the prospective comparative study (n = 13,846), iAorta demonstrated the capability to significantly shorten the time to correct diagnostic pathway for patients with initial false suspicion from an average of 219.7 (115-325) mins to 61.6 (43-89) mins. Furthermore, for the prospective pilot deployment that we conducted, iAorta correctly identified 21 out of 22 patients with AAS among 15,584 consecutive patients presenting with acute chest pain and under non-contrast CT protocol in the emergency department (ED) and enabled the average diagnostic time of these 21 AAS positive patients to be 102.1 (75-133) mins. Last, the iAorta can help avoid delayed or missed diagnosis of AAS in settings where non-contrast CT remains the unavoidable the initial or only imaging test in resource-constrained regions and in patients who cannot or did not receive intravenous contrast.

## Introduction

Despite recent advancements in diagnostic and therapeutic strategies, acute aortic syndrome (AAS) remains a catastrophic cardiovascular emergency associated with high mortality rates[1-2]. Approximately 40–50% of patients with AAS died within 48 hours of its onset, and the mortality increases by 1–2% per hour without appropriate management[3-4]. Delays in diagnosis and treatment substantially worsen prognosis, underscoring the need for rapid and accurate identification of AAS[2,5]. Clinical symptoms of AAS are often non-specific and variable, ranging from acute aortic chest pain that can mimics other acute conditions to atypical or mild presentations[6-8]. Additionally, physical exams and routine lab tests lack sensitivity and specificity for confirming or ruling out AAS, complicating timely diagnosis[9-11].

Aortic CT angiography (CTA) is the imaging protocol of choice in patients with suspected AAS, yielding high sensitivity and specificity[1-2]. However, CTA is costly and carries risks of contrast-induced complications, such as anaphylaxis and nephrotoxicity[11-13]. Furthermore, only 2.7% true incidence of AAS in patients who undergo CTA for suspected AAS were found[14-15]. In China, the cost of CTA is approximately five times higher than that of non-contrast CT, and healthcare resources remain significantly limited compared to developed countries[16-17]. Considering both cost and resource constraints, Chinese clinical guidelines and current practice recognize that non-contrast chest and abdominal CT imaging can provide a rapid and straightforward initial assessment for considered as less critical conditions, while CTA is reserved for patient cases with highly suspicious findings or a high clinical risk of AAS[18]. Similarly, in low- and middle-income countries, limited access to iodinated contrast media and financial constraints further restrict the use of CTA[19-20]. In this context, patients in China or low- and middle-income countries may frequently undergo non-contrast CT for AAS evaluation and triage, despite its limited sensitivity, which creates significant diagnostic challenges even for experienced radiologists and clinicians.

The potential of using non-contrast CT for screening AAS has been explored previously[21-22] and it is often performed as a standard practice along with contrast CT for the purpose of better visualized for intramural hematoma[23-24]. The diagnostic performance of non-contrast CT alone, however, has not been well characterized in prior studies. Recent advances in artificial intelligence (AI), particularly deep learning algorithms, have demonstrated superior capabilities in extracting and identifying latent and clinically relevant features from medical images. In the settings of cost-prohibitive or limited availability contrast CT, an AI tool with high sensitivity and negative predictive value can help improve the diagnostic confidence for extracting AAS patients from non-contrast CT exams, and direct the use of CTA to these patients with higher likelihood of AAS based on AI evaluation via non-contrast CT. Furthermore, it can help avoid delayed or missed diagnosis of AAS in settings where non-contrast CT remains the unavoidable the initial or only imaging test.

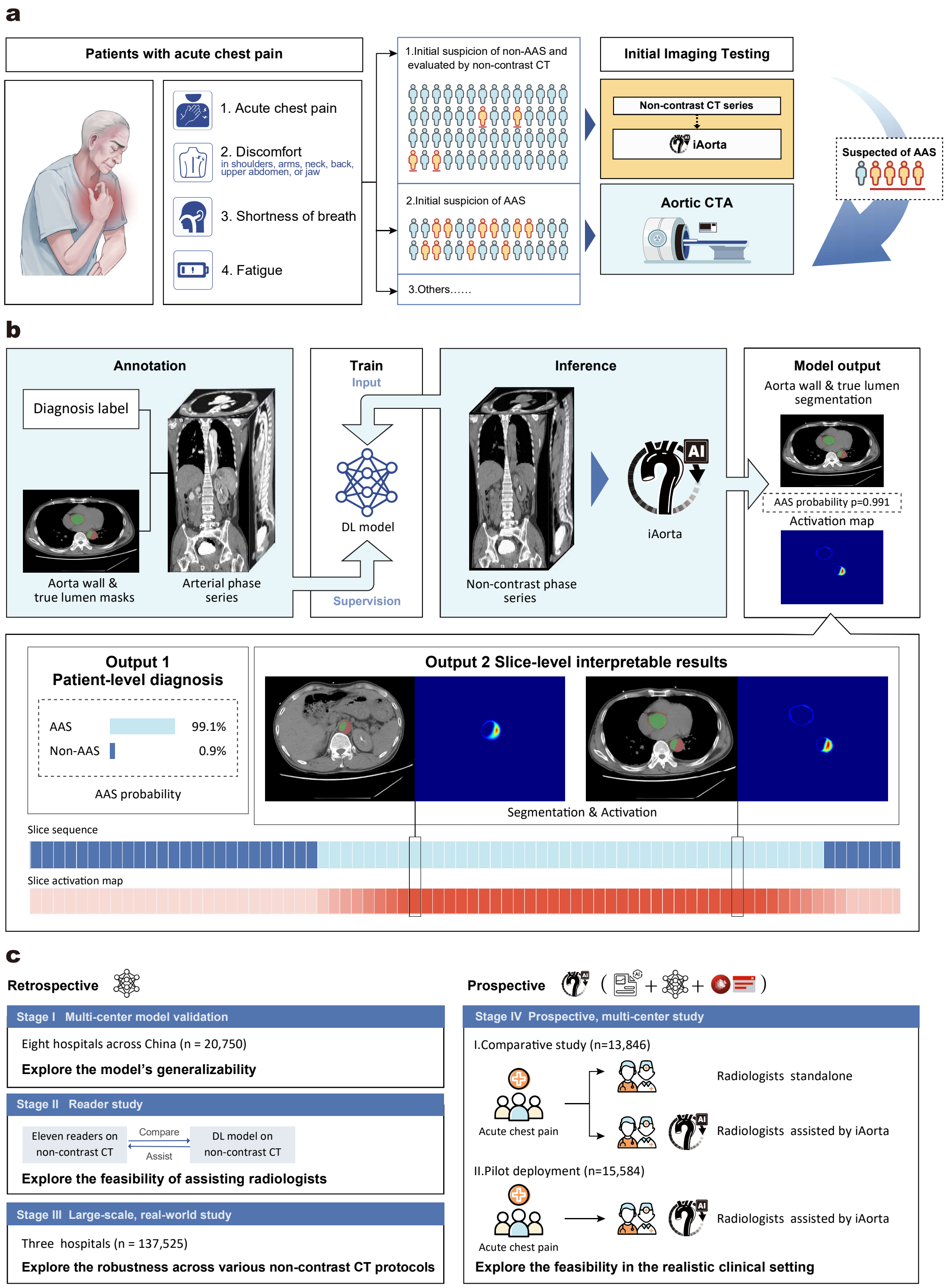

Fig. 1 | Overall study design and model pipeline. a. Clinical starting point for the study. The diagnosis of AAS poses a significant challenge within the ED due to its non-specific clinical symptoms. In China, more than half of the patients with acute chest pain are initially suspected of less critical illnesses and thus received non-contrast CT scans as the initial imaging test. Our goal is to develop an AI model that can rapidly and accurately identify patients with suspected AAS from this population of individuals undergoing non-contrast CT scans, while providing interpretable results to assist radiologists and physicians in making informed clinical decisions. b. Schematic overview of model. It was trained with patient-level diagnostic labels and segmentation masks

annotated on arterial phase series. The model takes non-contrast phase series as input and outputs the probability, the segmentation mask of aorta wall & true lumen, and activation map indicating possible lesion areas. c. Retrospective and prospective evaluation of model and iAorta. Retrospective evaluation for model consists of multi-center model validation (stage I, n = 20,750), reader study (stage II, n = 2,287), and large-scale real-world study (stage III, n = 137,525). Prospective evaluation (stage IV) for iAorta consists of comparative study (n = 13,846) and pilot deployment study (n = 15,584). iAorta incorporates a phase selection module, our model, and a popup warning module. AAS, acute aortic syndrome.

In this study, we developed and validated an AI-based system, iAorta, to identify AAS rapidly and accurately on the non-contrast CT scans in real-world emergency settings in China. Designed to automatically analyze non-contrast CT scans from patients presenting with acute chest pain, iAorta provides early warnings of AAS to radiologists, when necessary, as illustrated in Fig. 1a. To train the deep learning (DL) model, we collected aortic CTA scans with paired arterial and non-contrast CT phase series. In stage I, the diagnostic performance and generalizability of the model were evaluated across eight hospitals in China. In stage II, we assessed AI-assisted performance improvements in diagnostic accuracy by providing radiologists with interpretable model predictions. In stage III, we evaluated the model's robustness across various non-contrast CT protocols. In stage IV, we developed the iAorta system to achieve practical clinical applications in two prospective trials. A prospective multi-center study including comparative analysis, and a pilot deployment, were conducted to assess the feasibility of integrating iAorta into real-world emergency department (ED) workflows.

The study findings highlight the significant potential of our iAorta system as a vital decision-support tool in emergency care settings where contrast CT is either cost-prohibitive or unavailable to provide directly. On delivering accurate, rapid, and interpretable visual warnings, iAorta can facilitate and enable timely and informed clinical decisions. In sites with multiphase CT protocols for AAS evaluation, iAorta's high specificity and high negative predictive value on non-contrast CT scans may help avoid unnecessary contrast injections in low-risk patients, particularly younger individuals. This not only reduces the risks associated with contrast agents, such as nephrotoxicity and allergic reactions, but also decreases cumulative radiation exposures from additional post-contrast imaging, thereby enhancing patient safety.

## Results

### Overview of the study

We conducted this stepwise study in four stages with nine hospitals across China to develop and validate the iAorta system for effective detecting AAS, that is, Stanford Type A aortic dissection (TAAD), Stanford Type B aortic dissection (TBAD), intramural hematoma (IMH) and penetrating atherosclerotic ulcer (PAU), in real-world emergency settings. In stage I, we evaluated the model in an internal validation cohort (n = 2,287) and seven external validation cohorts (n = 18,463). In stage II, we compared the model performance with radiologists of varying expertise levels and radiologists assisted by AI on non-contrast CT. In stage III, we retrospectively tested the model in real-world emergency scenarios of three representative medical centers, involving a total of 145,201 various non-contrast CT scans from 137,525 consecutive patients presented with acute chest pain symptoms. In stage IV, we developed a real-time "hyperlink message" warning system, iAorta, and integrated it into the existing clinical workflow. To assess its impact on improving the identification of AAS on non-contrast CT and reducing the time to the correct diagnostic pathway, we further conducted a prospective, multi-center study. This study included a comparative study of 14,436 non-contrast CT scans from 13,846 consecutive patients, and more importantly, a pilot deployment study on 16,054 non-contrast CT scans from 15,584 consecutive patients (running from 12/20/2024 to 02/28/2025). The baseline demographic information and image characteristics of all cohorts are summarized in Extended Data Table 1. Patient and data sources of model development, multi-center model validation, reader study, retrospective large-scale real-world study, and prospective multi-center study are shown in Extended Data Fig. 1.

### Development of the iAorta System

**Model development.** To obtain accurate diagnosis labels and lesion regions, we collected 3,350 aortic CTA scans, including arterial and non-contrast phase series, to train the model. The arterial phase series were used as the gold standard for diagnosing AAS and suggesting lesion areas, while the non-contrast phase series were used for model training under the supervision of the diagnostic labels and the segmentation labels that are transferred by image registration technique from lesion annotations on paired arterial phase series. Our model can detect the presence or absence of AAS at the patient level, segmenting the aorta and true lumen precisely, and localizing and identifying potential lesion regions reliably to enhance slice-level interpretability. The final outputs generated by our model are fed back into the iAorta system to send timely alerts to radiologists, facilitating the rapid referral of high-risk patients for further aortic CTA examination and confirmation, ensuring prompt diagnosis and effective intervention. More details about the annotations and model architecture are provided in the Methods section, as shown in Extended Data Fig. 2. Quantitative analysis is provided in Supplementary Section 2.6 (Supplementary Fig. 2 and Supplementary Table 1).

**Model visualization and explanation.** We generated activation maps by mapping the normalized activation scores to their corresponding spatial location in each CT slice, as shown in Fig. 1b and Extended Data Fig. 3. Technical details on the generation of activation maps are provided in the Methods section. In our task, the regions with high activation scores typically corresponded to relevant abnormal areas, whereas low activation scores indicated healthy aortic areas. There is a visible difference in the appearance of activation maps between different subtypes. The high activation regions of AD and IMH are more refined and larger, whereas those of PAU are coarser and smaller. This observation is consistent with their representations of subtypes. The regions with high activation scores in each slice provide valuable visual information for AI-assisted detection of AAS in the ED. Further quantitative analysis revealed that the diagnostically relevant regions generated by the model improved the sensitivity of radiologists' assessments on AAS when using our **iAorta system.** We have developed a browser-server collaborative early warning system designed for rapid and accurate detection of AAS in real-world clinical medical settings. Integrated with a non-contrast CT phase selection tool, our developed model and warning popup plugin, the system can process and analyze large amounts of patient data in real-time, enabling radiologists to get timely alerts and make informed diagnosis and intervention. The browser-based interface ensures that the system is accessible from any device with an internet connection, enhancing its usability and flexibility. This system is designed to be highly secure, with strict data privacy and security protocols in place to protect sensitive patient information. The iAorta system both enhances the efficiency and accuracy of AAS detection and provides a scalable and secure solution for clinical use. Further details about the collaborative system are provided in the Methods section.

**Stage I: Multi-center model validation**

To assess the generalizability and robustness of our model to various patient populations, we collected 20,750 aortic CTA scans, including arterial and non-contrast phase series similar to those used in model training, from eight hospitals across China. The diagnosis labels were confirmed using the corresponding paired arterial phase series. In this internal validation cohort, the overall model performance in detecting AAS on non-contrast CT achieved an area under the receiver operating characteristic curve (AUC) of 0.980 (95% confidence interval [CI] 0.973-0.987) (Fig. 2a), a sensitivity of 0.984 (95% CI 0.972-0.990) and a specificity of 0.947 (95% CI 0.935-0.957). In the external validation cohorts, the performance of AAS detection obtained an AUC of 0.941-0.972, a sensitivity of 0.954-0.975, and a specificity of 0.929-0.946 (Fig. 2b and 2c). Furthermore, we conducted a subgroup analysis performance for different subtypes of AAS. We divided AAS into four subtypes, TAAD, TBAD, IMH and PAU, based on the severity of the condition. Details about the diagnostic criteria are provided in the Methods section. The model achieved a sensitivity of 0.971-0.995 for TAAD, 0.980-0.996 for TBAD, 0.953-0.980 for IMH and 0.912-0.955 for PAU (Fig. 2d), respectively. Detailed evaluation metrics, including accuracy, positive predictive value (PPV), negative predictive value (NPV) and F1 score, are shown in Extended Data Table 2.

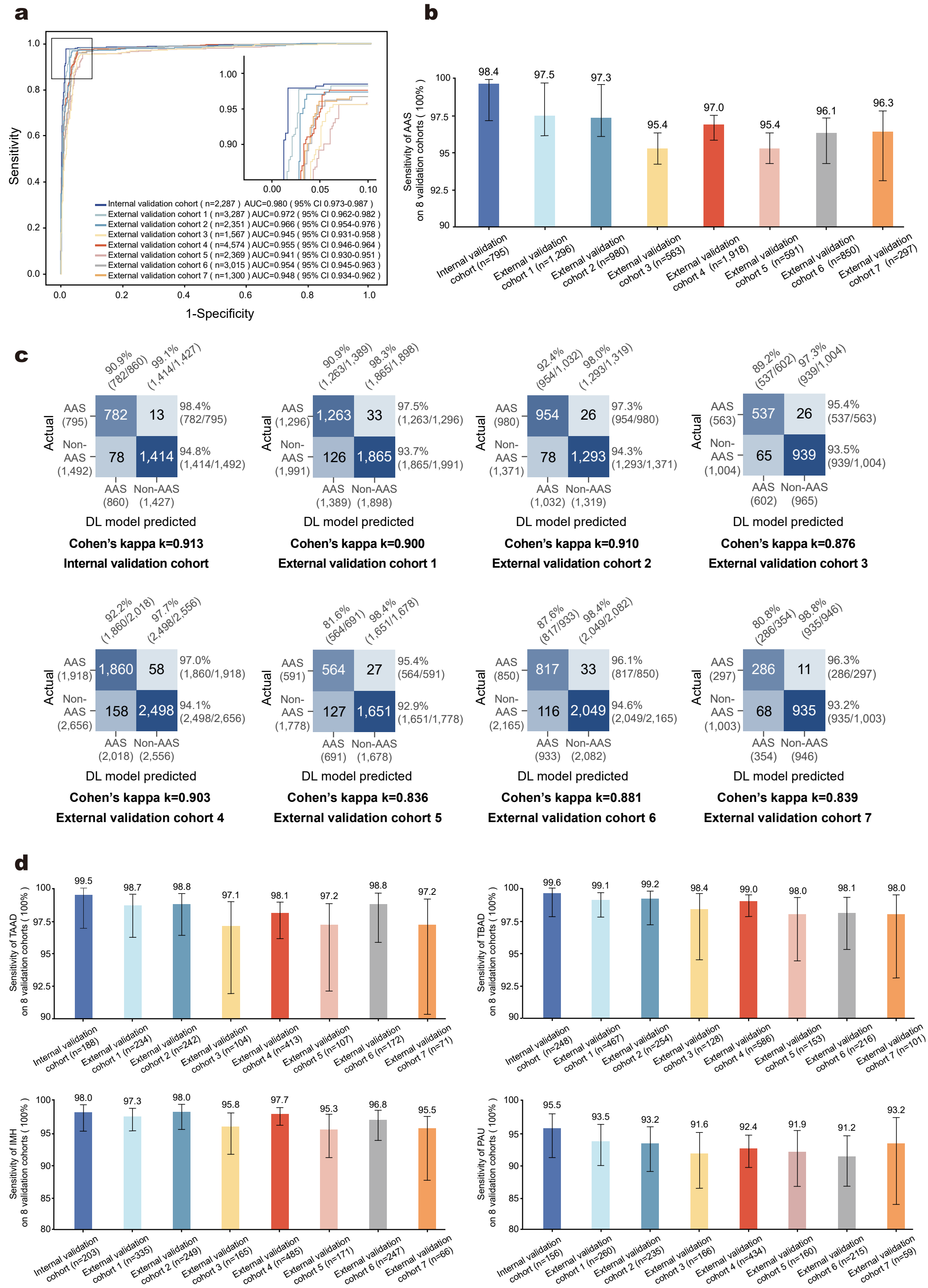

**Fig. 2 | Stage I multi-center model validation. a.** ROC curves of AAS detection on the internal and external validation cohorts. **b.** Sensitivity of AAS detection in the internal validation cohort (n = 795) and external validation cohorts (cohorts 1-7, n = 6,495). The error bars denote the two-sided 95% confidence interval computed from 1,000 bootstrapping iterations. **c.** Confusion matrices of AAS detection on the internal and external validation cohorts showing the TPs, TNs, FPs, FNs of AAS detection and the sensitivity, specificity, PPV, NPV calculated from the above. **d.** Sensitivity of four subtypes (TAAD, TBAD, IMH, PAU) of AAS detection in the internal cohorts (n = 795) and

external validation cohorts (cohorts 1-7, n = 6,495). The error bars denote the two-sided 95% confidence interval computed from 1,000 bootstrapping iterations. AAS, acute aortic syndrome; ROC, receiver operating characteristic; AUC, Area Under the Curve; FPs, false positives; FNs, false negatives; TPs, true positives; TNs, true negatives; PPV, positive predictive value; NPV, negative predictive value; TAAD, Stanford Type A dissection; TBAD, Stanford Type B dissection; IMH, intramural hematoma; PAU, penetrating atherosclerotic ulcer.

**Stage II: Reader study**

To further assess the model performance and explore the value of clinically interpretable results in assisting for the detection of AAS, we conducted a two-stage reader study using the internal validation cohort. We recruited eleven radiologists with varying degrees of expertise (specialty experts in cardiovascular, board-certified general radiologists, and medical trainees) to participate in the study. In the first stage of this study, radiologists were asked to independently diagnose each case as either AAS or non-AAS within a specified time limit, without any knowledge of the AI model produced results. After a 2-month washout period and randomization of the data, the same group of radiologists were given the interpretable results (in the form of distance maps) and diagnostic probabilities output from our AAS model to assist in their reassessment of patients' diagnoses again, without prior knowledge of the model's diagnostic accuracy. The detailed results are shown in Fig. 3 and Extended Data Table 3.

In general, the performance values of all 11 radiologists in either independent diagnosis or with AI assistance were lower than those indicated by ROC curves for model alone. As illustrated in Fig. 3a, in independent diagnoses, the sensitivity levels varied greatly among the 11 radiologists, with 0.807 (95% CI 0.795–0.819) for specialty experts, 0.778 (95% CI 0.764–0.796) for board-certified radiologists, and 0.603 (95% CI 0.588–0.619) for medical trainees, which were significantly lower than that of model (0.984, 95% CI 0.972–0.990, $p < 0.001$). When radiologists were assisted by model, the sensitivities of specialty experts, board-certified radiologists, and medical trainees all increased significantly, reaching 0.943 (95% CI 0.930–0.958, $p < 0.001$), 0.899 (95% CI 0.887–0.913, $p < 0.001$), and 0.919 (95% CI 0.906–0.933, $p < 0.001$), respectively. The improvement was especially notable among medical trainees, whose performance approached or even surpassed that of specialty experts when aided by AI. This highlights the model's ability to effectively bridge the diagnostic performance gap across varying levels of expertise.

For detecting AAS from its specific subtypes (Fig. 3b), the mean sensitivities of radiologists were 0.700 (95% CI 0.677-0.732) in TAAD, 0.741 (95% CI 0.701-0.796) in TBAD, and 0.647 (95% CI 0.599-0.693) in IMH, which were significantly greater than the score of 0.382 (95% CI 0.304-0.477) in PAU. With AI assistance, the mean sensitivity of radiologists improved significantly, reaching 0.893 (95% CI 0.865-0.927, $p < 0.001$) in TAAD, 0.941 (95% CI 0.896-0.982, $p < 0.001$) in TBAD, 0.859 (95% CI

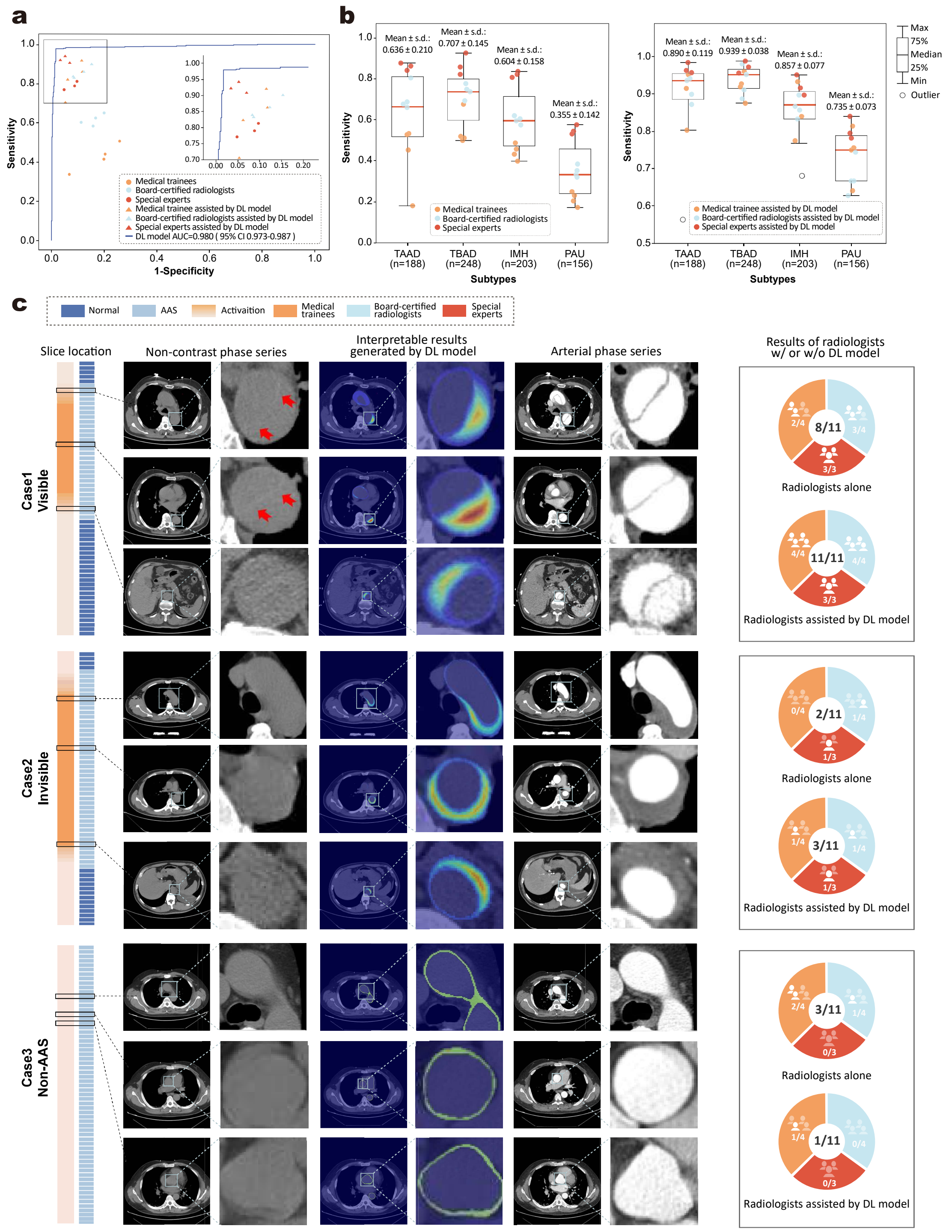

**Fig. 3 | Stage II reader study. a.** Comparison between DL model and 11 radiologists with different levels of expertise on non-contrast CT for AAS detection, with or without the assistance of model. **b.** Sensitivity of 11 radiologists with different levels of expertise on non-contrast CT for four different subtypes of AAS detection, with or without the assistance of model (n = 795 data samples). On each box in **b**, the central line indicates the median, and the bottom and top edges of the box indicate the 25th and 75th percentiles, respectively. The whiskers extend to 1.5 times the interquartile range. **c.** Examples of the "visible" positive case of TBAD, "invisible" positive case of IMH, and the negative case with ascending aorta artifact which was not detected by readers on non-contrast CT but well-classified with the assistance of model. ROC, receiver operating characteristic;

AUC, Area Under the Curve.

0.814-0.900, $p < 0.001$) in IMH and 0.738 (95% CI 0.664-0.789, $p < 0.001$) in PAU. These results demonstrate that AI assistance markedly enhanced radiologist performance in more clinically urgent subtypes (e.g., TAAD and TBAD) and diagnostically elusive cases such as PAU, for which sensitivity nearly doubled.

Remarkedly, our model has shown the ability to discern specific AAS images that remain elusive even with AI assistance to radiologists. As shown in Fig. 3c, for the "visible" AAS case, radiologists were able to detect distinctive lesion features on non-contrast CT with AI assistance, leading to a revision of their initial diagnosis. However, for the other highly "invisible" AAS case, even our model provided a very high AI prediction probability and visualized lesion results, but radiologists did not consider the case to be AAS.

**Stage III: Large-scale, real-world, retrospective study**

To further explore the model's robustness across various non-contrast CT protocols from real-world emergency, we performed a two-round, large-scale, real-world, retrospective study using a total of 145,201 non-contrast CT scans from 137,525 consecutive patients with acute chest pain in three hospitals. The details of the source of non-contrast CT scans are shown in the Fig. 4b and Extended Data Fig. 1. The diagnosis labels were confirmed by the conclusive diagnostic imaging or medical records within the two-week follow-up.

The first real-world evaluation cohort consisted of 23,094 non-contrast CT scans of 20,832 consecutive individuals from FAHZU, including 44 AAS scans. For AAS detection, the model achieved an overall sensitivity of 0.818 (95% CI 0.680-0.905) and a specificity of 0.994 (95% CI 0.993-0.995) (Fig. 4c). To comprehend the error patterns of the model across various non-contrast CT scans, a multidisciplinary team reviewed the error cases. The majority of false-negative cases occurred in cases with a limited FOV and small z-axis coverage. To improve the model performance in detecting AAS on non-contrast CT from different CT protocols, we implemented hard example mining and incremental learning techniques to upgrade the previously trained model. To evaluate the generalizability of the updated model, we conducted the second-round study within one internal real-world cohort ($n = 76,582$, 69 patients with AAS) and two external real-world cohorts ($n = 24,365$, 26 patients with AAS; and $n = 20,160$, 23 patients with AAS). For AAS detection, the updated model achieved a sensitivity of 0.913-0.942, and a specificity of 0.991-0.993. Compared to the original model, the updated one retains a similar level of specificity while achieving notably increased sensitivities by 11.6%-19.2% ($p < 0.001$). The ROC curves of both the original and upgraded models are shown in Fig. 4d. The significance test comparing the AUCs of the original and upgraded models is conducted using the Delong test. The result demonstrates that the observed differences in AUCs between the two models are significant ($p < 0.001$ in all 3 cohorts of RW2). Furthermore, Fig. 4e presents the model

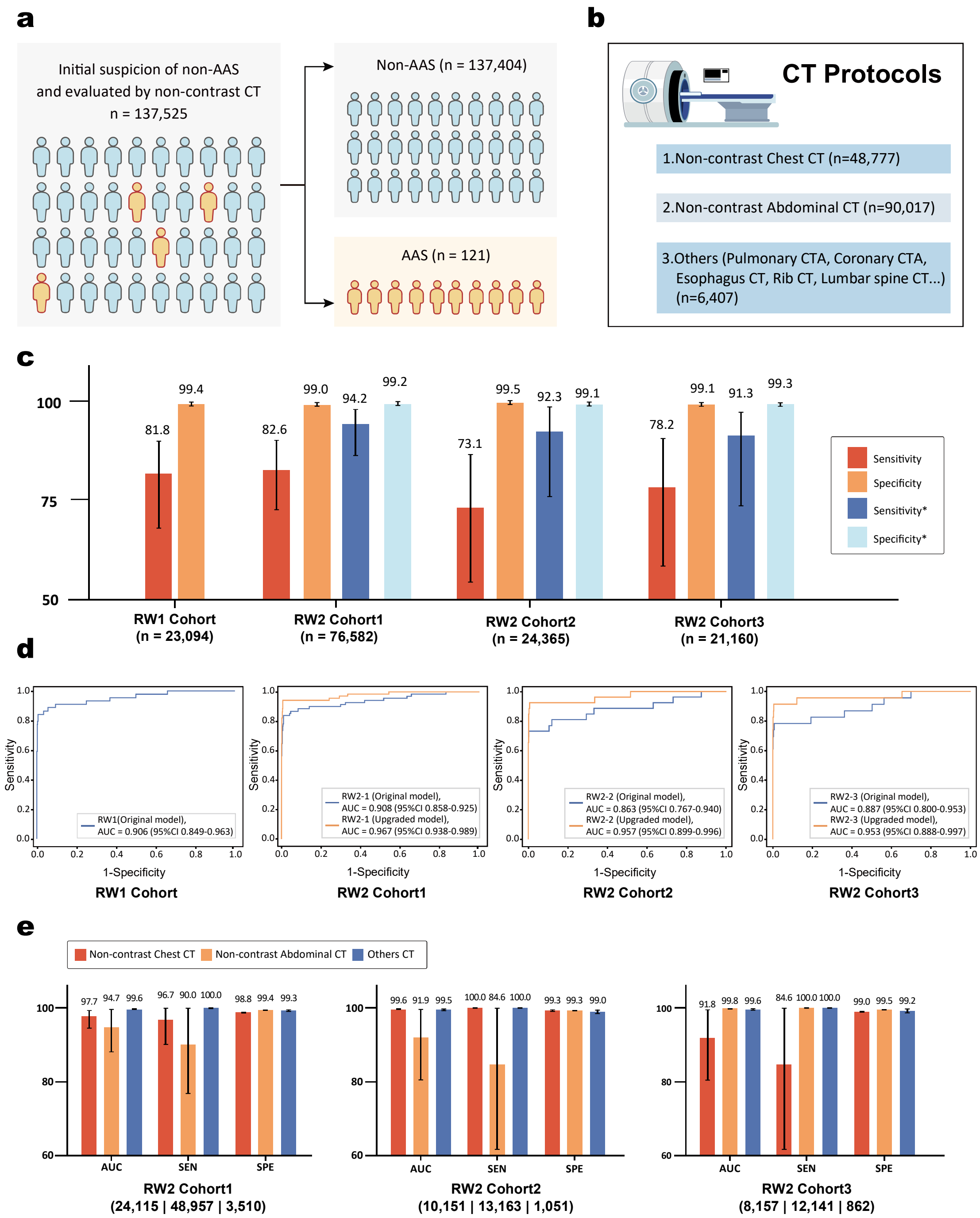

**Fig. 4 | Stage III large-scale real-world retrospective study. a.** The patients' distribution in the study. **b.** The distributions of CT protocols in the study. **c.** The sensitivity and specificity of original model on RW1(n = 23,094) and that of original and upgraded model on RW2 (cohorts 1-3, n = 122,107). The superscript * represents results of upgraded model. The error bars denote the two-sided 95% confidence interval computed from 1,000 bootstrapping iterations. **d.** ROC curves of AAS detection on the RW1 and RW2 cohorts. Blue and orange curves represent the performance of original model and upgraded model, respectively. e. Model performance metrics (AUC, accuracy, sensitivity, and specificity) for AAS detection across three subtypes (non-contrast chest CT [cohorts 1-3, n = 42,423], non-contrast abdominal CT [cohorts 1-3, n = 74,261], and others [cohorts 1-3, n = 5,423]) in the RW2 cohorts. The error bars denote the two-sided 95% confidence interval computed

from 1,000 bootstrapping iterations. AAS, acute aortic syndrome; ROC, receiver operating characteristic; AUC, Area Under the Curve; SEN, Sensitivity; SPE, Specificity.

performance metrics (AUC, accuracy, sensitivity, and specificity) for AAS detection across three subtypes (non-contrast chest CT, non-contrast abdominal CT, and others) in the RW2 cohorts. These findings demonstrate that our model consistently achieves robust numerical performance results under various CT protocols.

We evaluated the potential clinical value of our proposed AI model in identifying AAS patients. In China, initial diagnosis by emergency clinicians was based on preliminary investigation (using age, risk factors, history, pain characteristics, findings on physical examination, ECG and some laboratory tests). In this study, 127 patients with AAS (who were excluded from the cohort) were suspected of having AAS after the initial assessment, and the remaining 121 patients with AAS (31 with TAAD, 45 with TBAD, 21 with IMH and 24 with PAU) did not initially receive the proper or correct initial triage decisions from emergency clinicians and subsequently performed other CT protocols examination (mostly via non-contrast CT) as the initial imaging rule-out test. Applying our AI model helped correctly diagnose 90.1% (109/121) of patients with AAS in this subpopulation. On the other hand, a total of 9.9% of patients (RW1: n = 6, RW2: n = 6) were missed by the AI model (Supplementary Table 5-8); most of them (n = 11) had PAU, and the remaining patients had limited IMH. Three cases illustrating the potential benefits for patients with AAS are shown in Extended Data Fig. 4.

**Stage IV: Prospective multi-center study**

Given the rapid progression and high mortality rate of AAS, it is imperative to minimize the risk of inaccurate initial assessment, thereby enabling and assuring timely and appropriate clinical decision-making. To address this, we developed the iAorta system, which incorporates automatically computed analysis and early warning functionalities.

**Multi-center comparative study.** To investigate the effectiveness of iAorta in the ED as an early AAS warning tool, we conducted a prospective study involving a cohort of 13,846 consecutive patients (14,436 non-contrast CT scans) who were initially not suspected of having AAS and underwent non-contrast CT scans (or CT protocols that included a non-contrast phase) across three medical centers. The study aimed to compare the performance of two groups of radiologists in identifying early signs of AAS: Group A independently reviewing images, mirroring the current clinical workflow and Group B assisted by iAorta conducting the same task, as illustrated in Fig. 5a. In the first group, the radiologist team, consisting of the initial reporting (IR) radiologist and the report reviewing (RR) radiologist, independently evaluate images in a sequential manner within the current clinical workflow. This process generates two independent diagnostic records: one for the IR radiologist's diagnosis and the other for the RR radiologist's diagnosis, for each CT scan. In the second group, patients' CT scans are automatically fed into the iAorta system in real-time, which thus produces AI-based AAS detection results. If the AI detects any abnormalities, both the IR and RR

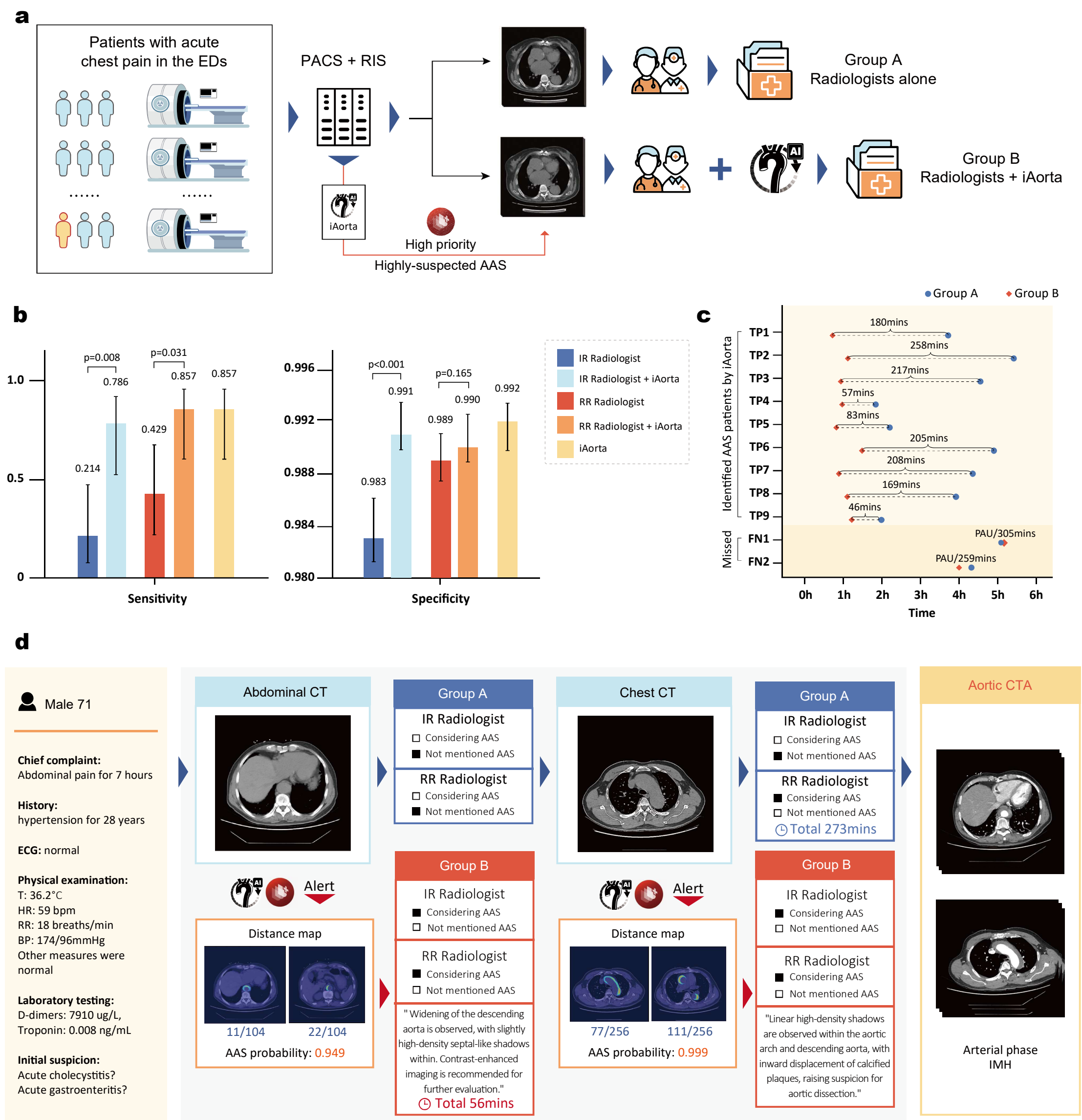

**Fig. 5 | Stage IV prospective multi-center study (comparative study). a.** The study involves two groups of radiologists. Group A, the radiologist team, consisting of the initial reporting (IR) radiologist and the report reviewing (RR) radiologist, who independently evaluate images in a sequential manner, mirroring the current conventional clinical workflow. Group B, the radiologist team evaluate images assisted by iAorta system. If iAorta detects any abnormalities, both the IR and RR radiologists receive sequential pop-up alerts, prompting them to prioritize the review of the AI flagged AAS positive image. **b.** The sensitivity and specificity of Group A (IR radiologist and RR radiologist), Group B (IR radiologist + iAorta and RR radiologist + iAorta), and iAorta system (n = 14,436 data samples). The error bars denote the two-sided 95% confidence interval computed from 1,000 bootstrapping iterations. The center shows the computed mean of the metric specified by its respective axis labels. *P*-value was calculated by two-sided McNamar test ($p < 0.001$ [$3.47 \times 10^{-21}$]). **c.** Time from presentation to correct diagnostic pathway (Group A vs Group B) in the realistic clinical settings. Nine AAS patients were identified by iAorta, yet two AAS patients (PAU) were missed. **d.** An illustrative example highlights the potential benefits of employing iAorta for AAS patients who initially were suspected of having other acute conditions and underwent non-contrast CT in real-world emergency settings. iAorta has demonstrated with the potential to significantly

reduce the time to reach a correct diagnostic pathway for this patient, from 273 mins to 56 mins.

radiologists receive subsequential pop-up alerts, prompting them to prioritize the review of the AI flagged (or AAS positive) CT images. After reviewing the AI results, the AAS-assisted diagnosis is recorded. This process yields three independent diagnostic records for each image: the IR radiologist's diagnosis (aided with AI), the RR radiologist's diagnosis (aided with AI), and the AI-generated results.

In the first group, IR radiologists identified 3 out of 14 AAS-related positive results in 14,436 non-contrast CT scans, while RR radiologists detected 6 out of 14 AAS-related positive results. In the second group, after reviewing the AI-generated results, IR radiologists reported 11 out of 14 positive results, and RR radiologists found 12 out of 14 positive results. The sensitivity and specificity of IR radiologists improved from 0.214 (95% CI 0.076-0.476) to 0.786 (95% CI 0.524-0.924; $p = 0.008$), and from 0.983 (95% CI 0.981-0.985) to 0.991 (95% CI 0.989-0.992; $p < 0.001$), respectively. For RR radiologists, their sensitivity and specificity improved from 0.429 (95% CI 0.214-0.674) to 0.857 (95% CI 0.601-0.960; $p = 0.031$), and from 0.989 (95% CI 0.987-0.990) to 0.990 (95% CI 0.989-0.992; $p = 0.165$), respectively (as shown in Fig. 5b). The AI model alone demonstrated a high sensitivity score of 0.857 (95% CI 0.601-0.960), and a specificity of 0.992 (95% CI 0.990-0.993), as illustrated in Fig. 5b and Extended Data Table 4. More importantly, integrating the iAorta system into the current diagnostic workflow reduced the average time “from admission to correct diagnosis” pathway for AAS patients with initial false suspicion, significantly from an average of 219.7 (115-325) mins down to 61.6 (43-89) mins (as shown in Fig. 5c).

**Pilot deployment.** To further assess the real operational performance of iAorta, we integrated the iAorta system into the current pilot clinical routine in Shanghai Changhai Hospital, as depicted in Fig. 6a. iAorta automatically processes large amount of input CT images from the hospital Picture Archiving and Communication System (PACS), then identifies patients at risk of AAS and provides pop-up alerts to the diagnostic radiology team if positive, prompting them to prioritize the review of these flagged AI-positive CT patient cases using an interactive visual interface. The detailed illustration is shown in Supplementary Video. With the aid of prompt and accurate warnings from iAorta, radiologist team successfully identified 21 out of 22 AAS patients among a cohort of consecutive 15,584 patients who underwent non-contrast CT as initial imaging test in the ED from 12/20/2024 until 02/28/2025. One PAU patient was missed. The AAS identification performance reached an exceptional level, with a sensitivity of 0.955 (95% CI 0.864-1.000) and a specificity of 0.994 (95% CI 0.993-0.995), as shown in Fig. 6b. Note that the average diagnostic time of these 21 patients was controlled within mean 102.1 (range of 75-133) mins.

Particularly, in one case, a patient was presented to the ED with the right upper abdominal pain, as illustrated in Fig. 6d. Following an initial assessment, the clinical suspicion of an abdominal pathology emergency was made and that led to the ordering

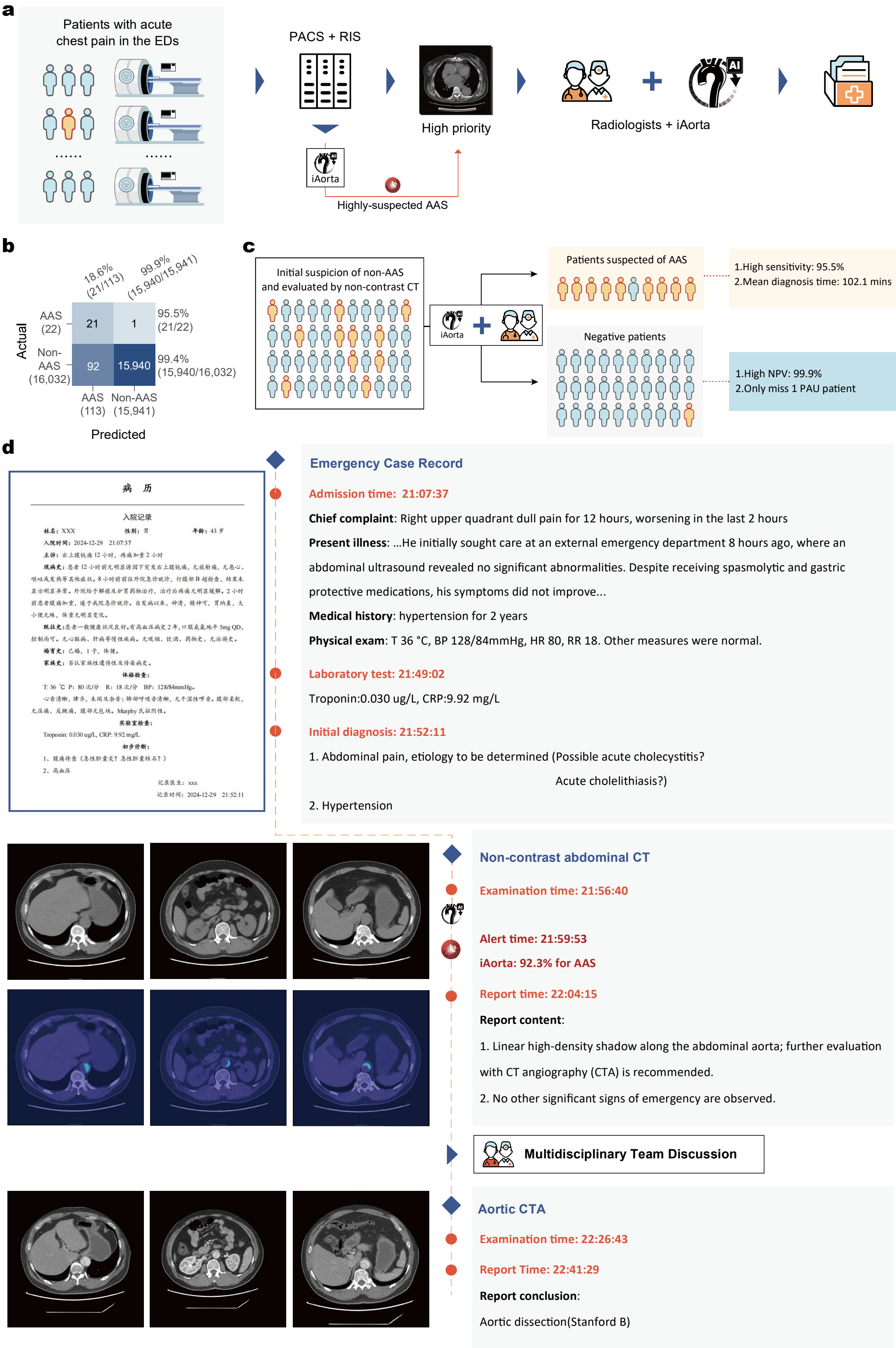

**Fig. 6 | Stage IV prospective multi-center study (pilot deployment study). a.** Seamlessly incorporate iAorta into the existing pilot clinical routine. Once iAorta identifies patients at risk of AAS, it generates pop-up alerts for the radiology team, prompting them to prioritize the review of the flagged images. **b.** Confusion matrices of AAS detection in this study showing the TPs, TNs,

FPs, FNs of AAS detection and the sensitivity (95.5%), specificity (99.4%), PPV (18.6%), NPV (99.9%) calculated from the above. **c.** Potential benefits of iAorta system in the realistic clinical settings of China. **d.** An illustrative example highlights the benefits of employing iAorta for AAS patients who initially presented with acute abdominal pain and were suspected of having other acute conditions in real-world emergency settings. While the patient initially entered an incorrect diagnostic pathway, the early warning provided by iAorta on the non-contrast CT enabled the definitive diagnosis of AAS within 94 minutes of hospital admission, ensuring both timely and appropriate management. FPs, false positives; FNs, false negatives; TPs, true positives; TNs, true negatives; PPV, positive predictive value; NPV, negative predictive value.

of a non-contrast abdominal CT. Approximately one minute after the CT scan were uploaded to the PACS (or three minutes after the CT scan examination), iAorta issued an early warning to the on-duty radiologist via a pop-up alert. Although the radiologist could not definitively identify or confirm signs of AAS on the non-contrast CT with the assistance of iAorta, the possibility of AAS diagnosis could not be excluded. After a multidisciplinary clinical team discussion and an informed consent was obtained from the patient, an aortic CTA exam was performed, which then revealed a Stanford type B aortic dissection. While the patient initially entered an incorrect diagnostic pathway (which is common), the early accurate warning provided by iAorta ran on the non-contrast CT enabled the final definitive diagnosis of AAS within 94 minutes from hospital admission, ensuring both timely and appropriate patient management.

## Discussion

In this study, we developed and validated iAorta, an AI-based real-time early warning system designed to assist radiologists and ED clinicians in accurately and rapidly detecting AAS using non-contrast CT scans alone. Through large-scale real-world retrospective and prospective validation across multiple centers in China, iAorta demonstrated robust performance, highlighting the significant potential of applying non-contrast CT as a reliable and alternative tool for AAS diagnosis. Notably, the interpretable output generated by iAorta significantly enhanced radiologists' diagnostic confidence, indicating its critical role in supporting clinical decision-making. Importantly, our study findings highlight several key benefits of iAorta: **1)** In resource-constrained settings where non-contrast CT is often the only available or primary initial imaging modality, iAorta can help reduce delayed or missed diagnoses of AAS. **2)** In clinical settings utilizing multiphase non-contrast and post-contrast CT protocols for AAS evaluation, iAorta's high specificity and NPV on non-contrast CT scans can potentially help avoid unnecessary contrast injection and reduce radiation doses from the post-contrast CT angiography.

The main strengths of iAorta are its validated high feasibility and reliability in real-world clinical settings. While several deep learning algorithms have been previously developed in this field, their performance varies significantly between internal and external validation datasets, and none have been thoroughly evaluated in large-scale real-world prospective scenarios[25-28]. In that situation, the low stability of the model design and bias from training data greatly limit the model generalizability. By leveraging a large-scale, high-quality dataset with precise pathology annotations and employing a refined effective two-stage training strategy, our model evidently outperformed other state-of-the-art algorithms, as demonstrated in Supplementary Fig. 2 and Supplementary Table 1. Our datasets primarily consisted of non-ECG-gated CT images and the non-contrast CT sourced from various scanning protocols, devices, and populations. Although artifacts in the ascending aorta on non-ECG-gated CT images can lead to false-positive diagnoses of aortic dissection[29,30] (as shown in Fig. 3c), iAorta effectively addresses these challenges and performs well. Moreover, previous studies have primarily focused on TAAD/TBAD alone, neglecting other critical conditions such as IMH and PAU. Recognizing the importance of these additional lesions is crucial for comprehensive coverage of complete AAS diagnosis and disease management. A key characteristic of AAS is the potential synchronous or metachronous appearance of these lesions in different aortic segments. These conditions, including TAAD, TBAD, IMH, and PAU, are all time-sensitive and can progress rapidly, with the risk of aortic rupture, especially during the initial stages of onset[31,22]. Our findings indicate that iAorta not only meets the substantially critical clinical needs associated with AAS satisfyingly but also demonstrates strong potentials for widely-adopted applications across non-contrast CT imaging with diverse patient scanning parameters, with no requirement needed for frequent adaptations or fine-tuning to facilitate future applications.

iAorta has demonstrated as a promising tool to assist radiologists and ED clinicians by providing effective and interpretable visual clues of acute aortic lesions to support the clinical decision-making. This feature is crucial in practice because visual clues help explain and validate AI system predictions, which is especially important for this acute life-threatening condition. Our visualization method is based on segmentation masks of the aorta and true lumen to generate the corresponding distance maps, making the presentation of lesion regions more spatially precise[33], increasing radiologists' confidence in evaluating and employing the iAorta system's results. In our study, differences in diagnostic accuracy among radiologists of various levels of experiences were evident, highlighting the challenges when detecting AAS on non-contrast CT scans. Since certain imaging features can be too subtle to notice without proper assistance, radiologists, particularly those with less experience, may often struggle to detect AAS on non-contrast CT images if not impossible. With the AAS flagging visual messages provided by our model/tool, radiologists' diagnostic accuracy can be greatly improved. Unlike existing medical imaging-assisted diagnostic systems, iAorta incorporates an additional alert function. By prioritizing patients with a high probability of being AAS positive through timely alerts, the system enables radiologists to allocate their time and necessary attention to critical cases earlier (at the earliest possible), thereby optimizing the overall clinical workflow efficiently and effectively.

iAorta has the capability to locate lesion regions that may not be detected by radiologists when only using visual inspection. This ability is likely due to the deep learning model's finer image feature representative capacity to detect visually subtle features in the data, such as subtle but critical gradient differences in pixel HU values[34]. In cases where the specific manifestations of AAS are not visually apparent on non-contrast CT imaging, especially this can be more often observed in younger patients (arguably this is a patient sub-cohort who can be more benefited from the high sensitivity of AAS detection). Visualized intimal flaps and thrombosed lumens appearing as high-attenuation areas are typical manifestations of AAS[35]. Older AAS patients, usually with chronic hypertension, may exhibit atherosclerotic changes in the arterial wall, such as intimal thickening and calcification. Conversely, AAS in young individuals, often caused by connective tissue disorders, may result in fewer atherosclerotic changes[36,37]. Young patients with AAS, especially when asymptomatic, may be noncompliant with follow-up appointments and could deny their medical conditions. Due to concerns about CT imaging contrast material risks and medical costs, these patients may choose not to undergo aortic CTA initially or even refuse doing it in the ED. However, importantly, young patients with AAS often experience severe or rapid disease progression[38]. iAorta has demonstrated with the potential to alleviate this diagnostic dilemma. If these patients had previously undergone routine non-contrast chest or abdominal CT scans, considering the critical nature of AAS, when radiologists cannot obtain consistent positive results with or without the help of iAorta, it should be advisable to proceed with the necessary aortic CTA examination immediately.

Our real-world clinical data highlight the critical role of non-contrast CT in the initial evaluation of patients presenting with acute chest pain in EDs across China, where the high cost and limited availability of CTA pose substantial barriers. Notably, the geographical distribution of tertiary hospitals in China varies significantly, with a substantial proportion of AAS patients first seeking care at community hospitals or non-tertiary centers[39,40]. In these clinical sites, CTA is often constrained by technical and equipment limitations, whereas non-contrast CT is maybe the unavoidable only imaging test. Integrating iAorta with inter-hospital cloud platforms could expedite the referral of AAS patients from non-tertiary to tertiary centers, ensuring timely and accurate treatment. Similarly, in low- and middle-income countries, the high cost and limited availability of iodinated intravenous contrast media further restrict access to CTA[19-20]. iAorta with its high sensitivity and NPV can help improve diagnostic confidence for excluding AAS from non-contrast CT. This enables CTA to be reserved for patients with a higher likelihood of AAS based on AI-driven evaluations of non-contrast CT scans. Moreover, delayed recognition of AAS remains a significant challenge. Even for TAAD, which often presents with typical symptoms, the median time from arrival at the emergency department to diagnosis is 4.3 hours[5]. The results from our prospective pilot deployment study demonstrated that when iAorta was integrated into routine non-contrast CT workflows, the average diagnostic time for AAS patients with atypical presentations was reduced to under two hours. These findings clearly underscore the good potential of iAorta to mitigate delayed or missed diagnoses of AAS in settings where non-contrast CT remains the unavoidable the initial or only imaging test. On the other hand, in settings where CTA is widely used, the application of iAorta does not imply the need to repeat or delay the contrast-enhanced CT scans. Many institutions (in Hong Kong SAR for example) adopt multiphase CT protocols, including both non-contrast and post-contrast phases, as the standard protocol for AAS evaluation.[41] Non-contrast CT is often performed prior to contrast administration to enhance the detection of IMH and other subtle radiological findings. In our prospective pilot deployment study, a standard server equipped with a single NVIDIA 3090 GPU was capable of completing computing inference for a non-contrast CT scan case within 40 seconds. In such settings with the multiphase CT protocols, iAorta's high specificity and high NPV on non-contrast CT could help avoid unnecessary contrast administration and reduce radiation exposure from post-contrast CT scans, particularly in low-risk patients (and possibly save healthcare cost).

Despite these remarkable results, there are several limitations of iAorta that are worth mentioning. First, our model was trained and validated on diverse cohorts within the Chinese population. Given the potential variability in aortic anatomy across different ethnicities, as well as differences in imaging equipment and acquisition protocols, further studies are needed to assess the model's generalizability across populations and scanning platforms. Second, the model's sensitivity for detecting PAU could be further improved. Integrating clinical symptoms and laboratory test results into a multi-modal deep learning model has the potential to enhance its detection capabilities. Third, more patients in the early stages of AAS seek initial care at EDs at non-tertiary centers or even community hospitals. We are planning to conduct another prospective, randomized, controlled trial extending to a broader range of medical centers and facilities to investigate the performance of iAorta in hospitals of various levels and its impact on enhancing the accuracy of the referral process. Also, through randomized controlled group comparisons, we can evaluate the time to definitive diagnosis and survival outcomes of AAS patients with and without iAorta deployment.

In conclusion, our findings demonstrate that iAorta is a highly valuable tool for critical emergency decision-making, facilitating significantly earlier and more accurate diagnosis of AAS through early warning, thereby enabling timely and appropriate patient treatment. Furthermore, iAorta may be extended to show good potential in detecting other life-threatening causes of chest pain beyond aortic conditions, such as non-ST-segment-elevation (NSTE)-ACS, pulmonary embolism (PE) and esophageal rupture[42]. Increased availability and improved scanning technology have led to drastic increases in CT utilization, particularly in the ED[43-45]. The integration of iAorta and its potential variations could offer valuable benefits into the clinical utility of non-contrast CT in the early and accurate diagnosis and management of high-risk acute chest pain patient cases. On the other hand, iAorta achieved excellent and robust specificity in several realistic and large-scale clinical settings. This showed that AI technology has the potential to make non-contrast CT as an effective rule-out tool and standardize the decision-making process to use advanced imaging means for life-threatening conditions, and to balance the risks of misdiagnosis and over testing. By leveraging AI technology to aid in the interpretation of CT scans, healthcare providers may be better equipped to promptly identify and address life-threatening causes of patients with acute chest pain, improve final patient outcomes and optimize the clinical decision-making in emergency department scenarios.

## Acknowledgments

We thank all the investigators and participants in this study. This study was supported by the Technical Innovation key project of Zhejiang Province (2024C03023) to H.Z.

## Contributors

Y.Hu, Y.Xiang, Y.Z., L.L., H.Z. designed the study. Y.Xiang, Yangyan He, D.Lang, S.Y., X.D., Y.Xu, Gaofeng Wang, J.H., W.Z., X.W., D.Li, Q.Z., Z.Li, C.Q., Z.W., Yunjun He, C.T., Y.Q., Z.Lin, X.Z., Yuan He, Z.Y., X.Z., R.F., R.C., X.L., Y.B. collected and organized data. W.G., J.X., J.Z., C.M., Z.L. carried out the data preprocessing. Y.Z., M.X. developed the AI model and the computational framework. Yangyan He, D.Lang, S.Y., C.D., L.H., W.X. analyzed and interpreted the data. W.G., C.S., Guofu Wang, L.W., H.Z. carried out the clinical deployment. Y.Hu, Y.Xiang, Z.D. carried out the statistical analysis. Y.Hu, Y.Xiang, Y.Z., M.X. wrote and revised the paper. Mannudeep K. Kalra, L.L., Z.H., M.X., H.Z. provided critical comments and reviewed the manuscript. All authors discussed the results and approved the final version before submission.

## Competing interests

Alibaba group has filed for patent protection (application numbers: CN 202311181343.8) on behalf of Y.Z., M.X. and L. Lu for the work related to the methods of detection of acute aortic syndrome on non-contrast CT. Y.Z., W.G., J.Z., C.M., Z.L., L. Lu and M.X. are employees of Alibaba Group and own Alibaba stock as part of the standard compensation package. All other authors have no competing interests.

## Methods

### Ethics approval

Data collection protocols and the use of images, radiological reports, and clinical information were approved by the ethical committee of each participating hospital. Informed consent was waived for retrospectively collected CT images and clinical information, and written informed consent was obtained from patients whose CT examinations, clinical information and follow-up records were prospectively collected. All procedures followed the tenets of the Declaration of Helsinki.

### Study design and study population

To train the model to learn the image features of AAS on non-contrast CT, we collected 3,350 consecutive patients who underwent aorta CTA scans, including both arterial and non-contrast phase CT series, at The First Affiliated Hospital of Zhejiang University School of Medicine (FAHZU; Zhejiang, China) between 2016 and 2020 as the internal training cohort. The detailed inclusion and exclusion criteria are presented in Supplementary Section 1.1. The internal training cohort consisted of 1,265 patients with AAS (296 with TAAD, 341 with TBAD, 321 with IMH and 307 with PAU) and 2,085 patients without AAS. The baseline demographic information and image characteristics of non-contrast phase CT series are summarized.

To evaluate the diagnostic performance of the model and its interpretability for radiologists, we collected data from 2,287 consecutive patients who underwent aorta CTA at FAHZU between 2021 and 2022 as the internal validation cohort under the same inclusion and exclusion criteria. The internal validation cohort consisted of 795 patients with AAS (188 with TAAD, 248 with TBAD, 203 with IMH and 156 with PAU) and 1,492 patients without AAS. We ensured that the internal validation dataset did not overlap with the internal training dataset, and the radiologists participating in the reader study had not previously reviewed the data from the internal validation dataset.

We enrolled seven independent multi-center cohorts for external validation to assess the generalizability and robustness of the developed model. The external multi-center validation cohorts were collected from seven centers across China: one top medical center in southern China (external validation cohort 1, Nanjing Drum Tower Hospital, NDTH, 3,287 patients), one top medical center in northern China (external validation cohort 2, Shandong Provincial Hospital Affiliated to Shandong First Medical University, SPH, 2,351 patients) and five regional medical centers in Zhejiang Province (external validation cohort 3, Taizhou Hospital of Zhejiang Province, TZH, 1,567 patients; external validation cohort 4, the First Affiliated Hospital of Wenzhou Medical University, FAHWMU, 4,574 patients; external validation cohort 5, Ningbo No. 2 Hospital, N2H, 2,369 patients; external validation cohort 6, Quzhou People's Hospital, QZH, 3,015 patients; external validation cohort 7, Shaoxing Central Hospital, SCH, 1,300 patients). The detailed inclusion and exclusion criteria were consistent with those described above, and the process of enrollment for each cohort is described in Supplementary Section 1.3 (as shown in Supplementary Fig. 4). The multi-center validation cohort, consisting of non-contrast CT scans of 6,495 patients with AAS (1,343 with TAAD, 1,905 with TBAD, 1,718 with IMH and 1,529 with PAU) and 11,968 patients without AAS, was used for independent validation when no model parameters were tuned or adjusted.

To further explore the model's robustness across various non-contrast CT protocols from real-world emergency, we performed a two-round, large-scale, real-world, retrospective study (RW1 and RW2) on three medical centers in Zhejiang Province, China. The original model and the upgraded model were evaluated in the RW1 and RW2 cohorts, respectively. Patients who were presented to ED with acute chest pain and underwent CT scans (for example, non-contrast chest CT, non-contrast abdominal CT, pulmonary CTA, coronary CTA, esophageal CT, lumbar CT or thoracic CT) covering part of the aortic region based on the initial suspicion of other acute diseases, were enrolled in this study. Note that some patients underwent multiple CT examinations with different protocols. We collected the non-contrast phase CT series images from the various CT examinations mentioned above. The detailed inclusion and exclusion criteria are described in Supplementary Section 4.1 (Supplementary Fig. 5-Fig. 9). The RW1 cohort comprised 23,094 non-contrast CT scans (44 AAS images, 23,050 non-AAS images) from 20,832 consecutive patients with acute chest pain at the FAHZU (32 AAS individuals, 20,800 non-AAS individuals). The RW2 cohort comprised 122,107 non-contrast CT scans (118 AAS images, 121,989 non-AAS images) of 116,693 consecutive individuals with acute chest pain at FAHZU, QPH and SCH (89 AAS individuals, 116,604 non-AAS individuals).

To further evaluate the implementation of iAorta in the realistic clinical settings, we conducted a two-stage, prospective and multi-center study. Patients who presented to the ED with acute chest pain symptoms and underwent CT scans covering part of the aortic region based on the initial suspicion of other acute diseases, were enrolled in this study. We collected the non-contrast phase series images from the various CT examinations. The detailed inclusion and exclusion criteria are described in Supplementary Section 5.1 (Supplementary Fig. 10-Fig. 11). (1) The comparative cohort comprised 14,436 non-contrast CT scans (14 AAS images, 14,422 non-AAS images) of 13,846 consecutive patients (11 AAS patients, 12,884 non-AAS patient) with acute chest pain at the FAHZU, QPH and SCH. (2) The deployment cohort contained 16,054 non-contrast CT scans (22 AAS images, 16,032 non-AAS images) of 15,584 consecutive patients (22 AAS patients, 15,562 non-AAS patients) with acute chest pain at the Shanghai Changhai Hospital, SHCH.

The CT images were retrieved from the Picture Archiving and Communication Systems (PACS) of each participating hospital and stored in Digital Imaging and Communications in Medicine (DICOM) format. For all aorta CTA scans in the training and validation datasets, both arterial and non-contrast phase CT series were included. The aorta CTA scan protocol started with a non-contrast CT scan from the thoracic inlet to the pubic symphysis, covering the entirety of the aorta. Thereafter, the arterial phase CT scan was performed over the same area as the systemic arterial phase scan. Due to the small-time gap between the two scans, the anatomical morphology depicted in the non-contrast phase series differed slightly from that observed in the arterial phase series taken "simultaneously". For the data in the clinical practicality study, non-contrast phase series were retrieved from various CT scans.

**Diagnostic criteria**

To evaluate the diagnostic accuracy of the AAS detection model, patients were initially categorized into non-AAS or AAS groups. Within the AAS category, patients were further classified into four subgroups based on the subtype of AAS as per the 2022 American Heart Association/American College of Cardiology guidelines[1]: TAAD, TBAD, IMH, and PAU. Patient cases where a patient presented with a PAU along with IMH were categorized under the IMH group due to the higher risk associated with this condition compared to isolated PAUs. This classification approach facilitated a more precise assessment of the model's ability to identify various emergent cases.

In the training and validation cohorts, corresponding arterial series CT images from the same aortic CTA were used as the gold standard to determine the presence of AAS. Each image was assigned a case-level diagnostic label, including the disease subtype, through a tiered annotation system. Moreover, two additional pixel-level segmentation labels were provided for the training datasets. The tiered annotation system consists of three cascades of trained radiologists at different levels for the verification and correction of CT image pixel-level labels. The details are described in Supplementary Section 1.4.

For both the retrospective real-world study and prospective multi-center study, patients' diagnoses were confirmed through the diagnostic data obtained during the ED visit and during the 14-day follow-up period. Details of the diagnosis standards are described in Supplementary Section 4.2 and 5.2, respectively.

**The development of DL model and iAorta system**

The DL model consists of two stages (Extended Data Fig. 2) and was trained using supervised deep learning. Given the input of a non-contrast CT scan, we localized the aorta at the coarse stage and then detected possible AAS at the fine stage. The output of model consists of three components, that is, the classification of the potential AAS with probabilities, the segmentation mask of the aorta wall and true lumen, and the activation result representing the potential lesion region in each CT slice.

The aim of the coarse processing stage is to localize the aorta. In this stage, we trained a lightweight nnU-Net[46-47] to segment the whole aorta from the input non-contrast CT scan. The aim of the fine processing stage was to detect the potential AAS lesion. The network is based on a multi-task learning strategy[48]. Specifically, we trained a joint segmentation and classification network to simultaneously segment the aortic wall and true lumen and classify the patient-level abnormalities, that is, AAS or non-AAS. The network output consists of the probabilities of AAS, the segmentation mask of the aorta wall and true lumen, and the activation map result indicating the potential lesion region for enhanced reader interpretability. The activation map result can indicate the potential lesion region spatially for enhanced slice-level and patient-level interpretability. Moreover, the network is supervised by a combination of focal loss[49], Dice loss[50] and voxel-wise cross-entropy loss[51]. More details on the training and inference of two stages are given in Supplementary Section 2.2-2.3. After the RW1 analysis, we collected both false-positive and false-negative non-contrast CT data from the RW1 cohort. This process aligns with machine learning strategies referred to as hard example mining[52] and incremental learning[53]. The evolved model was subsequently subjected to evaluation on the RW2 cohort. Details regarding the collection and annotation of these updated training datasets, along with the fine-tuning protocol, are described in Supplementary Section 4.4 (Supplementary Fig. 3).

We have developed a browser-server collaborative early warning system designed for rapid and accurate detection of AAS in two prospective real-world clinical settings. By integrating a non-contrast CT phase selection/identification tool, we developed the model and a warning popup plugin, so that the system can process and analyze large amounts of patient data in real-time, enabling radiologists to get timely alerts and make informed diagnosis and intervention. More details are given in Supplementary Section 2.7. Notice that we have provided a complimentary access to the browser interface of our iAorta system. (https://iaorta.medofmind.com/, Extended data Fig. 5). The browser interface provides an open-access CT image database containing typical cases, which may be a useful resource for training radiologists as well as familiarizing researchers in the field of analyzing aortic diseases and AI-assisted medical imaging.

**Model and system evaluation**

After training the DL model, we conducted a multi-center validation study, a reader study and a large-scale real-world retrospective study to evaluate the model's performance and its potential benefit for the current existing clinical workflows. After the real-world retrospective study, the DL model is developed and upgraded to iAorta, a browser-server collaborative early warning system, to enhance its application in real-world emergency clinical settings. We then conducted a prospective multi-center study to evaluate the implementation of iAorta in the realistic clinical setting.

**Stage I: Multi-center model validation.** We used eight independent multi-center validation patient cohorts to evaluate the model performance. Our model is formulated as a two-class classification task to distinguish AAS versus non-AAS. Having a AAS is defined as the "positive" class for calculation of the AUC, sensitivity, specificity, PPV, NPV, accuracy and F1-score. In addition, we calculated the diagnosis sensitivity of model in cases of four AAS subtypes (including TAAD, TBAD, IMH and PAU) separately because the disease progression and untreated mortality differed significantly among the four subtypes.

**Stage II: Reader study on the non-contrast CT.** We recruited a total of eleven radiologists with three levels of clinical specialty experience (specialty experts, board-certified general radiologists, and medical trainees) to compare the model performance with that of radiologists. This test was conducted in two phases or periods using non-contrast CT images from the internal validation dataset. Phase 1 took place over a 16-week period (January 6, 2023, to March 28, 2023) and was designed to evaluate the performance of the radiologists alone. During Phase 1, eleven radiologists interpreted scrambled non-contrast CT images on the edge-cloud collaborative platform mentioned above, while the widget displaying the AI results was intentionally set to be inactive. Therefore, the radiologists were blinded to the AI results in Phase 1. Following a two-month desensitization period, participating radiologists previously interpreting non-contrast CT images were retained and trained to use our model on the online platform with the widget activated, but they were not informed of the detection accuracy of our model. Phase 2 spanned a 16-week period (July 7, 2023, to October 10, 2023) and was designed to evaluate the performance of the clinicians assisted with AI image interpretation tool. The AI results were displayed in a visual interface concurrently during the image diagnostic evaluation. The details are provided in Supplementary Section 3.

**Stage III: Large-scale, real-world, retrospective study.** To explore the robustness across various non-contrast CT protocols in real-world emergency scenarios, we retrospectively collected four patient cohorts with acute chest pain symptoms presenting to the emergency departments in three hospitals and conducted two rounds of real-world evaluation. (RW1 and RW2). First, the model was evaluated on RW1. Then, after analyzing the reasons for bad cases based on the evaluation results, we incorporated the false positive and false negative cases from RW1 and upgraded our model by incremental learning techniques. Finally, the upgraded model was evaluated on RW2. At the same time, the original model was also evaluated on RW2 as a control group. For AAS, diagnosis within a short period is crucial to improve its prognosis, we evaluated the reduction of the risk of missed diagnoses and misdiagnoses, which serves as a critical reference for the potential benefit of our model.

**Stage IV: Prospective multi-center study.** To further evaluate the feasibility of the iAorta in realistic real-world clinical settings, we prospectively conducted both a comparative study and a pilot deployment study. We developed a real-time warning system, iAorta, which is integrated into the existing clinical workflow on PACS.

**The prospective comparative study** is a prospective, observational study. We recruited two groups of radiologists to explore the potential improvement in diagnostic performance and the reduction in final definitive diagnosis time when iAorta is integrated into the existing clinical workflow, as compared to the original clinical workflow. Each group included board-certified general radiologists per shift for the initial report and specialty experts for the final reviewing report. Group A independently reviewed the patients' CT scans in a sequential manner under the current clinical workflow using the local PACS and provided the final reviewing report to guide subsequent patient management. This process generates two independent diagnostic records: one for the IR radiologist's diagnosis and the other for the RR radiologist's diagnosis, for each patient CT scan. In the meanwhile, Group B reviewed the images with the assistance of a locally-deployed iAorta system. The images are automatically fed into iAorta system, which then produces AI-based results. If the AI detects any abnormalities, both the initial reporting (IR) radiologists and the report reviewing (RR) radiologists receive subsequential pop-up message alerts, prompting them to prioritize the review of the AI-flagged (AAS positive) patient scan. After reviewing the AI results, the AAS-related diagnosis is recorded. This process yields three independent diagnostic records for each image: the IR radiologist's diagnosis (aided with AI), the RR radiologist's diagnosis (aided with AI), and the AI-computed results. Additionally, we recorded the patients' arrival time to the hospital, the time of initial diagnosis and final definitive diagnosis, the time-to-results for laboratory testing and the time of ordering the CT examination, performing the CT scan, the initial report, and the final report. Here, we defined the time to the correct diagnostic pathway for AAS patients as the period from the patient's arrival in the ED until AAS was mentioned as a positive possibility in their non-contrast CT imaging report via AI associated with a recommendation for the aortic CTA examination, or the point at which they were advised to undergo conclusive diagnostic imaging for other reasons by clinicians.

**The prospective pilot deployment study** is an ongoing, single-arm, prospective study (12/20/2024 – 02/28/2025, ongoing)**.** We integrated iAorta system into the current pilot clinical daily routine and utilized AI-assisted or AI-alerted electronic imaging reports to guide subsequent patient management. We also recorded the patients' arrival time to the hospital, the time of initial diagnosis and final definitive diagnosis, the time-to-results for laboratory testing and the time of ordering the CT examination, performing the CT scan, the initial report, and the final report. We analyzed the time from the initial emergency department presentation to definitive diagnosis.

**Statistical analysis**

Continuous variables are presented as the means with standard deviations (SDs), while frequencies and percentages are noted to summarize classification variables. The accuracy, sensitivity, specificity, positive predictive value (PPV), and negative predictive value (NPV) of iAorta and radiologists for the detection of AAS were evaluated by calculating the 95% confidence intervals (CIs) using the Clopper-Pearson method. We used the receiver operating characteristic (ROC) curves as the main quantitative measure to demonstrate and evaluate the performance of the deep learning algorithm to discriminate patients with AAS from normal controls. ROC curves were generated by plotting the proportion of true positive cases (sensitivity) against the proportion of false positive cases (1-specificity) by varying the predictive probability threshold. A larger area under the ROC curve (AUC) indicated better diagnostic performance. The significance test comparing the AUCs is conducted using the Delong test. All the statistical tests were two-sided with a statistical significance level of 0.05. We used the scikit-learn package (version 1.0.2) to compute the evaluation metrics.

**Reporting summary**

Further information on the research design is available in the Nature Portfolio Reporting Summary linked to this article.

**Data availability**

The sample data and an interactive demonstration are provided at https://iaorta.medofmind.com/. The remaining datasets used in this study are currently not permitted for public release by the respective institutional review boards (IRBs). Data requests pertaining to the study may be made to the first author (Yujian Hu; huyujian@zju.edu.cn). Access will be granted subject to IRB approval, a signed data-use agreement and will be for noncommercial academic purposes only. All data provided were anonymized to protect the privacy of the patients who participated in the studies, in line with applicable laws and regulations. Requests will be processed within 6 weeks.

**Code availability**

The code used for the implementation of our model has dependencies on internal tooling and infrastructure, is under patent protection (application number: CN 202311181343.8), and thus cannot be publicly released. All experiments and implementation details are described in sufficient detail in the Methods and Supplementary Information (Details of model development) sections to support replication with non-proprietary libraries. Several major components of our work are available in open-source repositories: PyTorch (https://pytorch.org/) and nnU-NetV1 (https://github.com/MIC-DKFZ/nnUNet/tree/nnunetv1).

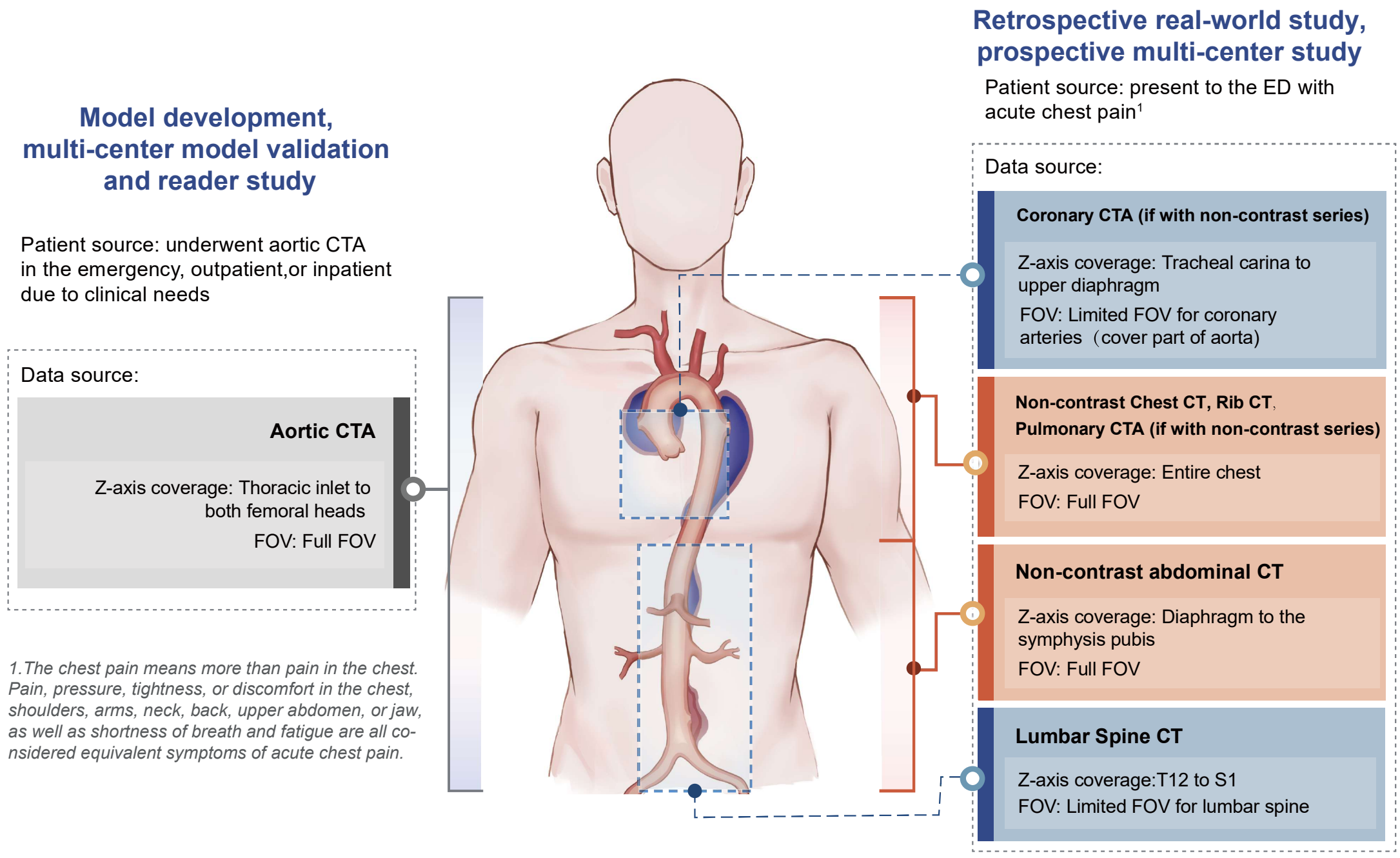

**Extended Data Fig. 1 | Patient and data sources of model development, multi-center model validation (stage I), reader study (stage II), large-scale real-world retrospective study (stage III), and prospective multi-center study (stage IV).** Data source includes different field of views (FOVs) and z-axis coverage in aortic CTA, coronary CTA, non-contrast chest CT, pulmonary CTA, rib CT, non-contrast abdominal CT, and lumbar spine CT.

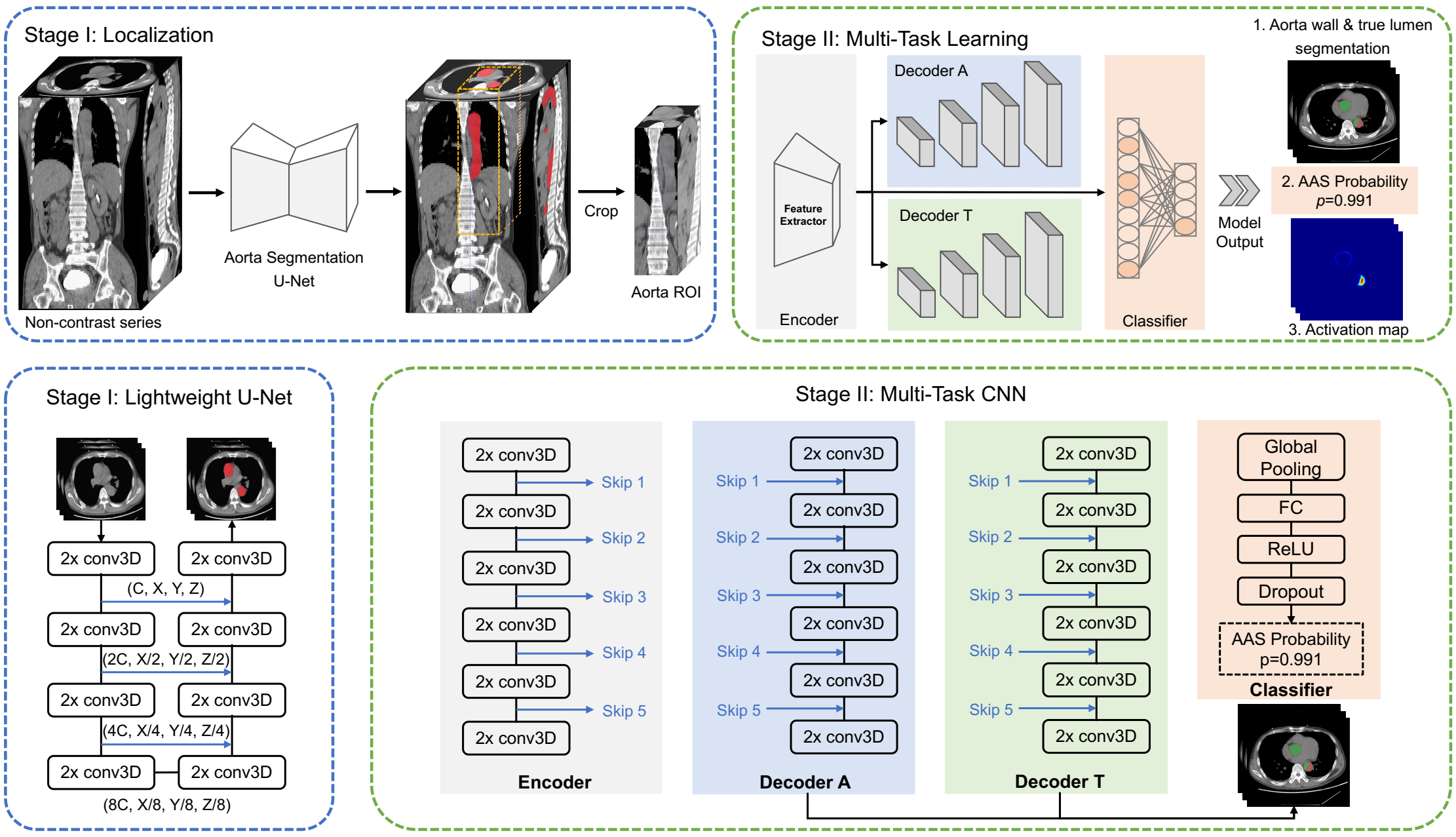

**Extended Data Fig. 2 | Network architecture.** (Top) Overview. Our deep learning framework consists of two stages: aorta localization using a lightweight U-Net, and abnormality detection using a multi-task CNN. (Bottom) Architectures of the lightweight U-Net and multi-task CNN. The features extracted from encoder are used for abnormal and normal classification. Decoder A and T are designed for the segmentation of aorta and true lumen, respectively.

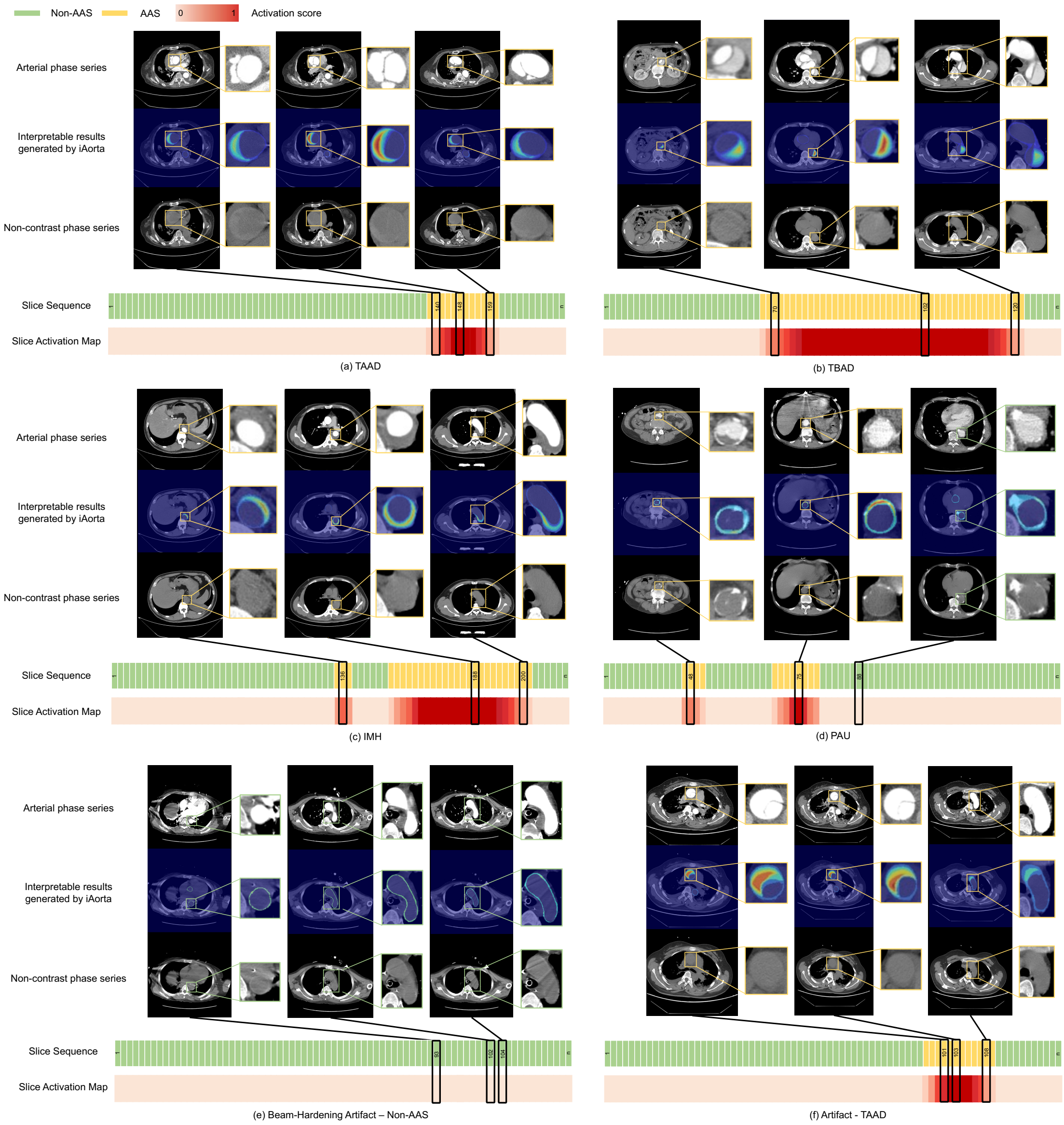

**Extended Data Fig. 3 | Interpretable results of iAorta in four subtypes of AAS and two artifact cases.** The two colors in the slice sequence (arranged by spatial location) represent two classes of slice-level prediction including AAS (yellow) and non-AAS (green). The slice activation map indicates the voxel-level localization of AAS. AAS, acute aortic syndrome; TAAD, Stanford Type A dissection; TBAD, Stanford Type B dissection; IMH, intramural hematoma; PAU, penetrating atherosclerotic ulcer.

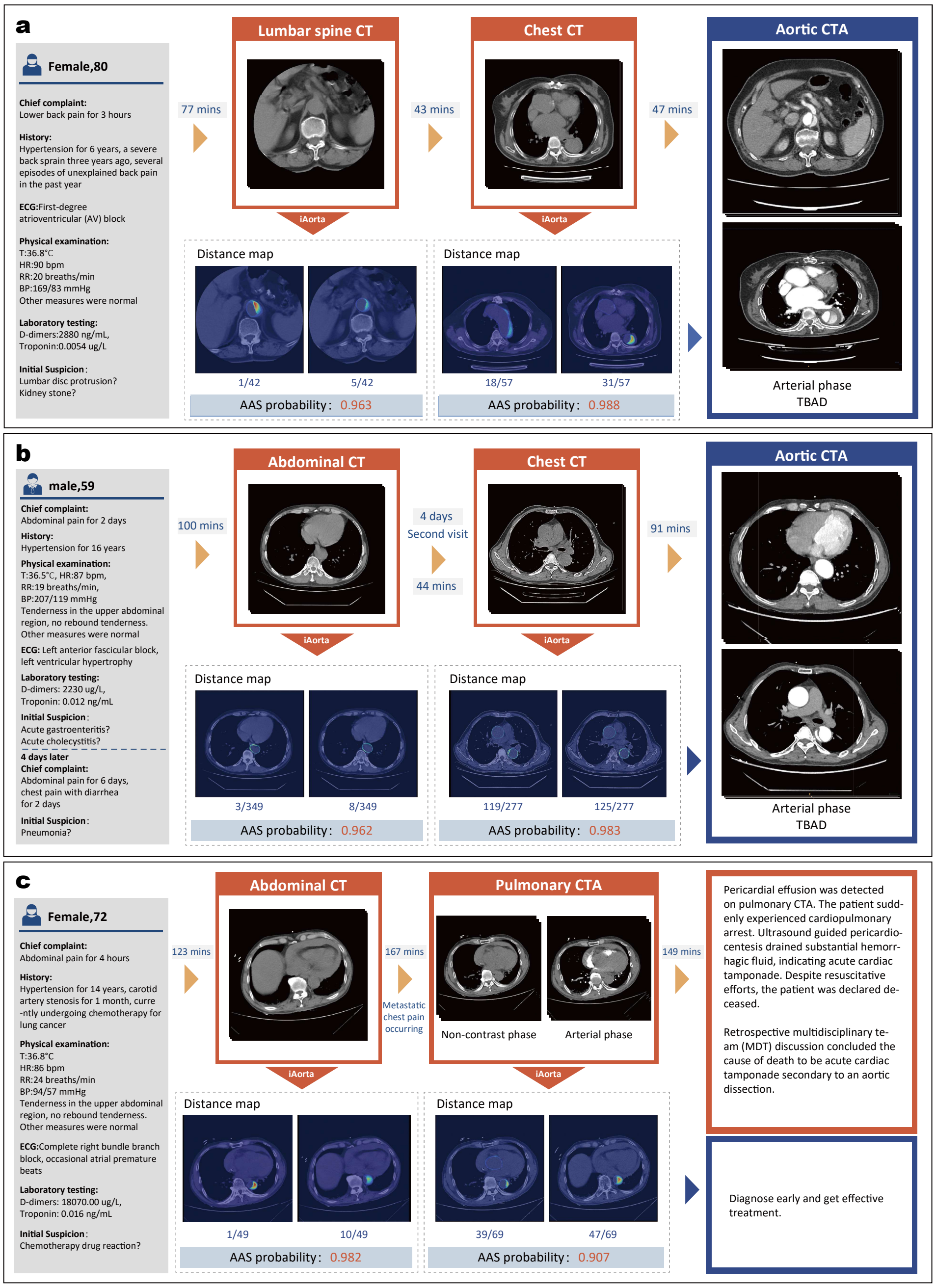

**Extended Data Fig. 4 | Flowchart illustrating the potential benefits for patients with Acute Aortic Syndrome (AAS) who were initially suspected to have other acute diseases in the retrospective real-world study (stage III), compared to the current clinical workflow. a.** A patient with lower back pain was initially suspected to have lumbar disc protrusion and underwent clinical investigations, including lumbar spine CT and non-contrast chest CT. **b.** A patient with abdominal pain was initially suspected to have acute gastroenteritis, underwent clinical investigations including a non-contrast abdominal CT, and was discharged from the ED. Four days later, he returned to the ED with worsening symptoms and underwent a non-contrast chest CT. Our model could have detected AAS during the first visit. **c.** A patient with abdominal pain was suspected to have a chemotherapy drug reaction and underwent clinical investigations, including non-contrast abdominal CT and pulmonary CTA. Before a definitive diagnosis, the patient experienced cardiopulmonary arrest. iAorta system could potentially save the patient's life.

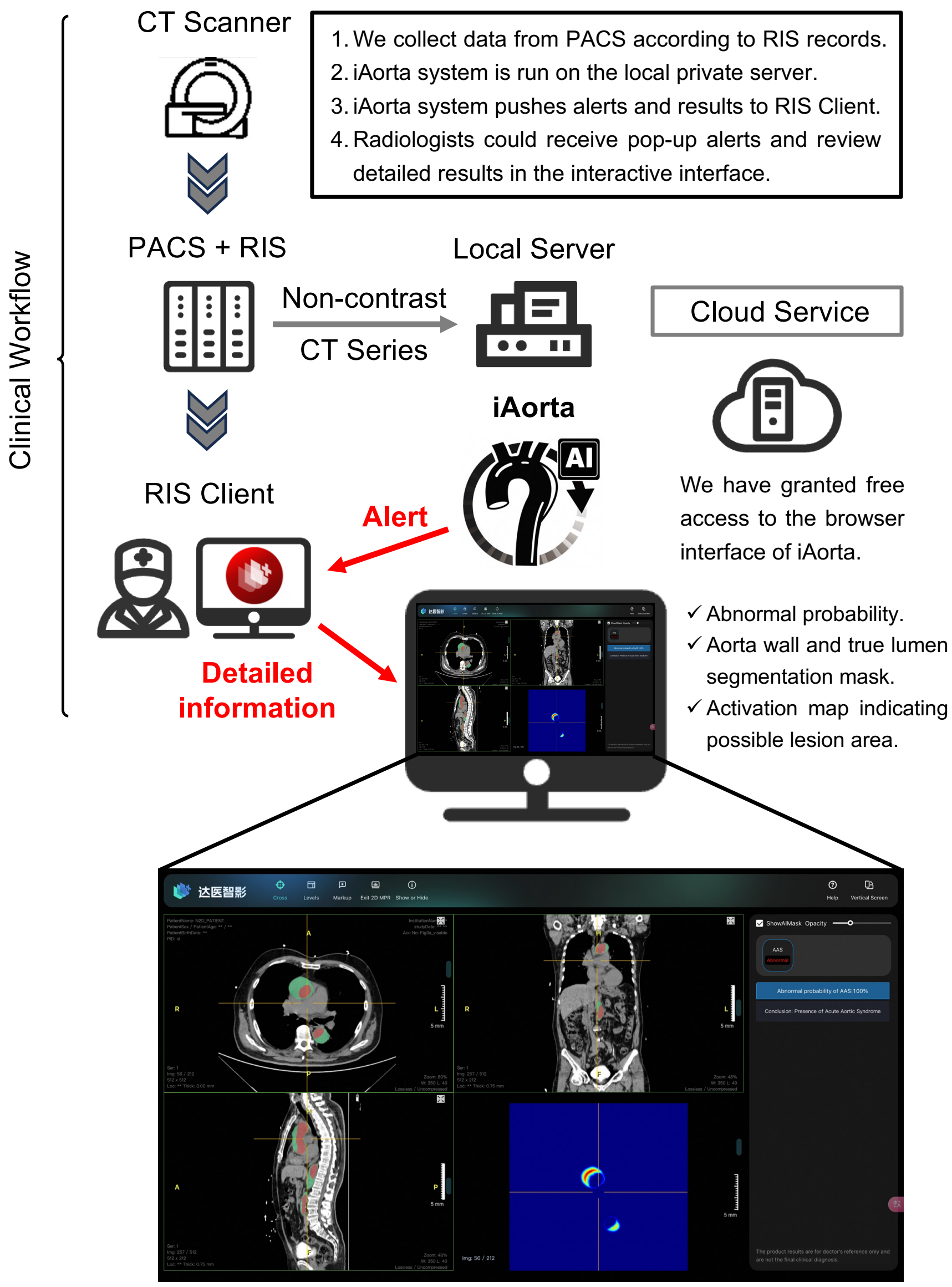

**Extended Data Fig. 5 | Flowchart describing the process of the seamless integration of iAorta into the existing clinical workflow.** PACS, picture archiving and communication system; RIS, radiology information system.

| | Model Training | Multi-center model validation | | | | | | | | Large-scale real-world retrospective study | | | |
|---|---|---|---|---|---|---|---|---|---|---|---|---|---|
| | Training dataset FAHZU | Internal cohort FAHZU | External cohort 1 NDTH | External cohort 2 SPH | External cohort 3 TZH | External cohort 4 FAHWMU | External cohort 5 N2H | External cohort 6 QPH | External cohort 7 SCH | RW1 FAHZU | RW2 FAHZU | RW2 SCH | RW2 QPH |
| **Number of participants** | 3,350 | 2,287 | 3,287 | 2,351 | 1,567 | 4,574 | 2,369 | 3,015 | 1,300 | 20,832 | 72,360 | 23,804 | 20,529 |
| **Patient Characteristics** | | | | | | | | | | | | | |
| Male, no. (%) | 2,414 (72.1%) | 1,636 (71.5%) | 2,231 (67.9%) | 1,699 (72.3%) | 1,086 (69.3%) | 3,229 (70.6%) | 1,647 (69.5%) | 2,088 (69.3%) | 913 (70.2%) | 10,963 (52.6%) | 37,068 (51.2%) | 13,679 (59.3%) | 11,403 (55.5%) |
| Age, years (SD) | 58.49±15.66 | 58.58±16.07 | 59.69±14.89 | 63.12±15.37 | 61.71±15.46 | 61.96±14.76 | 59.29±15.48 | 61.43±15.18 | 53.84±16.38 | 43.38±18.40 | 43.46±18.79 | 45.70±18.95 | 49.95±20.66 |
| **Number of images** | 3,350 | 2,287 | 3,287 | 2,351 | 1,567 | 4,574 | 2,369 | 3,015 | 1,300 | 23,094 | 76,582 | 24,365 | 21,160 |
| **Data source, no. (%)** | | | | | | | | | | | | | |
| Aortic CTA | 3,350 (100%) | 2,287 (100%) | 3,287 (100%) | 2,351 (100%) | 1,567 (100%) | 4,574 (100%) | 2,369 (100%) | 3,015 (100%) | 1,300 (100%) | 0 (0%) | 0 (0%) | 0 (0%) | 0 (0%) |
| Routine non-contrast chest CT | 0 (0%) | 0 (0%) | 0 (0%) | 0 (0%) | 0 (0%) | 0 (0%) | 0 (0%) | 0 (0%) | 0 (0%) | 6,354 (27.5%) | 24,115 (31.5%) | 10,151 (41.7%) | 8,157 (38.5%) |
| Routine non-contrast abdominal CT | 0 (0%) | 0 (0%) | 0 (0%) | 0 (0%) | 0 (0%) | 0 (0%) | 0 (0%) | 0 (0%) | 0 (0%) | 15,756 (68.2%) | 48,957 (63.9%) | 13,163 (54.0%) | 12,141 (57.4%) |
| Others | 0 (0%) | 0 (0%) | 0 (0%) | 0 (0%) | 0 (0%) | 0 (0%) | 0 (0%) | 0 (0%) | 0 (0%) | 984 (4.3%) | 3,510 (4.6%) | 1,051 (4.3%) | 862 (4.1%) |
| **Classes, no. (%)** | | | | | | | | | | | | | |
| Acute Aortic Syndrome(AAS) | 1,265 (37.8%) | 795 (34.8%) | 1,296 (39.4%) | 980 (41.7%) | 563 (35.9%) | 1,918 (41.9%) | 591 (24.9%) | 850 (28.2%) | 297 (22.8%) | 44 (0.2%) | 69 (0.1%) | 26 (0.107%) | 23 (0.11%) |
| TAAD | 296 (8.8%) | 188 (8.2%) | 234 (7.1%) | 242 (10.3%) | 104 (6.6%) | 413 (9.0%) | 107 (4.5%) | 172 (5.7%) | 71 (5.4%) | 15 (0.06%) | 15 (0.02%) | 9 (0.037%) | 3 (0.01%) |
| TBAD | 341 (10.2%) | 248 (10.8%) | 467 (14.2%) | 254 (10.8%) | 128 (8.2%) | 586 (12.8%) | 153 (6.5%) | 216 (7.2%) | 101 (7.8%) | 11 (0.05%) | 27 (0.04%) | 10 (0.041%) | 10 (0.05%) |
| IMH | 321 (9.6%) | 203 (8.9%) | 335 (10.2%) | 249 (10.6%) | 165 (10.5%) | 485 (10.6%) | 171 (7.2%) | 247 (8.2%) | 66 (5.1%) | 7 (0.04%) | 16 (0.03%) | 6 (0.025%) | 6 (0.03%) |
| PAU | 307 (9.2%) | 156 (6.9%) | 260 (7.9%) | 235 (10.0%) | 166 (10.6%) | 434 (9.5%) | 160 (6.7%) | 215 (7.1%) | 59 (4.5%) | 11 (0.05%) | 11 (0.01%) | 1 (0.004%) | 4 (0.02%) |
| Non-AAS | 2,085 (62.2%) | 1,492 (65.2%) | 1,991 (60.6%) | 1,371 (58.3%) | 1,004 (64.1%) | 2,656 (58.1%) | 1,778 (75.1%) | 2,165 (71.8%) | 1,003 (77.2%) | 23,050 (99.8%) | 76,513 (99.9%) | 24,339 (99.893%) | 21,137 (99.89%) |
| **CT characteristics** | | | | | | | | | | | | | |
| Non-ECG-Gated, no.(%) | 3,350 (100%) | 2,287 (100%) | 3,287 (100%) | 2,351 (100%) | 1,567 (100%) | 4,183 (91.5%) | 2,369 (100%) | 2,990 (99.2%) | 1,300 (100%) | 22,960 (99.4%) | 76,200 (99.5%) | 24,222 (99.4%) | 21,017 (99.3%) |
| Pixel Spacing, mm.(SD) | 0.729±0.059 | 0.762±0.082 | 0.789±0.085 | 0.747±0.074 | 0.696±0.083 | 0.755±0.069 | 0.773±0.046 | 0.783±0.052 | 0.732±0.103 | 0.730±0.079 | 0.728±0.077 | 0.726±0.073 | 0.729±0.074 |
| Slice Thickness, mm.(SD) | 4.925±0.377 | 3.929±1.067 | 1.113±0.245 | 2.279±1.890 | 3.954±0.994 | 3.856±0.835 | 3.486±0.919 | 4.821±0.860 | 3.628±1.151 | 3.798±1.818 | 3.712±1.798 | 3.184±1.900 | 3.351±1.896 |
| Peak Tube Voltage Range, kVp.(SD) | 111.85±9.92 | 109.62±10.28 | 117.21±9.84 | 105.50±13.43 | 116.43±8.67 | 108.53±12.65 | 103.21±9.44 | 112.89±7.11 | 113.47±10.48 | 120.05±1.32 | 120.05±1.26 | 120.03±1.16 | 120.06±1.09 |
| Tube Current, mA.(SD) | 294.23±89.96 | 333.58±76.17 | 262.13±151.56 | 343.08±159.44 | 315.36±98.63 | 309.86±129.36 | 288.37±84.91 | 299.68±115.98 | 319.29±42.65 | 301.13±83.51 | 294.55±86.02 | 281.46±90.59 | 287.11±87.72 |
| **CT Manufactures, no. (%)** | | | | | | | | | | | | | |
| GE Medical Systems | 1,564 (46.7%) | 1,161 (50.8%) | 768 (23.4%) | 0 (0%) | 675 (43.1%) | 0 (0%) | 77 (3.3%) | 10 (0.4%) | 268 (20.6%) | 9,653 (41.8%) | 33,302 (43.5%) | 4,824 (19.8%) | 693 (3.3%) |
| Siemens | 0 (0%) | 25 (1.1%) | 232 (7.1%) | 1,989 (84.6%) | 0 (0%) | 1,369 (29.9%) | 2,003 (84.6%) | 566 (18.8%) | 3 (0.2%) | 0 (0%) | 3,807 (5.0%) | 781 (3.2%) | 3,057 (14.5%) |
| Philips | 1,786 (53.3%) | 1,084 (47.4%) | 1,028 (31.3%) | 0 (0%) | 857 (54.7%) | 0 (0%) | 0 (0%) | 94 (3.1%) | 140 (10.8%) | 13,441 (58.2%) | 36,031 (47.0%) | 2,509 (10.3%) | 1,602 (7.6%) |
| Toshiba | 0 (0%) | 0 (0%) | 0 (0%) | 362 (15.4%) | 35 (2.2%) | 3,205 (70.1%) | 0 (0%) | 2,341 (77.6%) | 889 (68.4%) | 0 (0%) | 0 (0%) | 16,251 (66.7%) | 15,387 (72.7%) |
| UIH | 0 (0%) | 17 (0.7%) | 1,259 (38.3%) | 0 (0%) | 0 (0%) | 0 (0%) | 289 (12.1%) | 4 (0.1%) | 0 (0%) | 0 (0%) | 3,442 (4.5%) | 0 (0%) | 421 (1.9%) |

| | Prospective multi-center study | |
|---|---|---|
| | Comparative study FAHZU, SCH, QPH | Pilot deployment study SHCH |
| **Number of participants** | 13,846 | 15,584 |
| **Patient Characteristics** | | |
| Male, no. (%) | 7,353 (53.1%) | 9,069 (58.2%) |
| Age, years (SD) | 45.05±18.51 | 48.65±19.58 |
| **Clinical presentation, no. (%)** | | |
| Chest pain | 3,240 (23.4%) | 4,862 (31.2%) |
| Back pain | 2,105 (15.2%) | 2,899 (18.6%) |
| Abdominal pain | 8,418 (60.8%) | 8,384 (53.8%) |
| Lumbar pain | 1,786 (12.9%) | 1,621 (10.4%) |
| Shortness of breath | 2,575 (18.6%) | 3,537 (22.7%) |
| **Number of images** | 14,436 | 16,054 |
| **Data source, no. (%)** | | |
| Aortic CTA | 0 (0%) | 0 (0%) |
| Routine non-contrast chest CT | 4,361 (30.2%) | 6,148 (38.3%) |
| Routine non-contrast abdominal CT | 9,213 (63.8%) | 9,295 (57.9%) |
| Others | 862 (6.0%) | 611 (3.8%) |
| **Classes, no. (%)** | | |
| Acute Aortic Syndrome(AAS) | 14 (0.1%) | 22 (0.14%) |
| TAAD | 4 (0.03%) | 5 (0.03%) |
| TBAD | 4 (0.03%) | 9 (0.06%) |
| IMH | 4 (0.03%) | 5 (0.03%) |
| PAU | 2 (0.01%) | 3 (0.02%) |
| Non-AAS | 14,422 (99.9%) | 16,032 (99.86%) |
| **CT characteristics** | | |
| Non-ECG-Gated, no.(%) | 14,307 (99.1%) | 16,054 (100%) |
| Pixel Spacing, mm.(SD) | 0.729±0.069 | 0.766±0.080 |
| Slice Thickness, mm.(SD) | 3.423.±1.801 | 4.970±0.264 |
| Peak Tube Voltage Range, kVp.(SD) | 120.3±1.22 | 120.8±3.91 |
| Tube Current, mA.(SD) | 291.25±87.94 | 351.06±45.13 |
| **CT Manufactures,no.(%)** | | |
| GE Medical Systems | 3,904 (27.0%) | 0 (0%) |
| Siemens | 928 (6.4%) | 0 (0%) |
| Philips | 2,829 (19.6%) | 16,054 (100%) |
| Toshiba | 6,738 (46.7%) | 0 (0%) |
| UIH | 37 (0.3%) | 0 (0%) |

**Extended Data Table 1 | Cohort characteristics.** Retrospective cohort characteristics (left). Prospective cohort characteristics (right). Data are the mean±s.d. or number of individuals or scans (%). FAHZU, the First Affiliated Hospital of Zhejiang University School of Medicine; NDTH, Nanjing Drum Tower Hospital; SPH, Shandong Provincial Hospital Affiliated to Shandong First Medical University; TZH, Taizhou Hospital of Zhejiang Province; FAHWMU, the First Affiliated Hospital of Wenzhou Medical University; N2H, Ningbo No.2 Hospital; QPH, Quzhou People's Hospital; SCH, Shaoxing Central Hospital; SHCH, Shanghai Changhai Hospital; RW, real-world emergency scenario cohort; TAAD, Stanford Type A dissection; TBAD, Stanford Type B dissection; IMH, intramural hematoma; PAU, penetrating atherosclerotic ulcer; UIH, United Imaging Healthcare.

| | Internal validation cohort | External validation cohort 1 | External validation cohort 2 | External validation cohort 3 | External validation cohort 4 | External validation cohort 5 | External validation cohort 6 | External validation cohort 7 |
|---|---|---|---|---|---|---|---|---|
| Sensitivity (95% CI) | 0.984 (0.972-0.990) | 0.975 (0.964-0.982) | 0.973 (0.961-0.982) | 0.954 (0.933-0.968) | 0.970 (0.961-0.977) | 0.954 (0.934-0.968) | 0.961 (0.946-0.972) | 0.963 (0.935-0.979) |
| TAAD | 0.995 (0.970-0.999) | 0.987 (0.963-0.997) | 0.988 (0.964-0.996) | 0.971 (0.919-0.990) | 0.981 (0.962-0.990) | 0.972 (0.921-0.990) | 0.988 (0.959-0.997) | 0.972 (0.903-0.992) |
| TBAD | 0.996 (0.978-0.999) | 0.991 (0.978-0.997) | 0.992 (0.972-0.998) | 0.984 (0.945-0.996) | 0.990 (0.978-0.995) | 0.980 (0.944-0.993) | 0.981 (0.953-0.993) | 0.980 (0.931-0.995) |
| IMH | 0.980 (0.950-0.992) | 0.973 (0.950-0.986) | 0.980 (0.954-0.991) | 0.958 (0.915-0.979) | 0.977 (0.960-0.987) | 0.953 (0.910-0.976) | 0.968 (0.937-0.983) | 0.955 (0.875-0.984) |
| PAU | 0.955 (0.910-0.978) | 0.935 (0.898-0.959) | 0.932 (0.892-0.958) | 0.916 (0.863-0.949) | 0.924 (0.895-0.945) | 0.919 (0.866-0.952) | 0.912 (0.866-0.943) | 0.932 (0.838-0.973) |
| Specificity (95% CI) | 0.948 (0.935-0.958) | 0.937 (0.925-0.947) | 0.943 (0.930-0.954) | 0.935 (0.918-0.949) | 0.941 (0.931-0.949) | 0.929 (0.916-0.940) | 0.946 (0.936-0.955) | 0.932 (0.915-0.946) |
| Accuracy (95% CI) | 0.960 (0.951-0.967) | 0.952 (0.944-0.958) | 0.956 (0.947-0.963) | 0.942 (0.929-0.952) | 0.953 (0.946-0.959) | 0.935 (0.924-0.944) | 0.950 (0.942-0.958) | 0.939 (0.925-0.951) |
| AUC (95% CI) | 0.980 (0.973-0.987) | 0.972 (0.962-0.982) | 0.966 (0.954-0.976) | 0.945 (0.931-0.958) | 0.955 (0.946-0.964) | 0.941 (0.930-0.951) | 0.954 (0.945-0.963) | 0.948 (0.934-0.962) |
| PPV (95% CI) | 0.909 (0.888-0.927) | 0.909 (0.893-0.923) | 0.924 (0.907-0.940) | 0.892 (0.865-0.914) | 0.922 (0.909-0.933) | 0.816 (0.786-0.843) | 0.876 (0.853-0.895) | 0.808 (0.764-0.846) |
| NPV (95% CI) | 0.991 (0.984-0.995) | 0.983 (0.976-0.988) | 0.980 (0.971-0.987) | 0.973 (0.961-0.982) | 0.977 (0.971-0.982) | 0.984 (0.977-0.989) | 0.984 (0.978-0.989) | 0.988 (0.979-0.993) |
| F1-score (95% CI) | 0.944 (0.934-0.955) | 0.941 (0.932-0.949) | 0.948 (0.939-0.957) | 0.922 (0.907-0.937) | 0.945 (0.938-0.952) | 0.880 (0.864-0.896) | 0.916 (0.904-0.929) | 0.879 (0.856-0.901) |

**Extended Data Table 2 | Results of multi-center model validation.**

AUC, area under the curve; PPV, positive predictive value; NPV, negative predictive value; TAAD, Stanford Type A dissection; TBAD, Stanford Type B dissection; IMH, intramural hematoma; PAU, penetrating atherosclerotic ulcer.

| | Sensitivity (95% CI) | Specificity (95% CI) | Accuracy (95% CI) | PPV (95% CI) | NPV (95% CI) |
|---|---|---|---|---|---|
| Model Alone | 0.984 (0.972-0.990) | 0.948 (0.935-0.958) | 0.960 (0.951-0.967) | 0.909 (0.888-0.927) | 0.991 (0.984-0.995) |
| Radiologists Alone | | | | | |
| Trainee A | 0.416 (0.383-0.451) | 0.801 (0.780-0.820) | 0.667 (0.648-0.686) | 0.527 (0.488-0.566) | 0.720 (0.698-0.741) |
| Trainee B | 0.508 (0.473-0.543) | 0.741 (0.718-0.762) | 0.660 (0.640-0.679) | 0.511 (0.476-0.545) | 0.739 (0.716-0.760) |
| Trainee C | 0.442 (0.407-0.476) | 0.796 (0.774-0.815) | 0.672 (0.653-0.691) | 0.535 (0.497-0.573) | 0.728 (0.706-0.749) |
| Trainee D | 0.338 (0.306-0.372) | 0.932 (0.918-0.944) | 0.726 (0.707-0.744) | 0.727 (0.679-0.770) | 0.726 (0.705-0.745) |
| Trainee Mean | 0.426 (0.409-0.443) | 0.817 (0.807-0.827) | 0.681 (0.672-0.691) | 0.554 (0.534-0.574) | 0.728 (0.717-0.738) |
| Board-certificated A | 0.585 (0.550-0.619) | 0.842 (0.823-0.860) | 0.753 (0.735-0.770) | 0.664 (0.628-0.698) | 0.792 (0.771-0.811) |
| Board-certificated B | 0.652 (0.618-0.684) | 0.800 (0.779-0.819) | 0.748 (0.730-0.766) | 0.634 (0.600-0.666) | 0.812 (0.791-0.831) |
| Board-certificated C | 0.626 (0.592-0.659) | 0.832 (0.812-0.850) | 0.760 (0.742-0.777) | 0.665 (0.630-0.698) | 0.807 (0.786-0.826) |
| Board-certificated D | 0.604 (0.569-0.637) | 0.887 (0.870-0.902) | 0.789 (0.772-0.805) | 0.741 (0.706-0.773) | 0.808 (0.788-0.826) |
| Board-certificated Mean | 0.617 (0.600-0.633) | 0.840 (0.831-0.849) | 0.763 (0.754-0.771) | 0.673 (0.656-0.690) | 0.804 (0.794-0.814) |
| Expert A | 0.790 (0.760-0.817) | 0.912 (0.896-0.925) | 0.869 (0.855-0.882) | 0.826 (0.798-0.852) | 0.891 (0.874-0.905) |
| Expert B | 0.771 (0.741-0.799) | 0.952 (0.940-0.962) | 0.889 (0.875-0.901) | 0.895 (0.870-0.916) | 0.886 (0.870-0.901) |
| Expert C | 0.813 (0.784-0.838) | 0.903 (0.887-0.917) | 0.871 (0.857-0.885) | 0.817 (0.788-0.842) | 0.900 (0.884-0.915) |
| Expert Mean | 0.791 (0.774-0.807) | 0.922 (0.914-0.930) | 0.877 (0.869-0.884) | 0.844 (0.828-0.858) | 0.892 (0.883-0.901) |
| Radiologists w/ Model | | | | | |
| Trainee A | 0.863 (0.837-0.885) | 0.882 (0.865-0.897) | 0.875 (0.861-0.888) | 0.796 (0.768-0.821) | 0.924 (0.909-0.936) |
| Trainee B | 0.921 (0.900-0.938) | 0.885 (0.868-0.901) | 0.898 (0.885-0.909) | 0.811 (0.784-0.835) | 0.954 (0.942-0.964) |
| Trainee C | 0.824 (0.796-0.849) | 0.938 (0.925-0.949) | 0.899 (0.886-0.910) | 0.877 (0.851-0.898) | 0.909 (0.894-0.922) |
| Trainee D | 0.707 (0.674-0.738) | 0.947 (0.934-0.957) | 0.864 (0.849-0.877) | 0.877 (0.849-0.900) | 0.858 (0.841-0.874) |
| Trainee Mean | 0.829 (0.815-0.841) | 0.913 (0.906-0.920) | 0.884 (0.877-0.890) | 0.836 (0.822-0.848) | 0.909 (0.902-0.916) |
| Board-certificated A | 0.833 (0.805-0.857) | 0.914 (0.899-0.927) | 0.886 (0.872-0.898) | 0.838 (0.811-0.862) | 0.911 (0.896-0.925) |
| Board-certificated B | 0.864 (0.839-0.886) | 0.877 (0.860-0.893) | 0.873 (0.858-0.886) | 0.790 (0.761-0.815) | 0.924 (0.909-0.936) |
| Board-certificated C | 0.902 (0.879-0.921) | 0.847 (0.827-0.864) | 0.866 (0.851-0.879) | 0.758 (0.730-0.784) | 0.942 (0.928-0.953) |
| Board-certificated D | 0.840 (0.813-0.864) | 0.917 (0.902-0.930) | 0.890 (0.877-0.902) | 0.843 (0.816-0.867) | 0.915 (0.900-0.928) |
| Board-certificated Mean | 0.860 (0.847-0.871) | 0.889 (0.881-0.896) | 0.879 (0.872-0.885) | 0.805 (0.791-0.818) | 0.922 (0.915-0.929) |
| Expert A | 0.909 (0.887-0.927) | 0.930 (0.916-0.942) | 0.923 (0.911-0.933) | 0.874 (0.850-0.895) | 0.951 (0.938-0.961) |
| Expert B | 0.923 (0.903-0.940) | 0.966 (0.955-0.974) | 0.951 (0.941-0.959) | 0.935 (0.916-0.950) | 0.959 (0.948-0.968) |
| Expert C | 0.943 (0.925-0.957) | 0.948 (0.935-0.958) | 0.946 (0.936-0.955) | 0.906 (0.884-0.924) | 0.969 (0.959-0.977) |
| Expert Mean | 0.925 (0.914-0.935) | 0.948 (0.941-0.954) | 0.940 (0.934-0.945) | 0.905 (0.892-0.916) | 0.960 (0.954-0.965) |

**Extended Data Table 3 | Results of reader study.** PPV: positive predictive value. NPV: negative predictive value.

| | RW 1 cohort FAHZU | RW 2 cohort 1 FAHZU | RW 2 cohort 2 SCH | RW 2 cohort 3 QPH |
|---|---|---|---|---|
| Sensitivity (95% CI) Original model | 0.818 (0.680-0.905) | 0.826 (0.720-0.898) | 0.731 (0.539-0.863) | 0.782 (0.581-0.903) |
| **Sensitivity (95% CI) Upgraded model** | - | 0.942 (0.860-0.977) | 0.923 (0.759-0.979) | 0.913 (0.732-0.976) |
| Specificity (95% CI) Original model | 0.994 (0.993-0.995) | 0.990 (0.989-0.991) | 0.995 (0.994-0.996) | 0.991 (0.990-0.992) |
| **Specificity (95% CI) Upgraded model** | - | 0.992 (0.991-0.993) | 0.991 (0.990-0.992) | 0.993 (0.992-0.994) |
| AUC (95% CI) Original model | 0.906 (0.849-0.963) | 0.908 (0.858-0.925) | 0.863 (0.767-0.940) | 0.887 (0.800-0.953) |
| **AUC (95% CI) Upgraded model** | - | 0.967 (0.938-0.989) | 0.957 (0.899-0.996) | 0.953 (0.888-0.997) |

| | Group A IR Radiologist | Group A RR Radiologist | Group B IR Radiologist + AI | Group B RR Radiologist + AI | iAorta |
|---|---|---|---|---|---|
| Sensitivity (95% CI) | 0.214 (0.076-0.476) | 0.429 (0.214-0.674) | 0.786 (0.524-0.924) | 0.857 (0.601-0.960) | 0.857 (0.601-0.960) |
| Specificity (95% CI) | 0.983 (0.981-0.985) | 0.989 (0.987-0.990) | 0.991 (0.989-0.992) | 0.990 (0.989-0.992) | 0.992 (0.990-0.993) |

**Extended Data Table 4 | Results of retrospective real-world study (top) and prospective comparative study (bottom).** FAHZU, the First Affiliated Hospital of Zhejiang University School of Medicine; QPH, Quzhou People's Hospital; SCH, Shaoxing Central Hospital; AUC, Area Under the Curve; IR Radiologist, initial reporting radiologist; RR Radiologist, report reviewing radiologist; Group A, the radiologist team, consisting of the IR radiologist and the RR radiologist, independently evaluate images in a sequential manner, mirroring the current clinical workflow; Group B, the radiologist team evaluate images assisted by iAorta system.

## Table of contents-online supplement

## Section 1: Description of the training and validation cohorts

### *1.1 Inclusion and exclusion criteria*

We retrospectively retrieved 24,100 aortic computed tomography angiography (CTA) scans from eight medical centers. Consecutive adult patients (aged ≥ 18 years) who had undergone aortic CTA were considered for inclusion. Both the acute-phase imaging and subsequent follow-up images of AAS were included. We excluded patients if their aortic CTA scans did not include both non-contrast and arterial phase series, if they had undergone aortic stent implantation or were diagnosed with a traumatic aortic injury, or if their images exhibited severe motion artifacts. Supplementary Fig. 4 shows the detailed inclusion and exclusion process for each cohort. Because arterial series were mainly used for diagnostic labeling, we summarized the details of the included non-contrast series and demographic information, including age and sex, for each cohort.

### *1.2 Training and internal validation cohorts*

Between 2016 and 2022, 12,402 patients were assessed for eligibility at The First Affiliated Hospital of Zhejiang University School of Medicine (FAHZU). A total of 6,765 patients were excluded: 165 who were missing non-contrast series in the aortic CTA scans, 6,571 who had undergone aortic stent implantation, 12 who had traumatic aortic injuries, and 17 with low-quality images. To avoid overlap of the training cohort with the clinical practicality study cohort, the remaining 5,637 eligible patients were assigned to the training cohort (3,350 [59.4%]) or the internal validation cohort (2,287 [40.6%]) according to the year of data collection.

The training cohort comprised 3,350 patients with aortic CTA scans (1,265 patients with acute aortic syndrome (AAS)—including 296 with Stanford Type A Aortic Dissection (TAAD), 341 with Stanford Type B Aortic Dissection (TBAD), 321 with intramural hematoma (IMH) and 307 with penetrating atherosclerotic ulcer (PAU)—and 2,085 patients with non-AAS) collected from FAHZU between 2016 and 2020. The internal validation cohort comprised 2,287 patients with aortic CTA scans (795 patients with AAS—including 188 with TAAD, 248 with TBAD, 203 with IMH, and 156 PAU—and 1,492 patients with non-AAS) collected from the same hospital between 2021 and 2022.

### *1.3 External validation cohorts*

We used the same filtering strategy employed as above to obtain the external validation cohorts 1-7 and acquire valid aortic CTA scans. The arterial phase series were used as the gold standard for diagnosing AAS, while the non-contrast phase series were employed to validate the developed model.

**External validation cohort 1:** Between 2020 and 2022, we initially recruited 6,806 consecutive patients who underwent aortic CTA at Nanjing Drum Tower Hospital (NDTH). A total of 3,519 patients were excluded (28 missing non-contrast series from aortic CTA, 3,474 who had undergone aortic stent implantation, 7 who had undergone traumatic aortic injury, and 10 who had low-quality images), so that the external validation cohort 1 is composed of 3,287 patients (1,296 with AAS—including 234 with TAAD, 467 with TBAD, 335 with IMH and 260 with PAU—and 1,991 with non-AAS).

**External validation cohort 2:** Between 2018 and 2022, we initially recruited 3,376 consecutive patients who underwent aortic CTA at Shandong Provincial Hospital Affiliated with Shandong First Medical University (SPH). A total of 1,025 patients were excluded (16 missing non-contrast series from their aortic CTA scans, 985 who had undergone aortic stent implantation, 11 with traumatic aortic injury, and 13 with low-quality images), so that the external validation cohort 2 is composed of 2,351 patients (980 with AAS—including 242 with TAAD, 254 with TBAD, 249 with IMH and 235 with PAU—and 1,371 with non-AAS).

**External validation cohort 3:** Between 2016 and 2023, we initially recruited 2,080 consecutive patients who underwent aortic CTA at Taizhou Hospital of Zhejiang Province (TZH). A total of 513 patients were excluded (26 missing non-contrast series from their aortic CTA scans, 465 who had undergone aortic stent implantation, 13 with traumatic aortic injury, and 9 with low-quality images), which created the external validation cohort 3, comprising 1,567 patients (563 with AAS—including 104 with TAAD, 128 with TBAD, 165 with IMH, and 166 with PAU—and 1,004 with non-AAS).

**External validation cohort 4:** Between 2018 and 2022, we initially recruited 6,443 consecutive patients who underwent aortic CTA at the First Affiliated Hospital of Wenzhou Medical University (FAHWMU). A total of 1,869 patients were excluded (43 missing non-contrast series from their aortic CTA scans, 1,786 who had undergone aortic stent implantation, 19 with traumatic aortic injury, and 21 with low-quality images), which created the external validation cohort 4, comprising 4,574 patients (1,918 with AAS—including 413 with TAAD, 586 with TBAD, 485 with IMH, and 434 with PAU—and 2,656 with non-AAS).

**External validation cohort 5:** Between 2019 and 2023, we initially recruited 3,167 consecutive patients who underwent aortic CTA examination at Ningbo No.2 Hospital (N2H). A total of 798 patients were excluded (9 missing non-contrast series from their aortic CTA scans, 769 who had undergone aortic stent implantation, 4 with traumatic aortic injury, and 16 with low-quality images), resulting in the external validation cohort 5, consisting of 2,369 patients (591 with AAS—including 107 with TAAD, 153 with TBAD, 171 with IMH and 160 with PAU—and 1,778 with non-AAS).

**External validation cohort 6:** Between 2019 and 2023, we initially recruited 3,900 consecutive patients who underwent aortic CTA at Quzhou People's Hospital (QPH). A total of 885 patients were excluded (68 missing non-contrast series from their aortic CTA scans, 797 who had undergone aortic stent implantation, 11 with traumatic aortic injury, and 9 with low-quality images), which created the external validation cohort 6, consisting of 3,015 patients (850 with AAS—including 172 with TAAD, 216 with TBAD, 247 with IMH and 215 with PAU—and 2,165 with non-AAS).

**External validation cohort 7:** Between 2019 and 2023, we initially recruited 1,665 consecutive patients who underwent aortic CTA at Shaoxing Central Hospital (SCH). A total of 365 patients were excluded (26 missing non-contrast series from their aortic CTA scans, 315 who had undergone aortic stent implantation, 5 with traumatic aortic injury, and 19 with low-quality images), resulting in the external validation cohort 7, consisting of 1,300 patients (297 with AAS—including 71 with TAAD, 101 with TBAD, 66 with IMH and 59 with PAU—and 1,003 with non-AAS).

### *1.4 Image annotation system*

Each set of aortic CTA images imported into the database that was assigned with a case-level diagnostic label by a team of multiple radiologists. For the training datasets, two additional pixel-level segmentation labels, also provided by this radiologist team, were added. All labels were determined based on the arterial series CT images of the patients. The flowchart of annotation system is illustrated in Supplementary Fig. 1.

**Case-level annotation:** To better evaluate the diagnostic ability of the model for AAS, we divided the patient cases into five categories based on the 2022 American Heart Association/American College of Cardiology guidelines[1]: non-AAS, TAAD, TBAD, IMH, and PAU; if the case depicted a patient with PAU and IMH, it was classified into the IMH group. Because IMH and PAU with IMH are more dangerous than are isolated PAUs, this classification allows a more accurate assessment of the model's ability to identify patients in a potentially emergency situation.

First, the images were independently reviewed by two radiologists; if their results were inconsistent, a third radiologist with more than 5 years of clinical experience was asked to read the images, and their findings were considered the final result. Next, a radiologist with more than 10 years of clinical experience re-examined all the results as the final check-up.

**Pixel-level annotation:** To achieve better model performance, we performed a two-stage cascade of pixel-level annotations corresponding to the two stages of the model. First, we marked the extent of the whole aorta on the images as label one, which included the false lumen of the dissection, the area of the intramural hematoma, or the ulcerated vessel wall. The remaining background regions were annotated with label zero. Next, we marked the area of actual blood flow through the original lumen or the area of the true lumen (in cases of aortic dissection) as label two.

The pixel-level annotation process was similar to that described above. First, the images were independently reviewed by two radiologists. The accuracy of the labeled images was determined by calculating the Dice coefficient, which represents the consistency of the labelling between the two radiologists. Images with a Dice coefficient < 0.90 were revised and corrected by a third radiologist of more than 5 years of clinical experience. Finally, the 4$^{th}$ radiologist with more than 10 years of clinical experience re-examined the results as a final check-up.

## Section 2: Development of DL model and iAorta system

### *2.1 Image registration*

The image registration procedure involved registering the arterial series CT images with the non-contrast series images and mapping the segmentation mask annotated on the arterial series images onto the non-contrast series images. First, we employed DEEDs[2], a deformable medical registration method with discrete optimization, to align the arterial and non-contrast series images. This method adopts a contrast and modality-invariant similarity metric based on binarized SSC descriptors[3] and a dense displacement search over several iterations with different control-point spacings. The neighborhood relations of these control point grids are approximated using a minimum-spanning tree to efficiently find an optimal displacement field. In this work, we employed a five-level multi-resolution strategy, and the grid spacing for each level was set to 8 × 7 × 6 × 5 × 4. Second, segmentation masks annotated on the arterial series images were mapped to the non-contrast series images by the displacement field generated by DEEDs.

The image registration procedure involved registering the arterial series CT images with the non-contrast series images and propagating the segmentation mask annotated on the arterial series images to the non-contrast series images. For the initial cohort, we employed DEEDs[2], a deformable medical registration method with discrete optimization, to align the arterial and non-contrast series images. This method adopts a contrast and modality-invariant similarity metric based on binarized SSC descriptors[3] and a dense displacement search over several iterations with different control-point spacings. For subsequent datasets, to enhance computational efficiency while maintaining accuracy, we adopted a learning-based registration paradigm integrating SAMcoarse[4], a key point-based affine registration method, and a deformable registration network leveraging deep neighborhood self-similarity[5], to obtain the optimal displacement vector field for arterial and non-contrast series CT image alignment. Then, segmentation masks annotated on the arterial series images were mapped to the non-contrast series images by the optimal displacement field.

### *2.2 Deep learning framework*

The proposed deep learning model is based on a coarse-to-fine framework. At the coarse stage (localization), the aorta mask is generated by a lightweight 3D nnU-Net[6], which is built upon the classic U-Net[7] architecture and we introduce adaptive mechanisms to enhance the model’s performance and adaptability. Specifically, the adaptive mechanisms of nnU-Net employ two techniques: automatic network architecture adjustment and data preprocessing. First, in terms of network architecture, nnU-Net automatically adjusts the resolution and channel numbers to suit our segmentation task. Second, in terms of data preprocessing, nnU-Net employs an adaptive preprocessing pipeline to enhance the model's adaptability to new input images. This pipeline dynamically selects and applies a series of preprocessing steps, such as histogram equalization, contrast enhancement, and noise removal, based on the characteristics of the input data. Then, the input non-contrast CT images are cropped to a smaller size according to the bounding box coordinates of the aorta mask. The model can mitigate the influence of irrelevant content noise and minimize the computational resource use.

At the fine stage, the model employs a multi-task learning strategy[8]. In particular, the encoder consists of four down-sampling layers, which consist of pooling operations and convolutional layers.

All convolutional layers use a 3×3 convolutional kernel and a leaky rectified linear unit (ReLU)[9] as the activation function. For segmentation, two different decoders that have the same architecture sharing the same encoder, aiming to learn some general characteristics from the encoder, the unique characteristics of the aorta (Decoder A) and the true lumen (Decoder T). The architecture of the two decoders is the same as that of the decoder in 3D nnU-Net. The classification head is a linear layer, and its input is a feature map that concatenates the output of each down-sampling layer in the encoder.

### *2.3 Learning and optimization*

Model development was conducted in Python, using the PyTorch framework with an Nvidia Tesla V100 GPU on a Linux system. The input patch size was set to $192 \times 192 \times 192$. The epoch and training batch size were set to 100 and 2, respectively. The SGD optimizer[10] with momentum = 0.99 was employed, with an initial learning rate of $3 \times 10^{-4}$ and a weight decay of $3 \times 10^{-5}$, respectively. The focal loss function[11], defined as follows, was used for the classification task:

$$L_c = -\alpha(1-p)^{\gamma} y \, log(p) - (1-\alpha)p^{\gamma}(1-y) \, log(1-p)$$

where y represents the disease label corresponding to the image and p represents the predicted value from the model. $\alpha$ is a weight factor, and is set to 0.75 in this case, that helps balance the loss weight between positive and negative samples. $\gamma$ is a modulation factor, being set to 2, that controls the loss weight between difficult and easy samples.

The segmentation loss consists of two components: the aorta segmentation loss and the true lumen segmentation loss. Both segmentation loss functions utilize a combination of cross-entropy loss[12] and Dice loss[13], as shown below:

$$L_s = \alpha L_a + (1-\alpha) L_t$$

$$L_x = -y_x log(p_x) - (1-y_x) \, log(1-p_x) - (1 - 2p_x y_x/(p_x + y_x)), x \in \{a, t\}$$

where $\alpha$ is the weight of the aorta and true lumen segmentation loss functions, which is set to 0.4 in this case. y represents the label of each voxel, while p denotes the model's predicted value at each voxel. a and t are referenced to the aorta and true lumen, respectively.

### *2.4 Slice-level and voxel-level interpretability generation*

The procedure for generating and ensuring the interpretability is based on a distance map that depicts the distance from each voxel in an image to a specific object or feature. It assigns a numerical value to each voxel based on its proximity to the object of interest. Specifically, the construction of a distance map involves calculating the distance of each true lumen voxel to the nearest voxel representing the boundary of the aorta by applying the Euclidean distance transform[14]: voxels that are closer to the aortic wall have smaller values, whereas those farther away have larger values. By analyzing the distance map, the slice-level and voxel-level locations of the lesion within the aorta can be determined based on the different values assigned to different voxels. As a result, the model is capable of determining the specific slice corresponding to the AAS lesion and the precise location of the lesion within the slice, visualizing and enhancing the interpretability of the proposed model.

### *2.5 Evaluation metrics*

Evaluation metrics, including sensitivity, class-wise sensitivity, specificity, accuracy, area under the ROC curve (AUC), positive predictive value (PPV), negative predictive value (NPV), kappa value and F1 score, were computed. For a specific AAS class (TAAD, TBAD, IMH and PAU), the CT scans predicted as depicting AAS were regarded as positive, while the remaining other scans were regarded as negative. Moreover, the numbers of true positives (TPs), false positives (FPs), true negatives (TNs), and false negatives (FNs) were calculated, from which we can calculate the sensitivity, specificity, accuracy, PPV, NPV, and F1 score as follows:

$$Sensitivity = \frac{TP}{TP + FN}$$

$$Specificity = \frac{TN}{TN + FP}$$

$$Accuracy = \frac{TP + TN}{TP + TN + FP + FN}$$

$$PPV = \frac{TP}{TP + FP}$$

$$NPV = \frac{TN}{TN + FN}$$

$$F1\ Score = 2\frac{precision \cdot recall}{precision + recall} = \frac{2TP}{2TP + FP + FN}$$

ROC curves were created by plotting the proportion of true positives against the proportion of false positives at different predicted probability thresholds. The AUC was obtained by computing the area under the ROC curve. In the evaluation of the model, one non-contrast series CT image of a patient was considered a sample. Additionally, the segmentation was evaluated with the Dice coefficient, calculated as the sum of the areas of the predicted segmentation and the ground-truth segmentation divided by the area of the overlap between these two segmentation masks.

$$Dice = \frac{2*(pred \cap true)}{pred \cup true}$$

***2.6 Comparison to other state-of-the-art models***

A comparative experiment with other state-of-the-art models was conducted on the training dataset via five-fold cross validation. The compared methods included the following list.

CMTGF[15] is one of the best detection methods for detecting aortic dissection (AD) in non-contrast CT images. This method is a cascaded multi-task generative framework that includes a 3D nnU-Net and a 3D multi-task generative architecture (3D MTGA). Specifically, 3D nnU-Net is employed to segment aortas from non-contrast CT images. The 3D MTGA is then employed to simultaneously synthesize contrast-enhanced CT volumes, segment true and false lumens, and determine whether the patient has an AD.

Swin UNETR[16] is a neural network model designed for medical imaging, particularly for tasks such as segmentation, with the goal of identifying and delineating different regions or structures within medical images. Swin UNETR combines the Swin Transformer with a U-Net-like architecture in an attempt to leverage the strengths of both: the powerful representation capabilities of Transformers and the proven network structure of U-Nets for image segmentation. This model has been shown to perform well on tasks such as segmentation of tumors, organs, or other structures of interest in medical images, providing valuable assistance in medical diagnostics and research. Additionally, the multi-level features of the Swin UNETR decoder are employed as the input of the classification head.

Swin transformer[17] is an artificial intelligence model designed for computer vision tasks, such as image classification, detection, and segmentation. In Swin Transformer, the image is divided into small patches, and self-attention is computed within these local windows. To ensure that information can flow between different parts of the image, the position of the windows is shifted in subsequent layers, allowing a form of cross-window communication. This approach has lower computational complexity than applying self-attention across the entire image.

Vision transformer (ViT)[18] is an adaptation of the Transformer architecture that was originally invented for processing sequential data, such as text in natural language processing (NLP). Within Transformer, the self-attention mechanism allows the model to weigh the importance of different patches relative to one another for a given task, such as image classification. This means that the model can learn to focus on the most informative parts of the image for decision-making. ViT was one of the first models to demonstrate that the Transformer architecture could be directly applied to process raw images and achieve competitive results on image classification benchmarks, challenging the dominance of convolutional neural networks (CNNs).

ResNet[19] is an artificial neural network that is especially well suited for handling very deep architectures. The residual blocks in a ResNet architecture help to address the problem of vanishing gradients, which is a common issue when training very deep networks. Additionally, adding shortcuts or "skip connections" can allow the gradients to flow through the network more directly. Instead of attempting to learn the desired underlying mapping directly, each residual block in a ResNet learns the difference (or residual) between the input and output of the block. This process is easier and allows the network to be much deeper without running into training problems.

Notably, CMTGF and Swin UNETR were compared to the proposed method in terms of both classification and segmentation, whereas Swin transformer, ViT, and ResNet50 were only compared in terms of classification.

As shown in Supplementary Table 1, compared with these state-of-the-art approaches, our proposed model achieved the best classification performance. The mean sensitivity and specificity of our model reached 0.988 and 0.997, respectively. In particular, competing with those of the current best AD detection method (CMTGF), the mean obtained sensitivity and specificity are greater by the margins of 0.248 and 0.032, respectively. Our model achieved outstanding segmentation performance; the mean Dice coefficients for the aorta and true lumen obtained with our model were 0.972 and 0.920, respectively, which were 0.033 and 0.35, respectively, higher than those obtained with Swin UNETR. Similarly, our model achieved a high AUC of 0.997 (95% CI 0.991-1.000), as shown in Supplementary Fig. 2.

### *2.7 iAorta system*

We have developed an innovative browser-server collaborative early warning system being specifically designed for rapid and accurate detection of AAS in real-world clinical medical settings.

This system integrates an AI non-contrast phase selection tool, our developed AAS model, and a user-friendly warning popup plugin to process and analyze vast amounts of patient data in real-time. By leveraging these cutting-edge technologies, our system enables radiologists to receive timely and precise alerts, facilitating informed diagnosis and prompt intervention. Seamlessly integrating these components into iAorta system not only enhances the efficiency of our diagnostic workflow also significantly improves the patient outcomes by reducing the time to proper clinically critical decision-making. This comprehensive solution represents a significant advance in the field of medical informatics, offering a robust framework for the early detection and management of AAS, thereby contributing to improved patient care and safety.

The non-contrast phase selection tool is a pivotal component of our iAorta warning system, employing deep learning algorithms to automatically identify and classify distinct phases within a CT scan, including the non-contrast, arterial, venous, and delayed phases. By precisely segmenting and recognizing these different phases, this tool ensures that the non-contrast phase is always reliably selected as the input of our developed model. Automated phase recognition enhances the efficiency of the diagnostic process, and streamlines the workflow and reduces the potential human errors.

The warning popup plugin is another crucial component of our warning system. After the CT phase recognition tool has processed the images, our DL model meticulously evaluates the data for indications of AAS. Upon detection of any abnormalities, the plugin instantaneously generates a real-time alert, which is prominently displayed as a popup on the radiologist's screen. This immediate notification ensures that the radiologist is promptly informed of the cases with potential AAS, then facilitating swift and appropriate clinical action. The popup not only alerts the radiologist but also provides direct hyperlinks to an interactive interface. This visual interface offers detailed information about the detected AAS, including high-resolution images, a comprehensive summary of the findings, and relevant clinical data. This technology enables radiologists to quickly assess the situation, review the critical details, and make well-informed decisions, thereby enhancing the ultimate patient care and outcomes. The seamless integration of real-time alerts and detailed diagnostic information jointly supports a more efficient and effective workflow for diagnosing patients with acute chest pain in the radiology department.

The invented browser-server collaborative technology harnesses the power of cloud/local computing, providing a robust and highly scalable solution for data processing and storage. This architecture notably enhances the computational efficiency and accessibility, ensuring smooth integration with existing hospital information systems. Furthermore, our system is meticulously designed to uphold the highest standards of security, incorporating stringent data privacy and security protocols to safeguard the sensitive patient information. Note that we have granted the complimentary access to the browser interface of our iAorta system (available at https://iaorta.medofmind.com/). This browser interface offers an open-access CT image database containing typical patient cases. This resource would be very valuable for training radiologists and researchers in the field of aortic diseases, as well as for advancing AI-assisted medical imaging.

## Section 3: Details of the stage II reader study

### *3.1 Radiologists alone*

Eleven radiologists of varying levels of expertise (special experts, board-certified radiologists, and medical trainees) were recruited to independently complete the detection task using the CT data from the internal validation cohort, and their results were compared with those of our model. The enlisted 11 radiologists were all affiliated with SPH, thus ensuring that they had no prior exposure to the data of the internal validation cohort. The special experts were three medical school professors with more than 10 years of experience in cardiovascular imaging. The board-certified radiologists were four radiologists with more than 5 years of experience in cardiovascular imaging. The medical trainees were four radiology residents who had completed their residency training in medical imaging, specializing in cardiovascular imaging as their area of research interest.

These enrolled radiologists were brought together every Sunday afternoon to perform the classification task. Each participant was provided with a computer with randomly arranged, non-contrast CT images from the internal validation cohort. Each participant was asked to diagnose the patient with AAS or non-AAS and to provide their response on an answer card within 1 minute per case, simulating a busy emergency environment. The data were collected using a unified form and reviewed by two researchers (XYL and HYJ). All radiologists were compensated at an hourly rate comparable to the market rate. This radiologist-alone test started in January 2023 and ended in May 2023, spanning 20 Sundays (excluding the Sunday of the week of Chinese New Year).

### *3.2 Radiologists assisted by AI*

After a subsequent two-month desensitization period, we employed our model to inference the non-contrast CT images of the internal validation cohort to classify the images as containing AAS or non-AAS, and randomly shuffled the images again. The original CT images and the CT images combined with the distance map are shown to the radiologists who were enrolled in the radiologist-alone AAS detection test. They were then asked to diagnose the patients with the assistance of our model, but they were not informed of the detection performance of our model. This centralized testing was conducted on Sundays from July 2023 to November 2023.

**Section 4: Details of the stage III retrospective large-scale real-world study**

***4.1 Inclusion and exclusion criteria***

For the retrospective, real-world study, consecutive outpatients older than 18 years of age who presented to the ED were eligible if their principal complaint was non-traumatic acute chest pain that had begun within the previous 14 days. Here, chest pain referred to more than pain in the chest; pain, pressure, tightness, or discomfort in the chest, shoulders, arms, neck, back, upper abdomen, or jaw, as well as shortness of breath and fatigue, were considered equivalent symptoms of acute chest pain[18]. We excluded patients whose initial diagnosis was AAS, patients who did not undergo CT examinations that included the aorta (for example, non-contrast chest CT, non-contrast abdominal CT, pulmonary CTA, coronary CTA, esophageal CT, lumbar CT or thoracic CT) and patients whose CT images were compromised by severe motion artifacts, leading to low image quality. The initial diagnosis by emergency clinicians was based on their preliminary investigations (via age, risk factors, history, pain characteristics, findings on physical examination, ECG and some laboratory tests).

***4.2 Diagnostic criteria***

Diagnostic labels were confirmed through the conclusive diagnostic imaging, direct visualization of surgery, postmortem examination and other medical records within a 14-day timeframe following the onset of acute chest pain. Two expert vascular physicians would independently review these data and gave the case adjudication. The following advanced imaging methods were considered conclusive for the diagnosis of AAS: the aorta CTA, transesophageal echocardiography (TEE) or magnetic resonance angiography (MRA). For patients without the conclusive diagnostic imaging, the physicians would review the medical records, including patient history, radiology, laboratory results, treatment and follow-up history, during this ED visit or within the subsequent two-week follow-up. Patients with a positive AI prediction received an additional telephone follow-up to inquire about their physical condition within two weeks of the visit as a supplement to the clinical diagnosis. The following events were queried: diagnosis of AAS or any aortic disease, subsequent ED visit, hospital admission, and death.

***4.3 Real-world cohorts***

The first real-world evaluation (**RW1**) was based on a real-world, retrospective cohort from the FAHZU. We consecutively enrolled 36,422 patients who presented to the ED for eligibility assessment between 1 January and 31 December 2021. A total of 15,590 patients (364 patients whose initial diagnosis was AAS, 15,132 patients who did not undergo CT examinations, and 94 patients with low-quality images) were excluded. The RW1 cohort comprised 20,832 individuals (32 with AAS and 20,800 with non-AAS) with 23,094 non-contrast CT images (44 AAS images and 23,050 non-AAS images). A total of 828 patients had two or more non-contrast CT images.

The second real-world evaluation (RW2) was based on real-world, multi-center, retrospective cohorts from the FAZHU, QPH and SCH. In total, 116,693 consecutive individuals with acute chest pain (89 with AAS and 116,604 with non-AAS) with a total of 122,107 non-contrast CT images (118 AAS images and 121,989 non-AAS images) were included in the RW2.

**RW2-Cohort 1:** We consecutively enrolled 119,179 patients who presented to the ED at the FAZHU

for eligibility assessment between 1 January 2022 and 31 December 2023. A total of 46,819 patients (1,112 patients whose initial diagnosis was AAS, 45,472 patients who did not undergo CT examinations, and 235 patients with low-quality images) were excluded. Cohort 1 ultimately comprised 72,360 individuals (52 with AAS individuals and 72,308 with non-AAS) with 76,582 non-contrast CT images (69 AAS images and 76,513 non-AAS images). A total of 3,810 patients had two or more non-contrast CT images.

**RW2-Cohort 2:** We consecutively enrolled 41,052 patients who presented to the ED of the SCH for eligibility assessment between 1 January 2021 and 31 December 2023. A total of 17,248 patients (604 patients whose initial diagnosis was AAS, 16,528 patients who did not undergo CT examinations, and 116 patients with low-quality images) were excluded. Cohort 2 is composed of 23,804 individuals (21 with AAS and 23,783 with non-AAS) with 24,365 non-contrast CT images (26 AAS images and 24,339 non-AAS images). A total of 553 patients had two or more non-contrast CT images.

**RW2-Cohort 3:** We consecutively enrolled 35,289 patients who presented to the ED of the QPH for eligibility assessment between 1 January 2021 and 31 December 2023. A total of 14,760 patients (543 patients whose initial diagnosis was AAS, 14,125 patients who did not undergo CT examinations, and 92 patients with low-quality images) were excluded. Cohort 3 consisted of 20,529 individuals (16 with AAS and 20,513 with non-AAS) with 21,160 non-contrast CT images (23 AAS images and 21,137 non-AAS images). A total of 623 patients had two or more non-contrast CT images.

### *4.4 Model evolution*

The training set of the upgraded model included the data of 3,350 patients from the internal training cohort of the original model and 146 patients—including 138 with non-AAS (30 non-contrast chest CT images, 83 non-contrast abdominal CT images, 7 pulmonary CTA images, and 18 other CT images), 1 with IMH (1 non-contrast abdominal CT images), and 7 with PAU (3 non-contrast chest CT images and 4 non-contrast abdominal CT images)—from the RW1 cohort. These cases from the RW1 cohort were either *false positive* or *false negative* predictions of the original model. For the images of the new patients, the same radiologist team made the case-level and pixel-level annotations directly on the non-contrast CT scan images, referring to all existing clinical examinations and records, such as multi-phase contrast-enhanced CT scans and EMRs. With these extra training data, we finetuned the original Stage 2 model (previously trained only on the internal training dataset) using both the internal training data and the newly collected data for another 150 epochs using the same training hyperparameters as the original model. In the training (or retraining) process, we oversampled the newly collected CT data by a factor of 5, forcing the model to more heavily account for these challenging data (i.e., false positives and false negatives). The pipeline of model evolution is illustrated in Supplementary Fig. 3. In machine learning, these taxonomies are known as hard example mining and incremental learning. The evolved model was subsequently tested on three RW2 cohorts.

## Section 5: Details of the stage IV prospective multi-center study

### *5.1 Inclusion and exclusion criteria*

For the prospective study, consecutive outpatients older than 18 years of age who presented to the ED were eligible if their principal complaint was non-traumatic acute chest pain that had begun within the previous 14 days. Here, chest pain referred to more than pain in the chest; pain, pressure, tightness, or discomfort in the chest, shoulders, arms, neck, back, upper abdomen, or jaw, as well as shortness of breath and fatigue, were considered equivalent symptoms of acute chest pain[20]. We excluded patients whose initial diagnosis was AAS, patients who did not undergo CT examinations that included the aorta (for example, non-contrast chest CT, non-contrast abdominal CT, pulmonary CTA, coronary CTA, esophageal CT, lumbar CT or thoracic CT), patients who were unwilling to participate in the study, patients who were inadequate for follow-up and patients whose CT images were compromised by severe motion artifacts, leading to low image quality. The initial diagnosis by emergency clinicians was based on their preliminary investigations (using age, risk factors, history, pain characteristics, findings on physical examination, ECG and certain laboratory tests).

### *5.2 Diagnostic criteria*

Diagnostic labels were confirmed through the conclusive diagnostic imaging, direct visualization of surgery, postmortem examination and other medical records within a 14-day timeframe following the onset of acute chest pain. The advanced imaging methods considered as being conclusive for the diagnosis of AAS are: aorta CTA, transesophageal echocardiography (TEE) or magnetic resonance angiography (MRA). Patients not subjected to these tests and without surgical or autopsy data which either confirming or excluding AAS entered a 14-day clinical follow-up for case adjudication. For this purpose, patients or family members were interviewed by telephone with a structured questionnaire or underwent an outpatient visit after 14 days from ED discharge. The following events were queried: diagnosis of AAS or any aortic disease, subsequent ED visit, hospital admission, and death. Patients dismissed from the ED were instructed to return to the ED in case of new, worsening, or recurrent symptoms. Two expert vascular physicians would independently review these data and gave the case adjudication.

### *5.3 Prospective cohort*

The prospective multi-center study was based on 1) a comparative study cohort from the FAZHU, QPH and SCH, and 2) a pilot deployment study cohort from the Shanghai Changhai Hospital (SHCH). In the first comparative study cohort, 13,846 consecutive individuals with acute chest pain (11 with AAS and 13,835 with non-AAS) with a total of 14,436 non-contrast CT images (14 AAS images and 14,422 non-AAS images) were presented. In the pilot deployment study cohort, 15,584 consecutive individuals with acute chest pain (22 with AAS and 15,562 with non-AAS) with a total of 16,054 non-contrast CT images (22 AAS images and 16,032 non-AAS images) were included.

**Comparative study cohort:** We consecutively enrolled 23,567 patients who presented to the ED at the FAZHU, QPH and SCH for eligibility assessment between 20 August 2024 and 19 November 2024. A total of 9,721 (41.2%) patients were excluded from the study because of initial diagnosis as AAS (383 patients); no undergo CT examinations (9,021 patients); refusal to participate in the study (71 patients); unavailability for follow-up (207 patients); low-quality images (39 patients). The cohort ultimately consisted of 13,846 individuals (11 with AAS individuals and 13,835 with non-AAS) with 14,436 non-contrast CT images (14 AAS images and 14,422 non-AAS images).

**Pilot deployment study cohort:** This study is an ongoing, single-arm, prospective study. Between 20 December 2024 and 28 February 2025, we screen 28,151 acute chest pain patients who presented to the ED at the SHCH. A total of 12,567 (44.6%) patients were excluded from the study because of initial diagnosis as AAS (450 patients); no undergo CT examinations (11,912 patients); refusal to participate in the study (46 patients); unavailability for follow-up (127 patients); low-quality images (32 patients). The present cohort comprised 15,584 individuals (22 with AAS individuals and 15,562 with non-AAS) with 16,054 non-contrast CT images (22 AAS images and 16,032 non-AAS images).

***5.4 Prospective comparative study***

We recruited two groups of radiologists to explore the potential improvement in diagnostic performance and the reduction in final definitive diagnosis time when iAorta is integrated into the existing clinical workflow, as compared to the original clinical workflow. Each group included board-certified general radiologists per shift for the initial report and specialty experts for the final reviewing report. Group A consisted of 19 radiologists from three hospitals (Center 1: six initial reporting radiologists and three report reviewing radiologists; Center 2: three initial reporting radiologists and two report reviewing radiologists; Center 3: three initial reporting radiologists and two report reviewing radiologists). Group B consisted of 17 radiologists from three hospitals (Center 1: five initial reporting radiologists and two report reviewing radiologists; Center 2: three initial reporting radiologists and two report reviewing radiologists; Center 3: three initial reporting radiologists and two report reviewing radiologists). Each CTA diagnostic report was handled by one initial reporting radiologists and report reviewing radiologists in one of the local hospitals.

Both Group A and Group B analyzed the same set of scans. Group A independently reviewed the patients' CT scans in a sequential manner under the current clinical workflow using the local PACS and provided the final reviewing report to guide subsequent patient management. This process generates two independent diagnostic records: one for the IR radiologist's diagnosis and the other for the RR radiologist's diagnosis, for each patient CT scan. In the meanwhile, Group B reviewed the images with the assistance of a locally-deployed iAorta system. The images are automatically fed into iAorta system, which then produces AI-based results. If the AI detects any abnormalities, both the initial reporting (IR) radiologists and the report reviewing (RR) radiologists receive subsequential pop-up message alerts, prompting them to prioritize the review of the AI-flagged (AAS positive) patient scan. After reviewing the AI results, the AAS-related diagnosis is recorded. This process yields three independent diagnostic records for each image: the IR radiologist's diagnosis (aided with AI), the RR radiologist's diagnosis (aided with AI), and the AI-computed results. Additionally, we recorded the patients' arrival time to the hospital, the time of initial diagnosis and final definitive diagnosis, the time-to-results for laboratory testing and the time of ordering the CT examination, performing the CT scan, the initial report, and the final report. Here, we defined the time to the correct diagnostic pathway for AAS patients as the period from the patient's arrival in the ED until AAS was mentioned as a positive possibility in their non-contrast CT imaging report via AI associated with a recommendation for the aortic CTA examination, or the point at which they were advised to undergo conclusive diagnostic imaging for other reasons by clinicians.

### *5.5 Pilot deployment study*

We integrated iAorta system into the current pilot clinical daily routine and utilized AI-assisted or AI-alerted electronic imaging reports to guide subsequent patient management. We also recorded the patients' arrival time to the hospital, the time of initial diagnosis and final definitive diagnosis, the time-to-results for laboratory testing and the time of ordering the CT examination, performing the CT scan, the initial report, and the final report. We analyzed the time from the initial emergency department presentation to definitive diagnosis.

**Section 6: References for the Supplementary Appendix**

## Section 7: Supplementary Figures

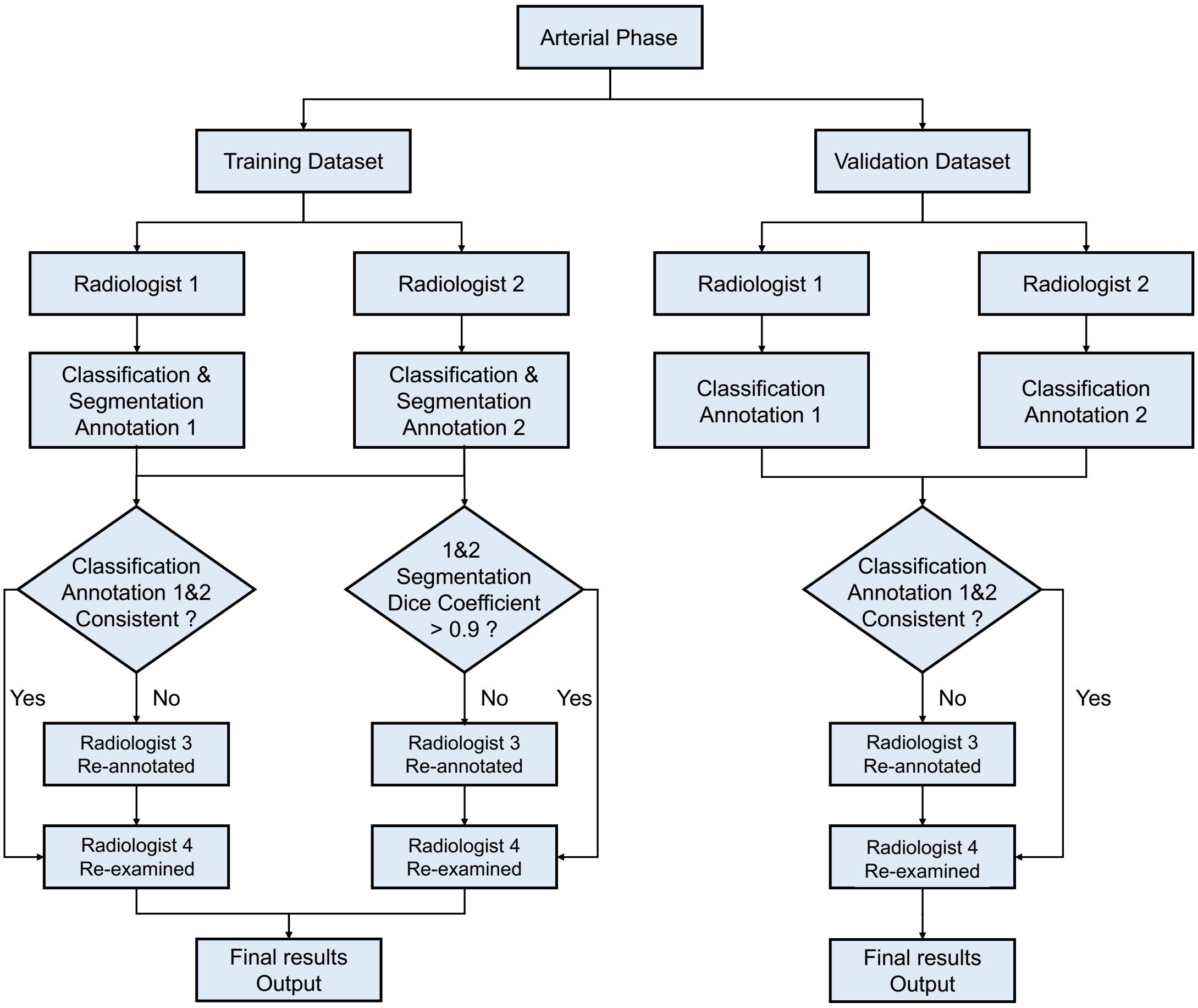

**Supplementary Fig. 1 | The flowchart of annotation system.** As for training datasets, radiologists 1 and 2 read the arterial phase images and independently perform case-level diagnostic labeling and voxel-level segmentation labeling. If the results from these two radiologists were inconsistent, a third radiologist with more than 5 years of clinical practice experience would be introduced to re-annotate the images as the confirmed result. At the same time, radiologist 4 with more than 10 years of clinical practice experience re-examined the results as the final check-ups. As for internal and external validation cohorts, the above procedure only contains case-level diagnostic annotations.

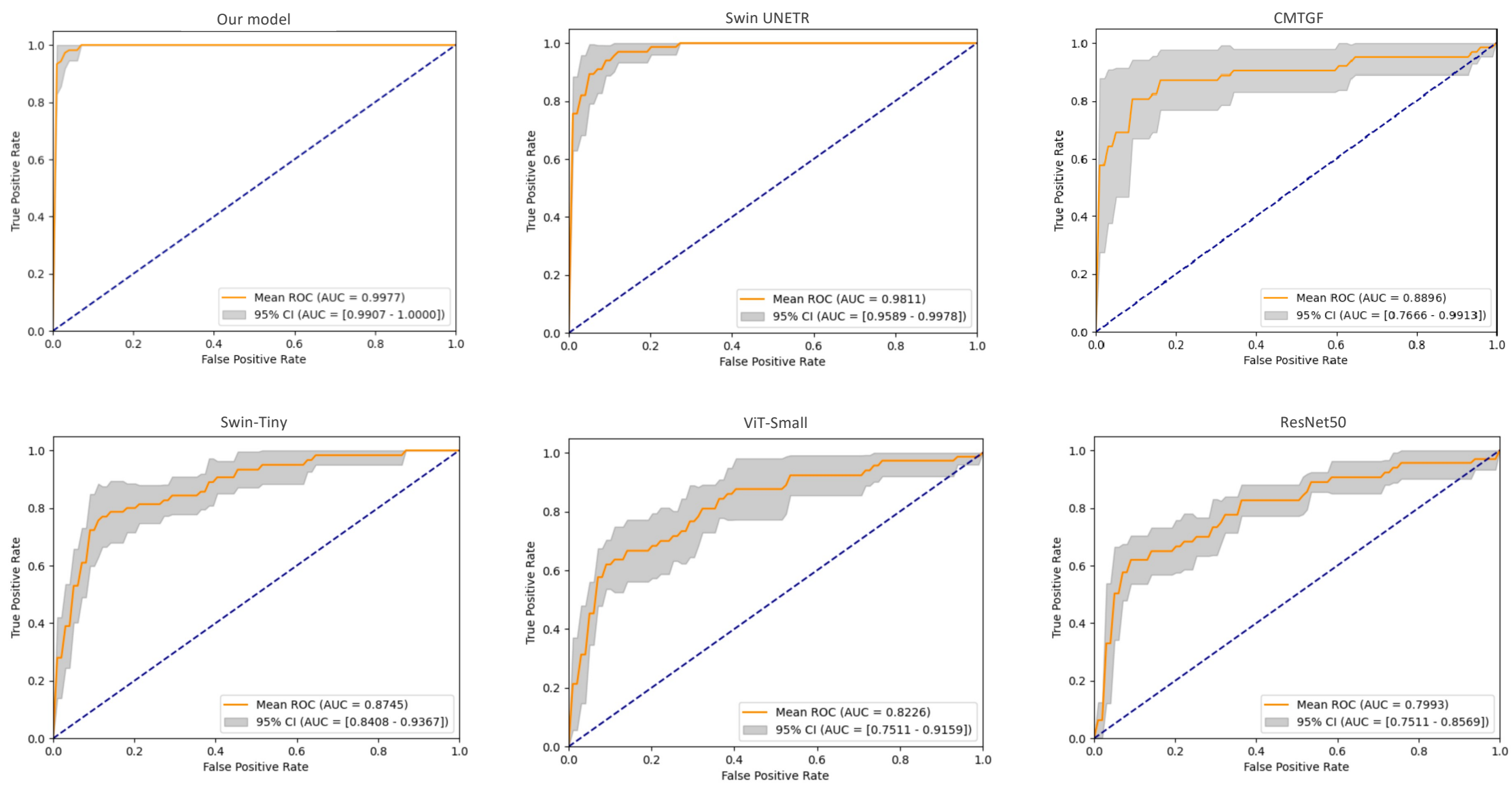

**Supplementary Fig. 2 | Performance of the proposed model and other state-of-the-art models to detect patients with AAS in five-fold cross validation of training dataset (n = 3,350).** ROC, Receiver Operating Characteristic Curve; AUC, Area Under the Curve; Swin UNETR, Swin U-Net Transformer; CMTGF, Cascaded Multi-Task Generative Framework; Swin-Tiny, Swin Transformer-Tiny; ViT-Small, Vision Transformer-Small.

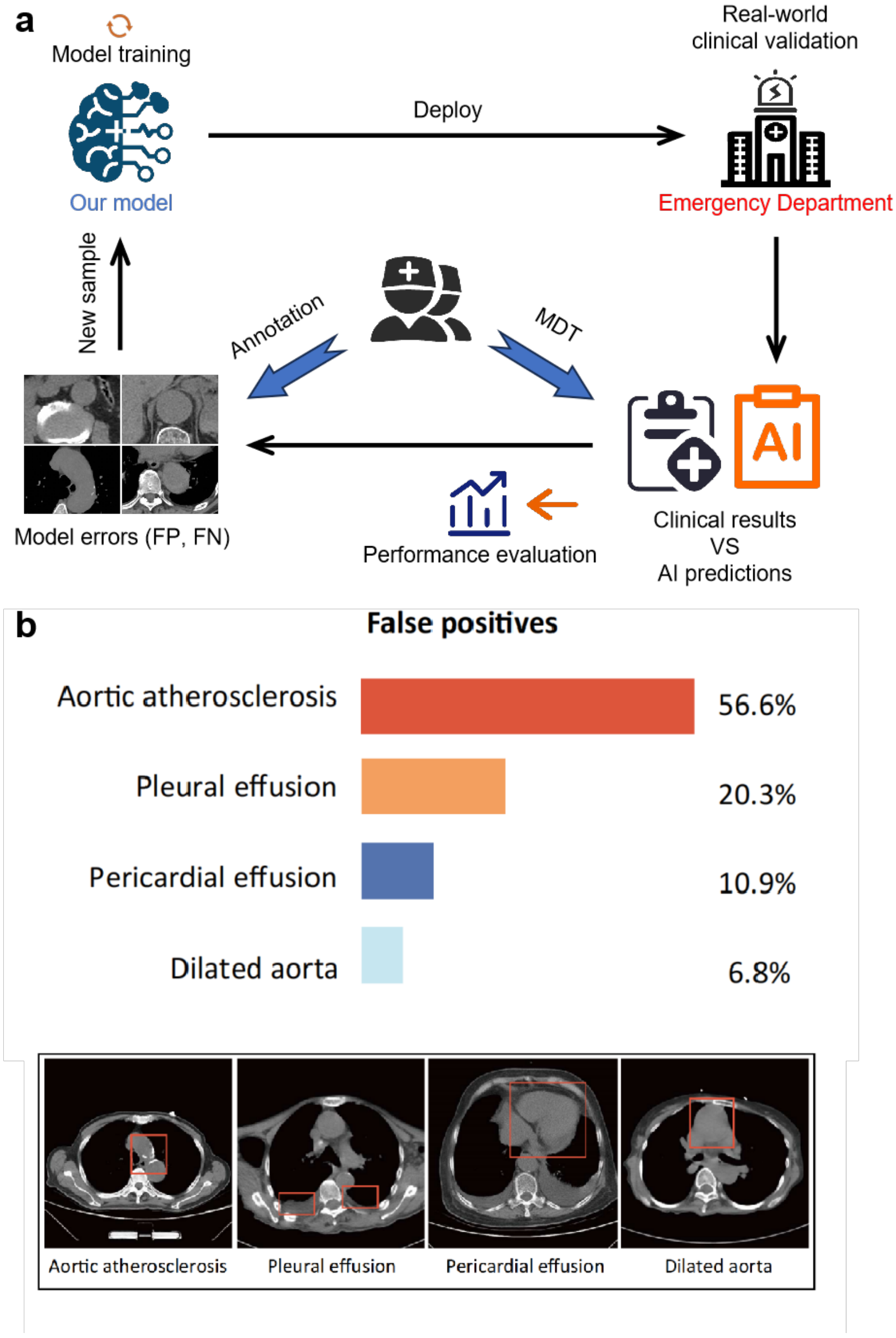

**Supplementary Fig. 3 | Model evolution. a.** We deployed our model for real-world retrospective study. The AI results were evaluated by the clinical results, e.g. standard-of-care (SOC) clinical decision, or multidisciplinary team (MDT) determination. The erroneous cases were further collected and annotated for model evolution. The upgraded model achieved a notable improvement in sensitivity by 11.6% - 19.2% while retaining a similar level of specificity. The upgraded model detected nine AAS cases that were missed by the original model. **b.** The distribution of the false-positive cases in real-world study (RW1) cohort. Aortic atherosclerosis, pleural effusion, pericardial effusion, and dilated aorta occupy the top four positions.

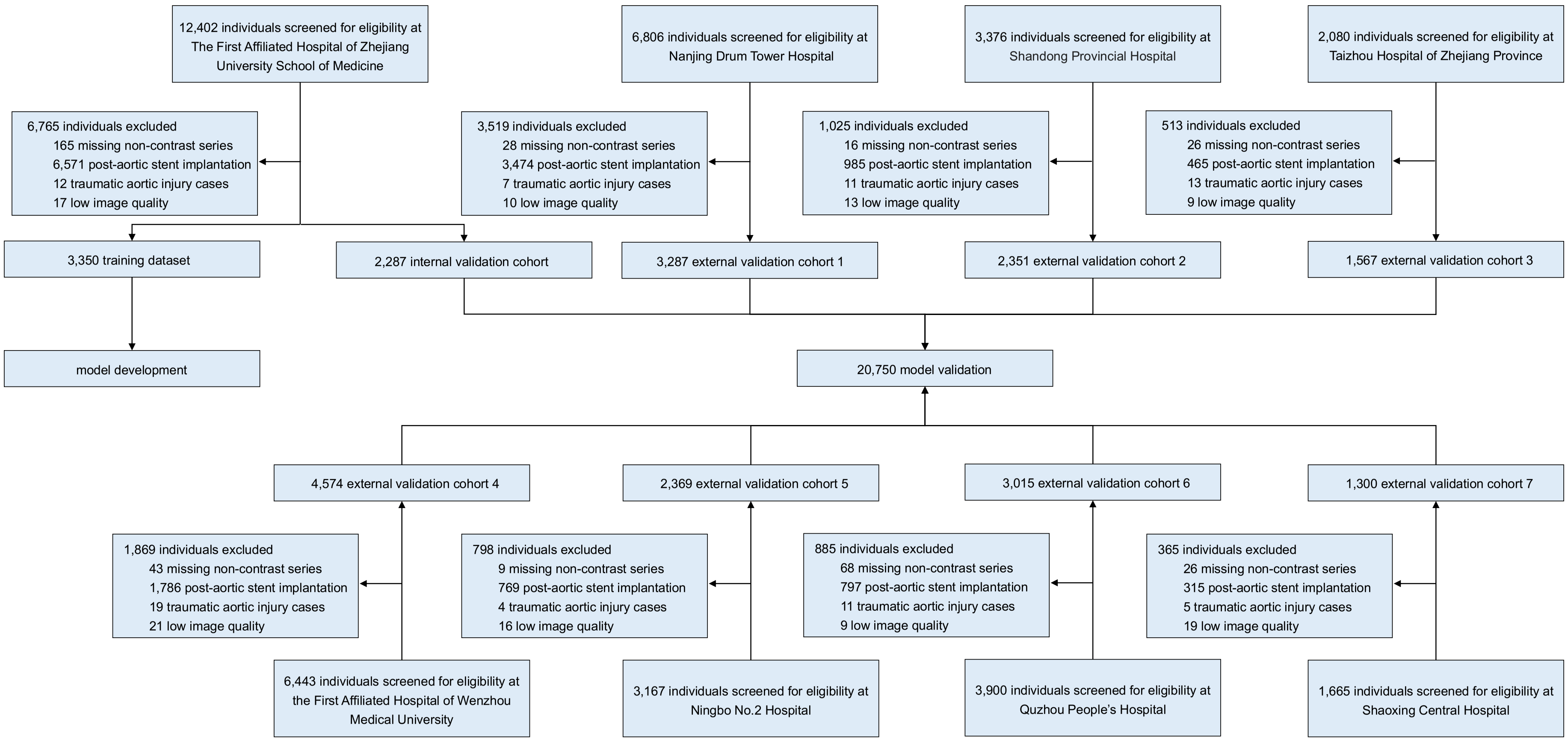
12,402 individuals screened for eligibility at The First Affiliated Hospital of Zhejiang University School of Medicine
6,765 individuals excluded
165 missing non-contrast series
6,571 post-aortic stent implantation
12 traumatic aortic injury cases
17 low image quality
3,350 training dataset
model development
2,287 internal validation cohort
6,806 individuals screened for eligibility at Nanjing Drum Tower Hospital
3,519 individuals excluded
28 missing non-contrast series
3,474 post-aortic stent implantation
7 traumatic aortic injury cases
10 low image quality
3,287 external validation cohort 1
3,376 individuals screened for eligibility at Shandong Provincial Hospital
1,025 individuals excluded
16 missing non-contrast series
985 post-aortic stent implantation
11 traumatic aortic injury cases
13 low image quality
2,351 external validation cohort 2
2,080 individuals screened for eligibility at Taizhou Hospital of Zhejiang Province
513 individuals excluded
26 missing non-contrast series
465 post-aortic stent implantation
13 traumatic aortic injury cases
9 low image quality
1,567 external validation cohort 3
20,750 model validation
4,574 external validation cohort 4
1,869 individuals excluded
43 missing non-contrast series
1,786 post-aortic stent implantation
19 traumatic aortic injury cases
21 low image quality
6,443 individuals screened for eligibility at the First Affiliated Hospital of Wenzhou Medical University
2,369 external validation cohort 5
798 individuals excluded
9 missing non-contrast series
769 post-aortic stent implantation
4 traumatic aortic injury cases
16 low image quality
3,167 individuals screened for eligibility at Ningbo No.2 Hospital
3,015 external validation cohort 6
885 individuals excluded
68 missing non-contrast series
797 post-aortic stent implantation
11 traumatic aortic injury cases
9 low image quality
3,900 individuals screened for eligibility at Quzhou People's Hospital
1,300 external validation cohort 7
365 individuals excluded
26 missing non-contrast series
315 post-aortic stent implantation
5 traumatic aortic injury cases
19 low image quality
1,665 individuals screened for eligibility at Shaoxing Central Hospital

**Supplementary Fig. 4 | Training, internal validation and external validation cohorts and sample selection.** This flow diagram shows patient inclusion and exclusion criteria in each cohort as well as the dataset partition for training, internal and external validation cohorts. The human-machine battle and human-machine assistance experiments were conducted on the internal validation cohort.

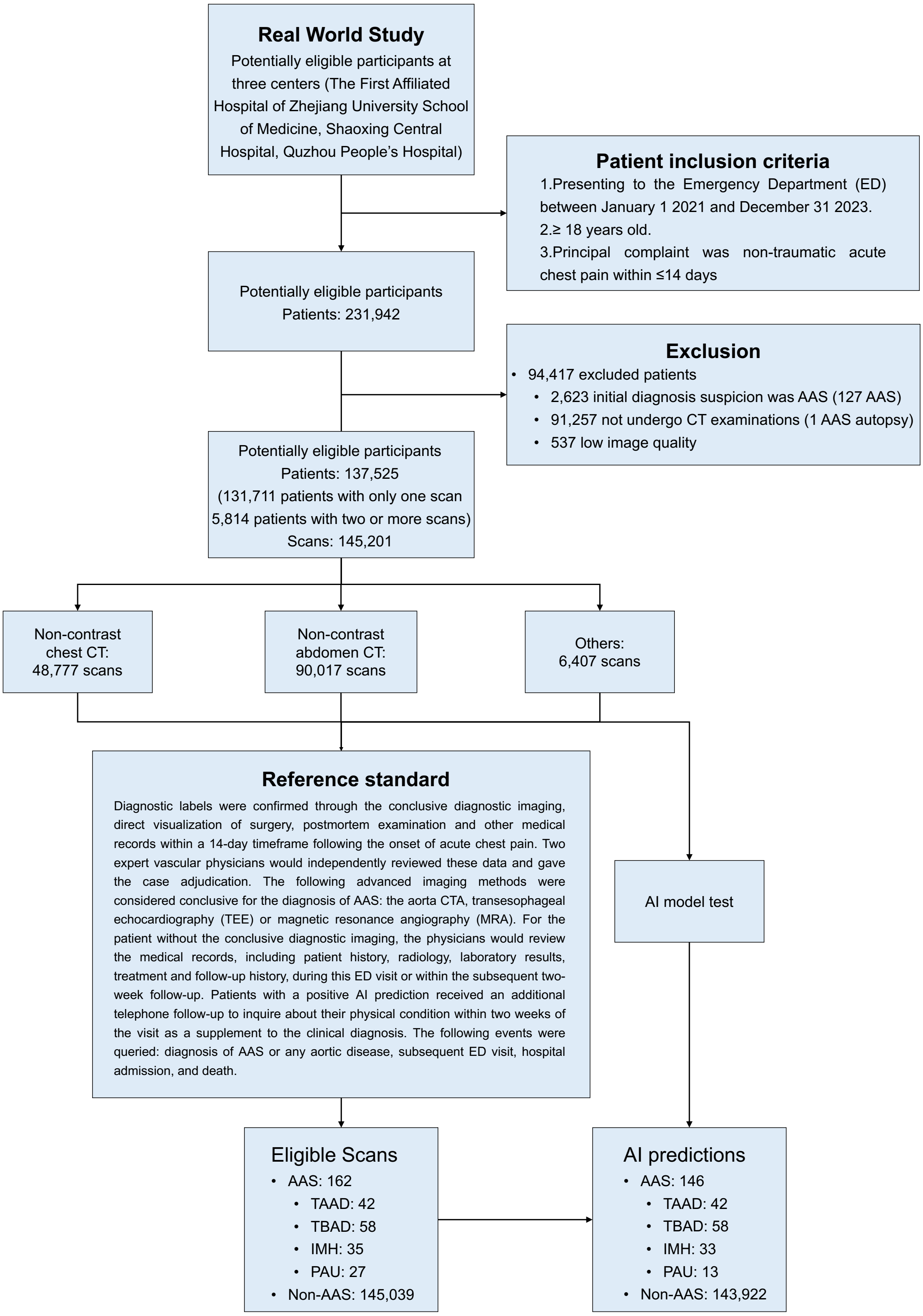

**Supplementary Fig. 5 | Overview of the cohort of the large-scale real-world retrospective study.** This flow diagram shows patients' inclusion and exclusion criteria, reference standard, and AI predictions.

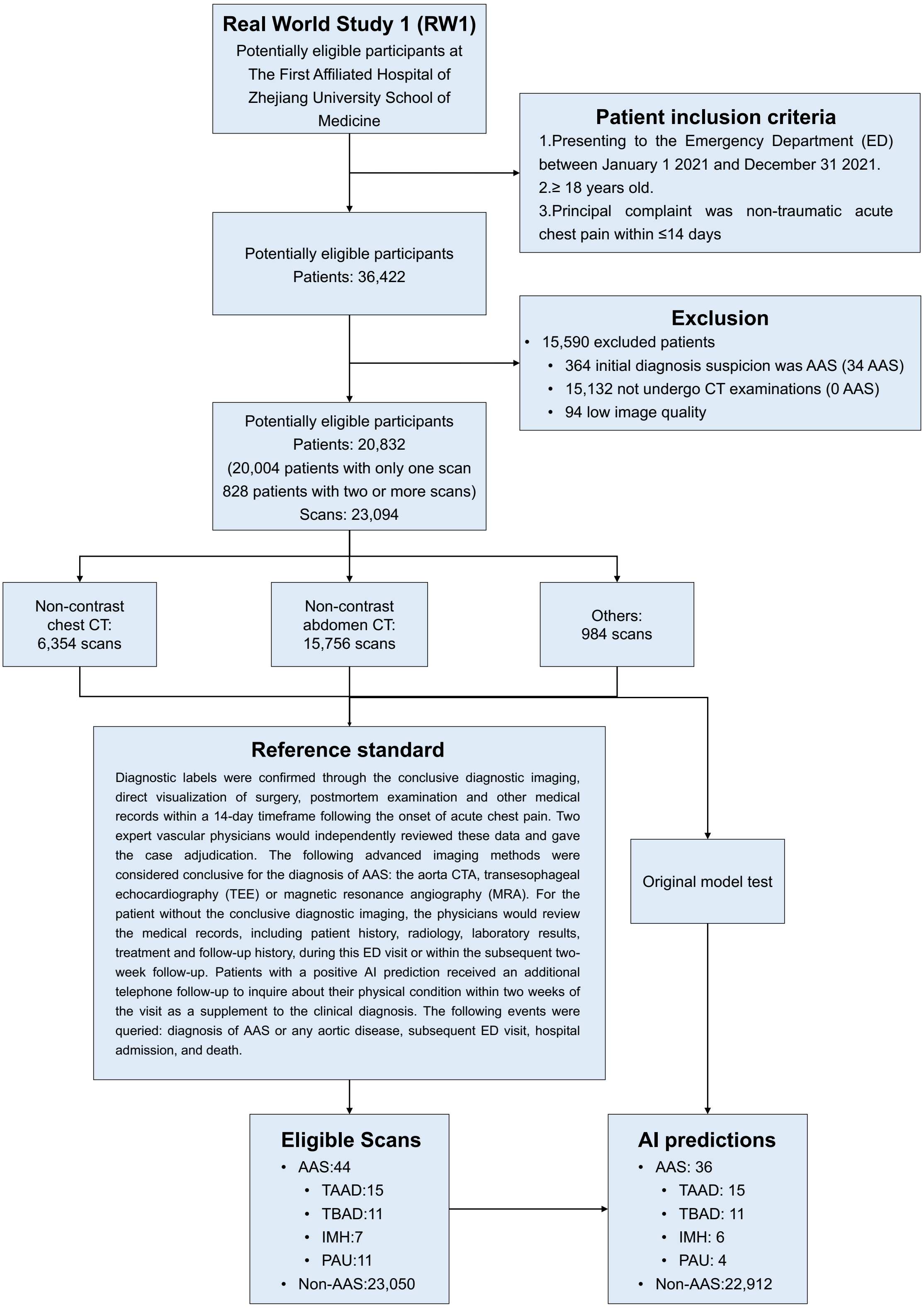

**Supplementary Fig. 6 | Overview of the workflow of the first real-world study (RW1) Cohort.**

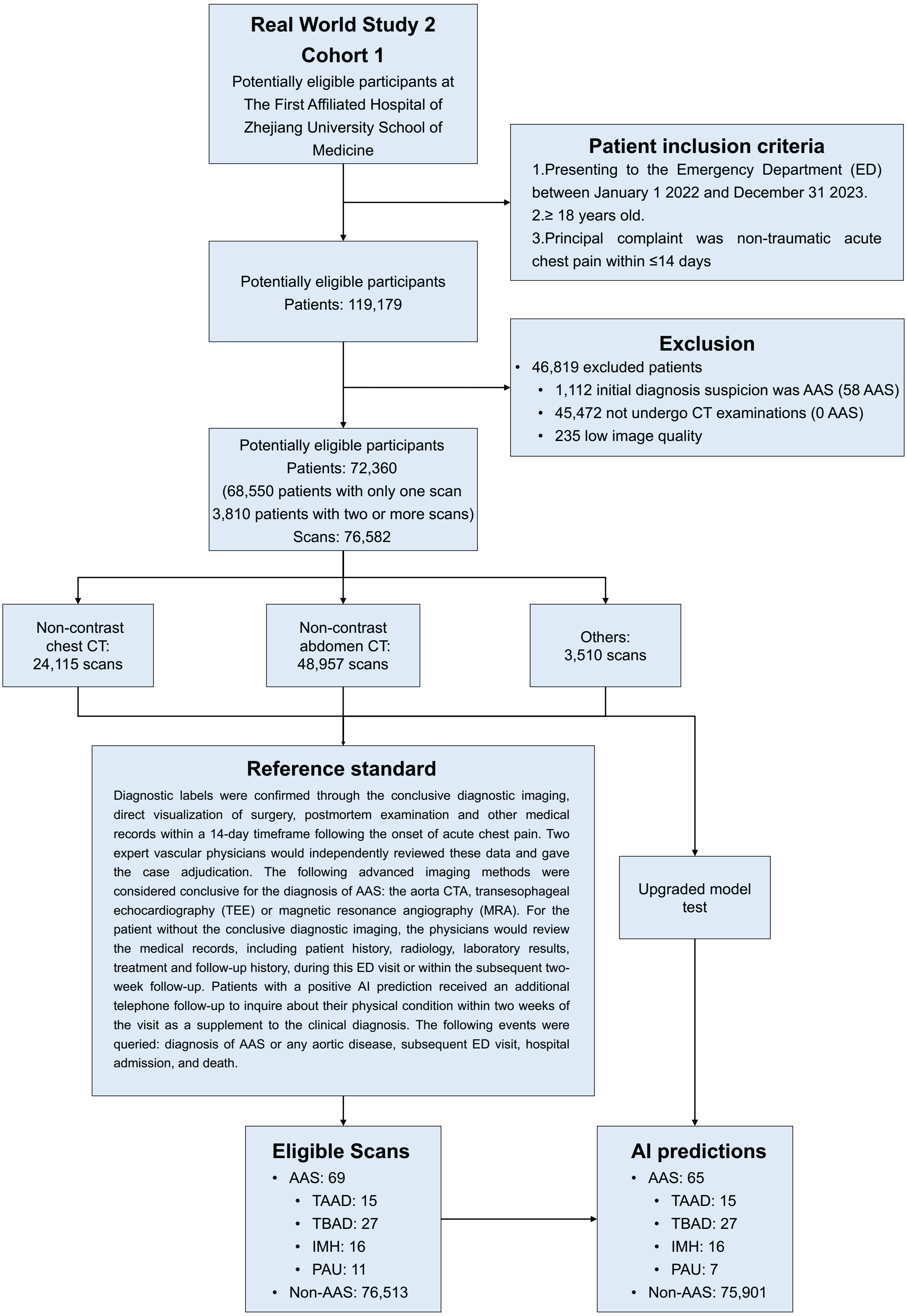

**Supplementary Fig. 7 | Overview of the workflow of the second real-world study (RW2) Cohort 1.**

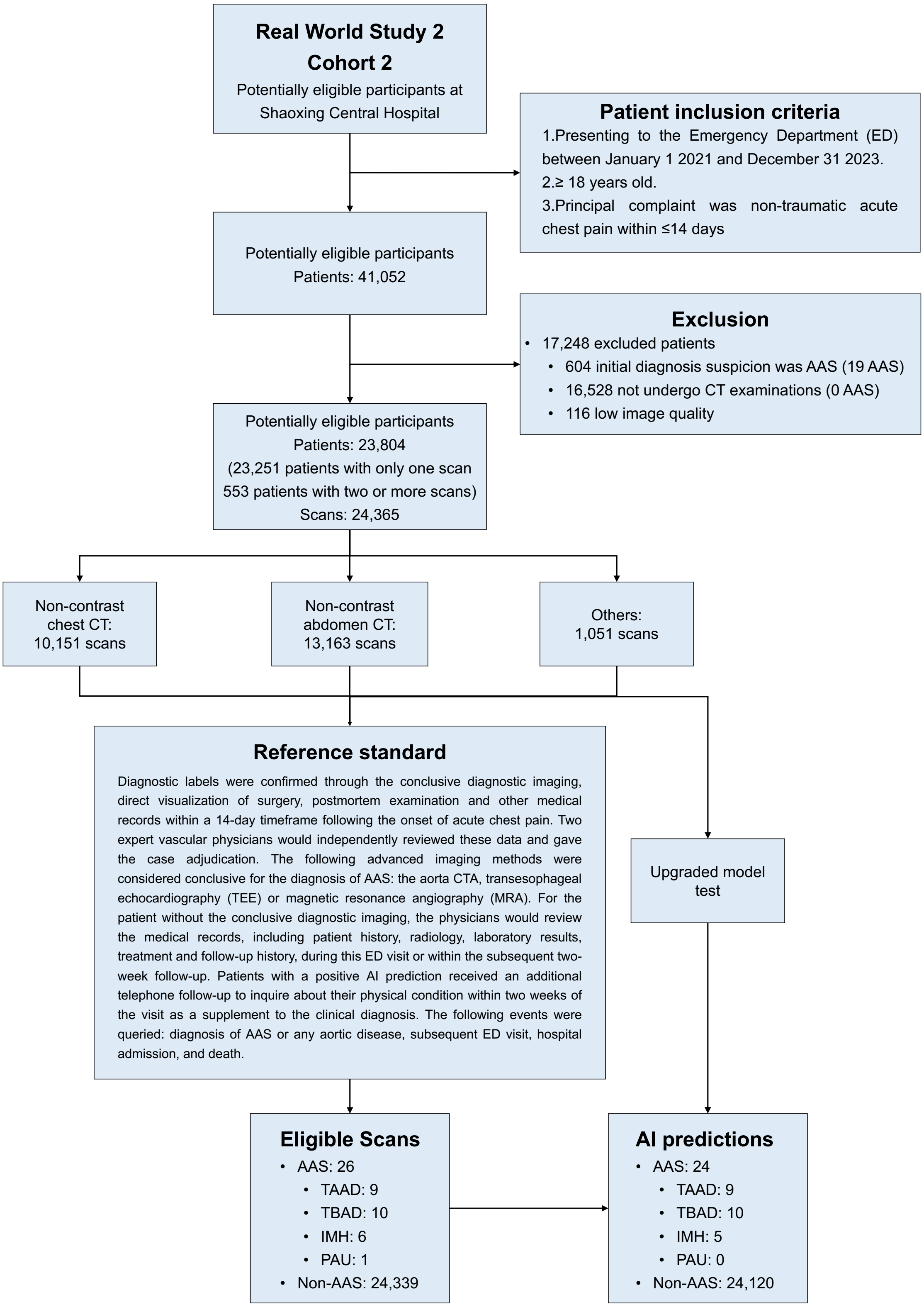

**Supplementary Fig. 8 | Overview of the workflow of the second real-world study (RW2) Cohort 2.**

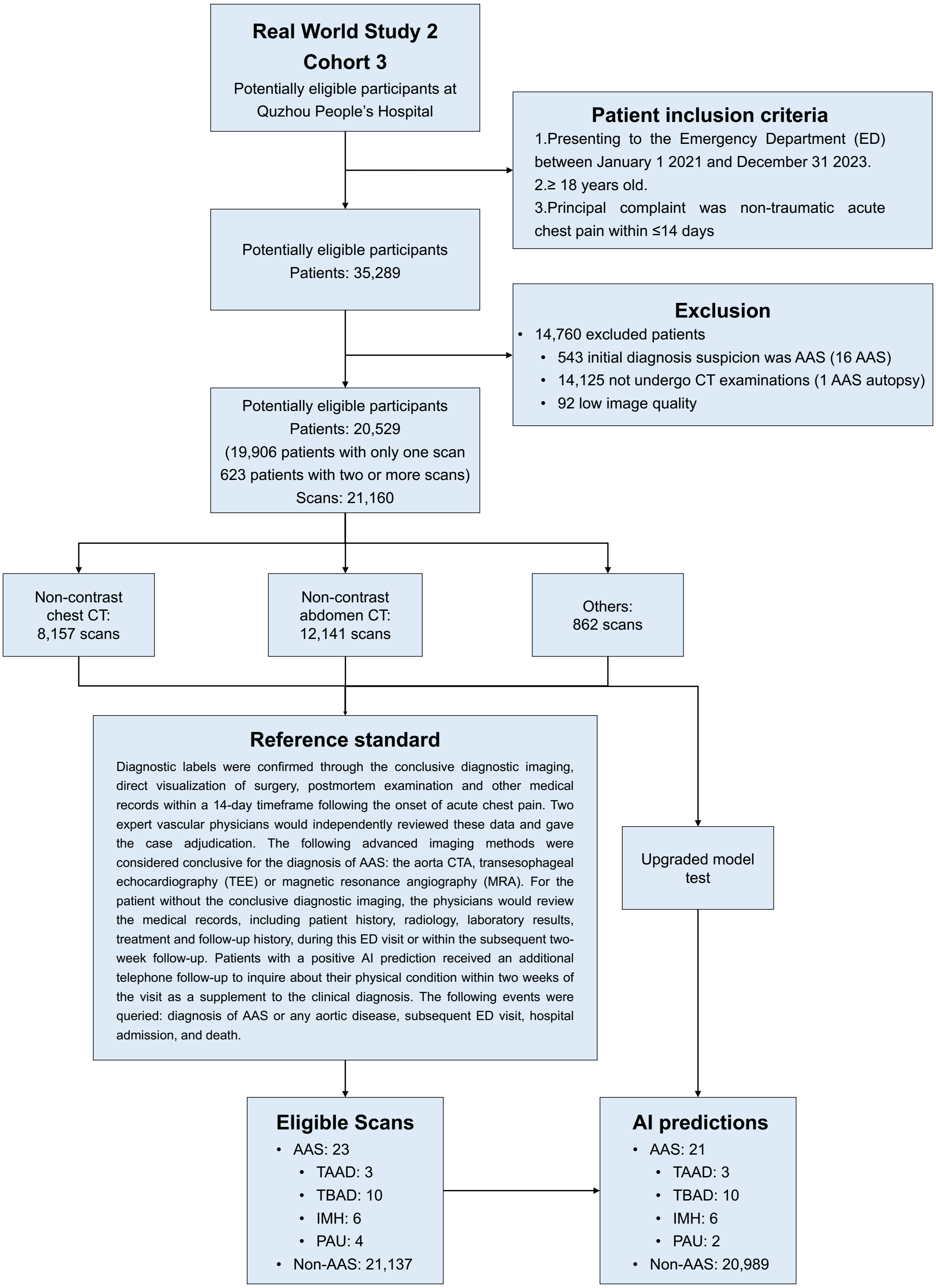

**Supplementary Fig. 9 | Overview of the workflow of the second real-world study (RW2) Cohort 3.**

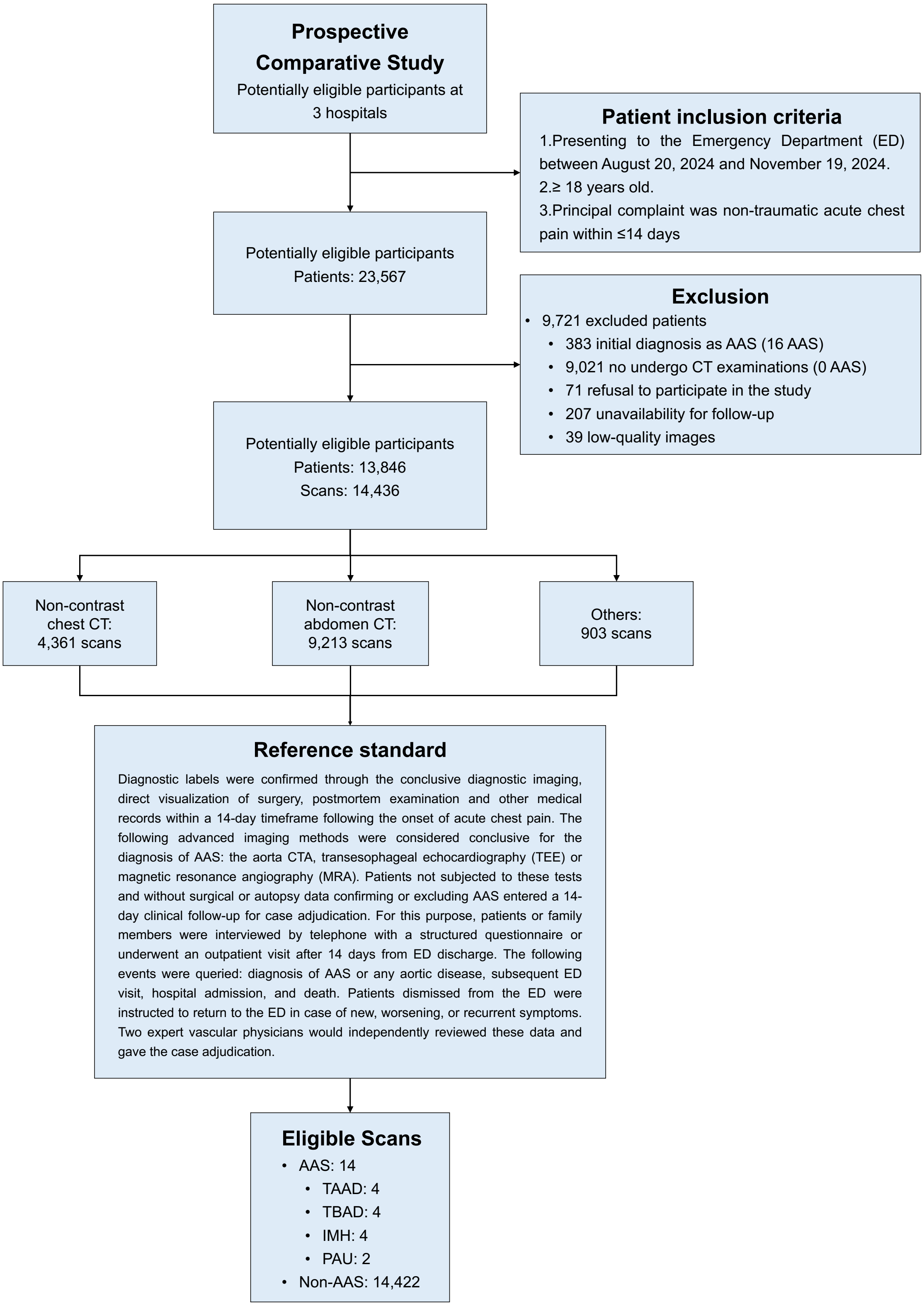

**Supplementary Fig. 10 | Overview of the comparative study cohort of the prospective multi-center study.** This flow diagram shows patients' inclusion and exclusion criteria, reference standard, and AI predictions.

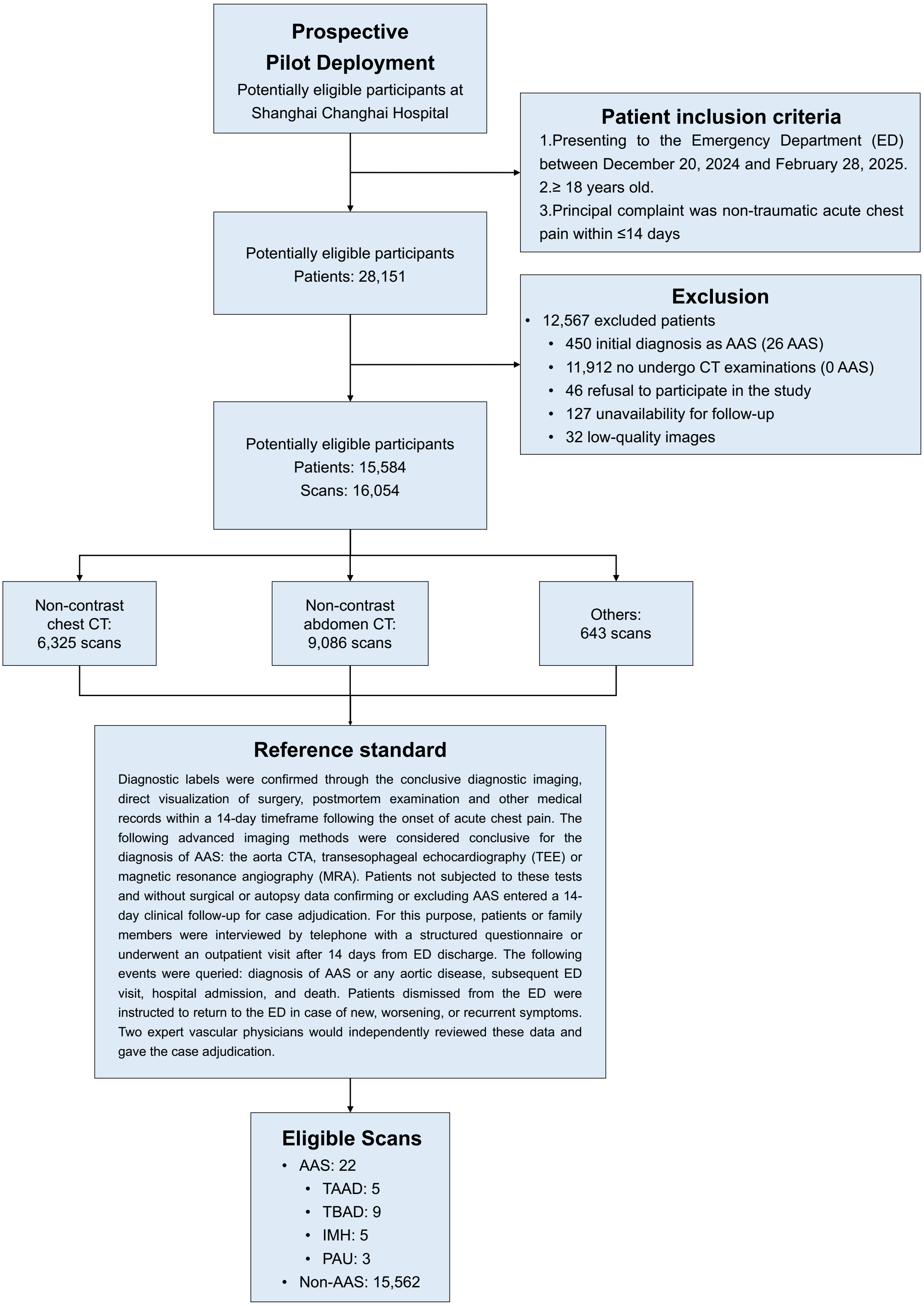

**Supplementary Fig. 11 | Overview of the pilot deployment study cohort of the prospective multi-center study.** This flow diagram shows patients' inclusion and exclusion criteria, reference standard, and AI predictions.

**Section 8: Supplementary Tables**

| Method | Sensitivity | Specificity | Accuracy | PPV | NPV | AUC |
|---|---|---|---|---|---|---|
| CMTGF | 0.740 (0.153) | 0.965 (0.034) | 0.916 (0.043) | 0.927 (0.070) | 0.859 (0.108) | 0.890 (0.089) |
| Swin UNETR | 0.877 (0.100) | 0.964 (0.046) | 0.945 (0.031) | 0.906 (0.082) | 0.927 (0.061) | 0.981 (0.017) |
| Swin-Tiny | 0.703 (0.123) | 0.906 (0.057) | 0.861 (0.061) | 0.819 (0.106) | 0.834 (0.083) | 0.875 (0.044) |
| ViT-Small | 0.677 (0.136) | 0.902 (0.060) | 0.851 (0.053) | 0.807 (0.115) | 0.821 (0.098) | 0.823 (0.074) |
| ResNet-50 | 0.670 (0.079) | 0.883 (0.143) | 0.838 (0.122) | 0.776 (0.064) | 0.815 (0.152) | 0.799 (0.045) |
| Ours | 0.988 (0.017) | 0.997 (0.006) | 0.996 (0.004) | 0.995 (0.005) | 0.993 (0.006) | 0.998 (0.003) |

**Supplementary Table 1 |** Results of the proposed model and other state-of-the-art models to detect patients with AAS in five-fold cross validation of training dataset (n = 3,350). Data are presented as the mean number (standard deviation). PPV, Positive Predictive Value; NPV, Negative Predictive Value; AUC, Area Under the Curve; CMTGF, Cascaded Multi-Task Generative Framework; Swin UNETR, Swin U-Net Transformer; Swin-Tiny, Swin Transformer-Tiny; ViT-Small, Vision Transformer -Small.

| | Internal Cohort FAHZU | External Cohort 1 NDTH | External Cohort 2 SPH | External Cohort 3 TZH | External Cohort 4 FAHWMU | External Cohort 5 N2H | External Cohort 6 QPH | External Cohort 7 SCH |
|---|---|---|---|---|---|---|---|---|
| **BRIER SCORE** | **0.069** | **0.072** | **0.071** | **0.079** | **0.073** | **0.084** | **0.077** | **0.082** |
| RELIABILITY | 0.037 | 0.032 | 0.033 | 0.030 | 0.031 | 0.037 | 0.036 | 0.039 |
| RESOLUTION | 0.194 | 0.199 | 0.204 | 0.181 | 0.202 | 0.140 | 0.161 | 0.133 |
| UNCERTAINTY | 0.227 | 0.239 | 0.243 | 0.230 | 0.244 | 0.187 | 0.202 | 0.176 |

**Supplementary Table 2 |** Calibration analysis (brier score) of our model to detect patients with AAS on internal and external cohorts.

| Reader | Sensitivity | Δ | 95% CI | p-value | Specificity | Δ | 95% CI | p-value |
|---|---|---|---|---|---|---|---|---|
| DL model | 0.984 | - | - | - | 0.948 | - | - | - |
| T1 | 0.416 | 0.568 | (0.531-0.603) | < 0.001 | 0.801 | 0.147 | (0.124-0.170) | < 0.001 |
| T2 | 0.508 | 0.476 | (0.440-0.511) | < 0.001 | 0.741 | 0.207 | (0.182-0.232) | < 0.001 |
| T3 | 0.442 | 0.542 | (0.507-0.577) | < 0.001 | 0.796 | 0.152 | (0.129-0.176) | < 0.001 |
| T4 | 0.338 | 0.646 | (0.611-0.679) | < 0.001 | 0.932 | 0.016 | (-0.002-0.032) | 0.474 |
| B1 | 0.585 | 0.399 | (0.365-0.434) | < 0.001 | 0.842 | 0.106 | (0.084-0.128) | < 0.001 |
| B2 | 0.652 | 0.332 | (0.298-0.366) | < 0.001 | 0.800 | 0.148 | (0.125-0.172) | < 0.001 |
| B3 | 0.626 | 0.358 | (0.322-0.392) | < 0.001 | 0.832 | 0.116 | (0.095-0.138) | < 0.001 |
| B4 | 0.604 | 0.380 | (0.345-0.415) | < 0.001 | 0.887 | 0.061 | (0.041-0.080) | 0.041 |
| E1 | 0.790 | 0.194 | (0.164-0.224) | < 0.001 | 0.912 | 0.036 | (0.018-0.054) | 0.114 |
| E2 | 0.771 | 0.213 | (0.182-0.244) | < 0.001 | 0.952 | -0.004 | (-0.019-0.012) | 0.923 |
| E3 | 0.813 | 0.171 | (0.143-0.200) | < 0.001 | 0.903 | 0.045 | (0.027-0.064) | 0.052 |
| Mean T | 0.426 | 0.558 | (0.538-0.576) | < 0.001 | 0.817 | 0.131 | (0.115-0.145) | < 0.001 |
| Mean B | 0.617 | 0.367 | (0.348-0.386) | < 0.001 | 0.840 | 0.108 | (0.093-0.122) | < 0.001 |
| Mean E | 0.791 | 0.193 | (0.174-0.211) | < 0.001 | 0.922 | 0.026 | (0.012-0.039) | 0.203 |

(a) Reader (non-contrast CT) vs. DL model (non-contrast CT) by the evaluation of sensitivity and specificity for AAS identification.

| Reader | Sensitivity (TAAD) | Δ | 95% CI | p-value | Sensitivity (TBAD) | Δ | 95% CI | p-value |
|---|---|---|---|---|---|---|---|---|
| DL model | 0.995 | - | - | - | 0.996 | - | - | - |
| T1 | 0.441 | 0.554 | (0.479-0.622) | < 0.001 | 0.500 | 0.496 | (0.435-0.556) | < 0.001 |
| T2 | 0.521 | 0.474 | (0.399-0.548) | < 0.001 | 0.677 | 0.319 | (0.262-0.379) | < 0.001 |
| T3 | 0.516 | 0.479 | (0.404-0.553) | < 0.001 | 0.520 | 0.496 | (0.434-0.558) | < 0.001 |
| T4 | 0.181 | 0.814 | (0.761-0.867) | < 0.001 | 0.512 | 0.484 | (0.419-0.548) | < 0.001 |
| B1 | 0.649 | 0.346 | (0.277-0.415) | < 0.001 | 0.694 | 0.302 | (0.246-0.359) | < 0.001 |
| B2 | 0.793 | 0.202 | (0.144-0.261) | < 0.001 | 0.738 | 0.258 | (0.202-0.315) | < 0.001 |
| B3 | 0.676 | 0.319 | (0.255-0.388) | < 0.001 | 0.778 | 0.218 | (0.165-0.270) | < 0.001 |
| B4 | 0.665 | 0.330 | (0.261-0.399) | < 0.001 | 0.750 | 0.246 | (0.194-0.302) | < 0.001 |
| E1 | 0.878 | 0.117 | (0.069-0.165) | < 0.001 | 0.859 | 0.137 | (0.093-0.181) | < 0.001 |
| E2 | 0.851 | 0.144 | (0.090-0.197) | < 0.001 | 0.823 | 0.173 | (0.125-0.222) | < 0.001 |
| E3 | 0.830 | 0.165 | (0.112-0.223) | < 0.001 | 0.927 | 0.069 | (0.036-0.105) | 0.008 |
| Mean T | 0.415 | 0.580 | (0.543-0.617) | < 0.001 | 0.552 | 0.444 | (0.411-0.476) | < 0.001 |
| Mean B | 0.695 | 0.300 | (0.265-0.334) | < 0.001 | 0.740 | 0.256 | (0.228-0.284) | < 0.001 |
| Mean E | 0.853 | 0.142 | (0.112-0.174) | < 0.001 | 0.870 | 0.126 | (0.101-0.152) | < 0.001 |

(b) Reader (non-contrast CT) vs. DL model (non-contrast CT) by the evaluation of sensitivity for TAAD and TBAD identification.

| Reader | Sensitivity (IMH) | Δ | 95% CI | p-value | Sensitivity (PAU) | Δ | 95% CI | p-value |
|---|---|---|---|---|---|---|---|---|
| DL model | 0.980 | - | - | - | 0.955 | - | - | - |
| T1 | 0.433 | 0.547 | (0.473-0.616) | < 0.001 | 0.231 | 0.724 | (0.647-0.801) | < 0.001 |
| T2 | 0.488 | 0.492 | (0.419-0.567) | < 0.001 | 0.250 | 0.705 | (0.628-0.776) | < 0.001 |
| T3 | 0.458 | 0.522 | (0.448-0.596) | < 0.001 | 0.205 | 0.750 | (0.679-0.821) | < 0.001 |
| T4 | 0.399 | 0.581 | (0.512-0.650) | < 0.001 | 0.173 | 0.782 | (0.712-0.846) | < 0.001 |
| B1 | 0.596 | 0.384 | (0.315-0.453) | < 0.001 | 0.321 | 0.634 | (0.558-0.712) | < 0.001 |
| B2 | 0.621 | 0.359 | (0.291-0.429) | < 0.001 | 0.385 | 0.570 | (0.487-0.654) | < 0.001 |
| B3 | 0.606 | 0.374 | (0.305-0.443) | < 0.001 | 0.353 | 0.602 | (0.519-0.686) | < 0.001 |
| B4 | 0.576 | 0.404 | (0.335-0.473) | < 0.001 | 0.333 | 0.622 | (0.538-0.699) | < 0.001 |
| E1 | 0.823 | 0.157 | (0.103-0.217) | < 0.001 | 0.532 | 0.423 | (0.340-0.506) | < 0.001 |
| E2 | 0.808 | 0.172 | (0.113-0.232) | < 0.001 | 0.545 | 0.410 | (0.321-0.494) | < 0.001 |
| E3 | 0.837 | 0.143 | (0.089-0.197) | < 0.001 | 0.577 | 0.378 | (0.295-0.462) | < 0.001 |
| Mean T | 0.445 | 0.534 | (0.495-0.574) | < 0.001 | 0.215 | 0.740 | (0.692-0.785) | < 0.001 |
| Mean B | 0.600 | 0.380 | (0.341-0.419) | < 0.001 | 0.348 | 0.607 | (0.558-0.655) | < 0.001 |
| Mean E | 0.823 | 0.157 | (0.122-0.194) | < 0.001 | 0.551 | 0.404 | (0.348-0.457) | < 0.001 |

(c) Reader (non-contrast CT) vs. DL model (non-contrast CT) by the evaluation of sensitivity for IMH and PAU identification.

**Supplementary Table 3** | Reader (non-contrast CT) vs. DL model (non-contrast CT) by the evaluation of sensitivity and specificity for AAS and four subtypes identification. Two-sided permutation tests were used to compute the statistical difference. T, trainee; B, Board-certificated; E, expert; TAAD, Stanford Type A dissection; TBAD, Stanford Type B dissection; IMH, intramural hematoma; PAU, penetrating atherosclerotic ulcer.

| Reader | Sensitivity | Sensitivity-A | Δ | 95% CI | p-value | Specificity | Specificity-A | Δ | 95% CI | p-value |
|---|---|---|---|---|---|---|---|---|---|---|
| T1 | 0.416 | 0.863 | 0.447 | (0.404-0.488) | < 0.001 | 0.801 | 0.882 | 0.081 | (0.055-0.107) | 0.002 |
| T2 | 0.508 | 0.921 | 0.413 | (0.372-0.450) | < 0.001 | 0.741 | 0.885 | 0.144 | (0.117-0.173) | < 0.001 |
| T3 | 0.442 | 0.824 | 0.382 | (0.338-0.425) | < 0.001 | 0.796 | 0.938 | 0.142 | (0.119-0.166) | < 0.001 |
| T4 | 0.338 | 0.707 | 0.369 | (0.322-0.415) | < 0.001 | 0.932 | 0.947 | 0.015 | (-0.002-0.032) | 0.413 |
| B1 | 0.585 | 0.833 | 0.248 | (0.204-0.291) | < 0.001 | 0.842 | 0.914 | 0.072 | (0.049-0.095) | 0.004 |
| B2 | 0.652 | 0.864 | 0.212 | (0.172-0.253) | < 0.001 | 0.800 | 0.877 | 0.077 | (0.052-0.104) | 0.003 |
| B3 | 0.626 | 0.902 | 0.276 | (0.235-0.314) | < 0.001 | 0.832 | 0.847 | 0.015 | (-0.011-0.041) | 0.458 |
| B4 | 0.604 | 0.840 | 0.236 | (0.194-0.279) | < 0.001 | 0.887 | 0.917 | 0.030 | (0.008-0.051) | 0.228 |
| E1 | 0.790 | 0.909 | 0.119 | (0.084-0.153) | < 0.001 | 0.912 | 0.930 | 0.018 | (-0.001-0.038) | 0.495 |
| E2 | 0.771 | 0.923 | 0.152 | (0.118-0.187) | < 0.001 | 0.952 | 0.966 | 0.014 | (0.000-0.029) | 0.408 |
| E3 | 0.813 | 0.943 | 0.130 | (0.099-0.162) | < 0.001 | 0.903 | 0.948 | 0.045 | (0.026-0.064) | 0.062 |
| Mean T | 0.426 | 0.829 | 0.403 | (0.381-0.424) | < 0.001 | 0.817 | 0.913 | 0.096 | (0.083-0.108) | < 0.001 |
| Mean B | 0.617 | 0.860 | 0.243 | (0.223-0.264) | < 0.001 | 0.840 | 0.889 | 0.049 | (0.036-0.061) | 0.048 |
| Mean E | 0.791 | 0.925 | 0.134 | (0.115-0.153) | < 0.001 | 0.922 | 0.948 | 0.026 | (0.016-0.036) | 0.204 |

(a) Reader (non-contrast CT) vs. Reader + DL model assistance (non-contrast CT) by the evaluation of sensitivity and specificity for AAS identification.

| Reader | Sensitivity (TAAD) | Sensitivity-A (TAAD) | Δ | 95% CI | p-value | Sensitivity (TBAD) | Sensitivity-A (TBAD) | Δ | 95% CI | p-value |
|---|---|---|---|---|---|---|---|---|---|---|
| T1 | 0.441 | 0.915 | 0.474 | (0.394-0.553) | < 0.001 | 0.500 | 0.919 | 0.419 | (0.347-0.492) | < 0.001 |
| T2 | 0.521 | 0.952 | 0.431 | (0.351-0.505) | < 0.001 | 0.677 | 0.960 | 0.283 | (0.218-0.347) | < 0.001 |
| T3 | 0.516 | 0.803 | 0.287 | (0.197-0.378) | < 0.001 | 0.520 | 0.927 | 0.407 | (0.339-0.476) | < 0.001 |
| T4 | 0.181 | 0.564 | 0.383 | (0.293-0.473) | < 0.001 | 0.512 | 0.875 | 0.363 | (0.286-0.435) | < 0.001 |
| B1 | 0.649 | 0.899 | 0.250 | (0.170-0.330) | < 0.001 | 0.694 | 0.887 | 0.193 | (0.125-0.262) | < 0.001 |
| B2 | 0.793 | 0.941 | 0.148 | (0.080-0.218) | < 0.001 | 0.738 | 0.952 | 0.214 | (0.153-0.274) | < 0.001 |
| B3 | 0.676 | 0.936 | 0.260 | (0.186-0.335) | < 0.001 | 0.778 | 0.980 | 0.202 | (0.149-0.258) | < 0.001 |
| B4 | 0.665 | 0.872 | 0.207 | (0.122-0.287) | < 0.001 | 0.750 | 0.911 | 0.161 | (0.097-0.226) | < 0.001 |
| E1 | 0.878 | 0.957 | 0.079 | (0.027-0.133) | 0.016 | 0.859 | 0.956 | 0.097 | (0.048-0.145) | 0.005 |
| E2 | 0.851 | 0.963 | 0.112 | (0.053-0.170) | < 0.001 | 0.823 | 0.988 | 0.165 | (0.117-0.218) | < 0.001 |
| E3 | 0.830 | 0.984 | 0.154 | (0.101-0.213) | < 0.001 | 0.927 | 0.972 | 0.045 | (0.008-0.085) | 0.019 |
| Mean T | 0.415 | 0.809 | 0.394 | (0.348-0.438) | < 0.001 | 0.552 | 0.920 | 0.368 | (0.333-0.403) | < 0.001 |
| Mean B | 0.695 | 0.912 | 0.217 | (0.177-0.255) | < 0.001 | 0.740 | 0.932 | 0.192 | (0.161-0.225) | < 0.001 |
| Mean E | 0.853 | 0.968 | 0.115 | (0.083-0.149) | < 0.001 | 0.870 | 0.972 | 0.102 | (0.075-0.129) | < 0.001 |

(b) Reader (non-contrast CT) vs. Reader + DL model assistance (non-contrast CT) by the evaluation of sensitivity and specificity for TAAD and TBAD identification.

| Reader | Sensitivity (IMH) | Sensitivity-A (IMH) | Δ | 95% CI | p-value | Sensitivity (PAU) | Sensitivity-A (PAU) | Δ | 95% CI | p-value |
|---|---|---|---|---|---|---|---|---|---|---|
| T1 | 0.433 | 0.833 | 0.400 | (0.310-0.483) | < 0.001 | 0.231 | 0.750 | 0.519 | (0.423-0.609) | < 0.001 |
| T2 | 0.488 | 0.926 | 0.438 | (0.360-0.517) | < 0.001 | 0.250 | 0.814 | 0.564 | (0.474-0.654) | < 0.001 |
| T3 | 0.458 | 0.768 | 0.310 | (0.217-0.399) | < 0.001 | 0.205 | 0.756 | 0.551 | (0.462-0.641) | < 0.001 |
| T4 | 0.399 | 0.685 | 0.286 | (0.192-0.379) | < 0.001 | 0.173 | 0.641 | 0.468 | (0.372-0.564) | < 0.001 |
| B1 | 0.596 | 0.833 | 0.237 | (0.148-0.320) | < 0.001 | 0.321 | 0.667 | 0.346 | (0.244-0.449) | < 0.001 |
| B2 | 0.621 | 0.867 | 0.246 | (0.163-0.325) | < 0.001 | 0.385 | 0.628 | 0.243 | (0.135-0.353) | < 0.001 |
| B3 | 0.606 | 0.897 | 0.291 | (0.212-0.369) | < 0.001 | 0.353 | 0.744 | 0.391 | (0.288-0.487) | < 0.001 |
| B4 | 0.576 | 0.857 | 0.281 | (0.197-0.365) | < 0.001 | 0.333 | 0.667 | 0.334 | (0.231-0.442) | < 0.001 |
| E1 | 0.823 | 0.897 | 0.074 | (0.010-0.143) | 0.095 | 0.532 | 0.795 | 0.263 | (0.160-0.365) | < 0.001 |
| E2 | 0.808 | 0.916 | 0.108 | (0.044-0.172) | 0.001 | 0.545 | 0.782 | 0.237 | (0.135-0.340) | < 0.001 |
| E3 | 0.837 | 0.951 | 0.114 | (0.054-0.172) | < 0.001 | 0.577 | 0.840 | 0.263 | (0.167-0.359) | < 0.001 |
| Mean T | 0.445 | 0.803 | 0.358 | (0.315-0.403) | < 0.001 | 0.215 | 0.740 | 0.525 | (0.479-0.572) | < 0.001 |
| Mean B | 0.600 | 0.863 | 0.263 | (0.223-0.304) | < 0.001 | 0.348 | 0.676 | 0.328 | (0.276-0.381) | < 0.001 |
| Mean E | 0.823 | 0.921 | 0.098 | (0.062-0.136) | 0.002 | 0.551 | 0.806 | 0.255 | (0.199-0.312) | < 0.001 |

(c) Reader (non-contrast CT) vs. Reader + DL model assistance (non-contrast CT) by the evaluation of sensitivity and specificity for IMH and PAU identification.

**Supplementary Table 4 |** Reader (non-contrast CT) vs. Reader + DL model assistance (non-contrast CT) by the evaluation of sensitivity and specificity for AAS and four subtypes identification. Two-sided permutation tests were used to compute the statistical difference. T, trainee; B, Board-certificated; E, expert; Sensitivity-A, sensitivity with DL model assistance; Specificity-A, specificity with DL model assistance; TAAD, Stanford Type A dissection; TBAD, Stanford Type B dissection; IMH, intramural hematoma; PAU, penetrating atherosclerotic ulcer.

| Patient | Definitive diagnosis | Age | Gender | Chief complaint | Primary CT protocol | Rounds of investigations |
|---|---|---|---|---|---|---|
| 1 | PAU | 88 | Male | Chest tightness and shortness of breath for 1 hour | Chest CT | / |
| 2 | TBAD | 69 | Male | Shock for 1 hour | Pulmonary CT Angiography* | Preliminary diagnosis of pulmonary embolism, symptomatic treatment given, recheck with chest CT* after 5 hours |
| 3 | IMH | 71 | Female | Abdominal pain and bloating 10 days ago, with chest tightness and shortness of breath for over 2 days | Chest CT* | Abdominal CT |
| 4 | TAAD | 66 | Male | Chest pain and tightness for half an hour | Pulmonary CT Angiography* | / |
| 5 | PAU | 66 | Male | Chest pain accompanied by upper abdominal pain for 12 hours | Chest CT | / |
| 6 | PAU | 79 | Female | Upper abdominal pain for 2 days | Abdominal CT* | / |
| 7 | PAU | 84 | Male | Abdominal pain for 1 day | Abdominal CT | Abdominal CT with contrast |
| 8 | PAU | 83 | Female | Upper abdominal pain for 3 days | Abdominal CT | / |
| 9 | TAAD | 70 | Female | Right shoulder pain and  and chest tightness for 1 hour | Chest CT* | / |
| 10 | TBAD | 68 | Male | Abdominal pain for 15 hours | Abdominal CT* | / |
| 11 | TBAD | 48 | Female | Bloating for 12 hours, abdominal pain for 1 hour | Abdominal CT* | / |
| 12 | PAU | 89 | Female | Chest tightness and shortness of breath for 3 days | Chest CT* | / |
| 13 | TBAD | 46 | Male | Abdominal pain for 24 hours | Abdominal CT* | / |
| 14 | IMH | 70 | Male | Chest pain for 2 hours | Chest CT* | Pulmonary CT Angiography* |
| 15 | TBAD | 58 | Male | Discomfort in abdomen and back for 1 hour | Abdominal CT* | Chest CT* |
| 16 | TAAD | 71 | Male | Chest tightness for 12 hours | Chest CT* | Abdominal CT* |
| 17 | IMH | 74 | Male | Chest pain for half an hour | Chest CT* | / |
| 18 | IMH | 92 | Female | Abdominal pain for 23 hours | Abdominal CT* | Chest CT* |
| 19 | TAAD | 62 | Male | Chest pain for half an hour | Pulmonary CT Angiography* | / |
| 20 | TBAD | 75 | Male | Chest and back pain for 2 hours | Chest CT* | / |
| 21 | TAAD | 42 | Male | Chest tightness and shortness of breath for 6 hours | Chest CT* | Abdominal CT* |
| 22 | TAAD | 41 | Male | Abdominal pain with weakness in the lower limbs for half an hour | Abdominal CT* | The bilateral lower extremities CT Angiography* |
| 23 | TAAD | 49 | Male | Chest tightness  for 17 hours | Chest CT* | / |
| 24 | PAU | 78 | Male | Abdominal pain for 8 hours | Abdominal CT | / |
| 25 | PAU | 85 | Male | Chest tightness for 1 week | Chest CT* | / |
| 26 | TAAD | 80 | Female | Chest pain for half an hour | Chest CT* | / |
| 27 | PAU | 76 | Male | Chest pain for 2 hours | Chest CT | / |
| 28 | TBAD | 72 | Male | Chest tightness and shortness of breath for 6 hours | Chest CT* | Abdominal CT* |
| 29 | TBAD | 43 | Male | Chest tightness for 4 days | Chest CT* | / |
| 30 | TAAD | 46 | Male | Chest pain,tightness and shortness of breath for 3 hours | Chest CT* | Abdominal CT* |
| 31 | PAU | 62 | Male | Chest  pain for 1 day | Coronary CT Angiography* | / |
| 32 | TAAD | 72 | Male | Abdominal pain for 4 hours | Abdominal CT* | Pulmonary CT Angiography* (and Ultrasound Cardiogram) |

**Supplementary Table 5 |** Cases with initial false suspicion in the clinical evaluations of RW1. The superscript * represents the cases that were successfully detected by model. Details of case ID 32 are shown in Extended Data Fig. 4c of the manuscript. Chest CT represents non-contrast chest CT; Abdominal CT represents non-contrast abdominal CT.

| Patient | Definitive diagnosis | Age | Gender | Chief complaint | Primary CT protocol | Rounds of investigations |
|---|---|---|---|---|---|---|
| 1 | TAAD | 80 | Male | Nausea and vomiting for 2 hours, syncope for 1 hour | Abdominal CT*# | Chest CT# |
| 2 | TAAD | 49 | Male | Abdominal pain for 8 hours | Abdominal CT*# | / |
| 3 | IMH | 75 | Male | Upper abdominal pain with back pain for 5 days | Abdominal CT*# | / |
| 4 | PAU | 73 | Female | Abdominal pain for 9 hours, vomiting 3 times | Abdominal CT | / |
| 5 | TBAD | 68 | Male | Lower back pain for 2 hours | Lumbar CT# | / |
| 6 | IMH | 69 | Male | Lower back pain for 1 hour | Abdominal CT# | Chest CT*# |
| 7 | TAAD | 54 | Male | Upper abdominal pain with nausea and chest tightness for 3 hours | Abdominal CT*# | / |
| 8 | TBAD | 64 | Male | Abdominal pain for 1 day | Abdominal CT*# | Mesenteric artery and vein CT Angiography*# |
| 9 | TBAD | 57 | Male | Chest and back pain for 2 hours | Chest CT*# | / |
| 10 | TBAD | 32 | Male | Chest and back pain with abdominal pain for 2 hours | Chest CT*# | / |
| 11 | TAAD | 54 | Male | Abdominal pain for 1 day | Abdominal CT*# | Abdominal CT with contrast*# |
| 12 | IMH | 75 | Female | Chest and back pain for 1 week | Chest CT*# | Abdominal CT#, Pulmonary CT Angiography*# |
| 13 | TBAD | 45 | Male | Right side abdominal and lower back pain for 2 hours | Abdominal CT*# | Abdominal CT with contrast*# |
| 14 | TAAD | 54 | Female | Back pain for 12 days | Chest CT*# | / |
| 15 | PAU | 71 | Male | Chest tightness for 4 days | Chest CT# | / |
| 16 | PAU | 60 | Male | Chest pain for 1 hour | Coronary CT Angiography*# | / |
| 17 | PAU | 62 | Male | Chest tightness for 1 day, worsening for 4 hours | Chest CT | / |
| 18 | TBAD | 63 | Female | Abdominal pain with bloating for 11 days | Abdominal CT*# | Esophageal CT*# |
| 19 | PAU | 58 | Male | Left lower back pain for 1 day | Abdominal CT*# | / |
| 20 | TBAD | 71 | Male | Paroxysmal right upper abdominal pain for 10 days,, worsening for 2 hours | Abdominal CT*# | Pulmonary CT Angiography*# |
| 21 | IMH | 76 | Female | Chest pain for 3 hours | Chest CT*# | |
| 22 | TBAD | 73 | Male | Chest tightness and shortness of breath for 4 hours | Chest CT*# | / |
| 23 | PAU | 88 | Male | Lower back pain for 6 days | Abdominal CT# | / |
| 24 | PAU | 67 | Male | Upper abdominal pain for 11 hours | Abdominal CT*# | / |
| 25 | TAAD | 74 | Female | Abdominal pain with diarrhea and vomiting for 13 hours | Abdominal CT*# | / |
| 26 | TBAD | 45 | Male | Chest tightness for 8 days | Chest CT*# | / |
| 27 | PAU | 89 | Female | Chest pain and tightness for 1 day | Chest CT# | Abdominal CT |
| 28 | IMH | 73 | Female | Upper abdominal pain with lower back pain for 2 hours | Abdominal CT# | / |
| 29 | IMH | 70 | Male | Chest pain for 2 hours | Chest CT*# | / |
| 30 | TBAD | 64 | Female | Chest pain for 7 days | Chest CT*# | / |
| 31 | PAU | 74 | Male | Abdominal pain for 10 hours | Abdominal CT | / |
| 32 | IMH | 61 | Male | Chest pain for 2 days | Chest CT*# | Abdominal CT*# |
| 33 | TAAD | 50 | Male | Chest pain for 1 day | Chest CT*# | / |
| 34 | PAU | 74 | Male | Chest pain for 7 days, worsening for 1 day | Chest CT*# | / |
| 35 | TBAD | 56 | Female | Upper left abdominal pain with vomiting for 1 day | Abdominal CT*# | Chest CT*# |
| 36 | TBAD | 31 | Male | Chest and back pain for 1 day, difficulty breathing for 1 hour | Chest CT*# | Coronary CT Angiography*# |
| 37 | TAAD | 59 | Female | Chest pain with vomiting for 5 hours | Chest CT*# | / |
| 38 | IMH | 59 | Male | Upper abdominal pain for 1 day | Abdominal CT*# | Abdominal CT with contrast*# |
| 39 | TAAD | 55 | Male | Abdominal pain for 8 hours | Abdominal CT*# | / |
| 40 | TAAD | 34 | Male | Chest and abdominal pain with back pain for 2 hours | Abdominal CT*# | / |
| 41 | TBAD | 59 | Male | Chest tightness for 1 hour | Chest CT*# | / |
| 42 | TBAD | 54 | Male | Chest tightness for 1 day | Chest CT*# | / |
| 43 | IMH | 45 | Male | Chest tightness and shortness of breathe for 2 hours | Chest CT*# | / |
| 44 | TBAD | 46 | Male | Chest and back pain for 7 hours | Chest CT*# | / |
| 45 | TBAD | 56 | Male | Pain below the sternum for 2 hours | Chest CT*# | / |
| 46 | TAAD | 57 | Male | Chest tightness with dizziness and vomiting for 5 hours | Chest CT*# | / |
| 47 | IMH | 93 | Female | Chest tightness for over 3 hours | Chest CT*# | Abdominal CT*# |
| 48 | TBAD | 59 | Male | Abdominal pain for 2 days | Abdominal CT*# | / |
| 49 | TBAD | 59 | Male | Abdominal pain for 6 days, chest pain with diarrhea for 2 days | Chest CT*# | / |
| 50 | TBAD | 64 | Female | Chest and back pain for 2 hours | Chest CT*# | / |
| 51 | TBAD | 64 | Female | Chest tightness and pain for 1 day, abdominal pain with nausea and vomiting for 4 hours | Abdominal CT*# | Lumbar CT# |
| 52 | TAAD | 64 | Male | Chest pain for 3 hours | Chest CT*# | Pulmonary CT Angiography*# |

**Supplementary Table 6 |** Cases with initial false suspicion in the clinical evaluations of RW2 Cohort 1. The superscript * and # represents the cases that were successfully detected by original and upgraded model, respectively. Details of case ID 48&49 are shown in Extended Data Fig. 4b of the manuscript. Chest CT represents non-contrast chest CT; Abdominal CT represents non-contrast abdominal CT.

| Patient | Definitive diagnosis | Age | Gender | Chief complaint | Primary CT protocol | Rounds of investigations |
|---|---|---|---|---|---|---|
| 1 | TBAD | 80 | Female | Lower back pain for 3 hours,a severe back sprain three years age, several episodes of unexplained back pain in the past year | Lumbar CT# | Chest CT*# |
| 2 | TAAD | 58 | Female | Pain in the chest and back for 30 minutes | Chest CT*# | / |
| 3 | TBAD | 71 | Female | Abdominal pain for 1 week, worsened in the past day | Abdominal CT# | / |
| 4 | TAAD | 40 | Male | Chest pain for 2 hours, lower back pain for 30 minutes | Chest CT*# | Pulmonary CT angiography*# |
| 5 | IMH | 66 | Male | Chest tightness and pain for 1 day | Chest CT*# | Pulmonary CT angiography# |
| 6 | TBAD | 48 | Male | Chest tightness and shortness of breath for 3 days | Chest CT# | / |
| 7 | TBAD | 61 | Male | Abdominal pain for 2 hours | Abdominal CT*# | / |
| 8 | TBAD | 85 | Male | Lower back pain for 0.5 hours | Abdominal CT*# | / |
| 9 | IMH | 52 | Male | Abdominal pain with lower back pain for 1 hour | Abdominal CT | / |
| 10 | TAAD | 57 | Male | Chest pain for 0.5 hours | Chest CT*# | / |
| 11 | TAAD | 30 | Male | Lower back pain with chest tightness for 1 hour | Abdominal CT*# | / |
| 12 | TBAD | 82 | Female | Abdominal pain with poor appetite for 5 days | Abdominal CT*# | / |
| 13 | TAAD | 28 | Male | Right-sided lower back pain for 0.5 hours | Abdominal CT*# | Abdominal CT with contrast*# |
| 14 | TAAD | 56 | Male | Lower back pain for 3 hours | Abdominal CT*# | / |
| 15 | TBAD | 57 | Male | Upper abdominal pain for 1 hour | Abdominal CT*# | / |
| 16 | TAAD | 17 | Male | Chest tightness and shortness for 30 minutes | Chest CT*# | / |
| 17 | IMH | 83 | Male | Back pain for 1 hour | Chest CT# | / |
| 18 | PAU | 73 | Male | Abdominal pain for 1 day | Abdominal CT | / |
| 19 | IMH | 65 | Female | Chest pain for 3 days | Chest CT*# | Pulmonary CT angiography*# |
| 20 | TBAD | 72 | Male | Abdominal pain with lower back pain for 3 day | Abdominal CT*# | / |
| 21 | TBAD | 46 | Male | Abdominal pain for 4 days, worsened for 1 day | Abdominal CT*# | / |

**Supplementary Table 7 |** Cases with initial false suspicion in the clinical evaluations of RW2 Cohort 2. The superscript * and # represents the cases that were successfully detected by original and upgraded model, respectively. Details of case ID 1 are shown in Extended Data Fig. 4a of the manuscript. Chest CT represents non-contrast chest CT; Abdominal CT represents non-contrast abdominal CT.

| Patient | Definitive diagnosis | Age | Gender | Chief complaint | Primary CT protocol | Rounds of investigations |
|---|---|---|---|---|---|---|
| 1 | PAU | 96 | Male | Chest and abdominal pain for 1 day | Chest CT | Abdominal CT# |
| 2 | TBAD | 77 | Male | Chest pain for 4 days | Chest CT*# | Abdominal CT# |
| 3 | IMH | 54 | Male | Abdominal pain with backache for 4 days | Abdominal CT*# | Chest CT#, Abdominal CT with contrast*# |
| 4 | PAU | 64 | Male | Chest pain for 12 hours | Chest CT# | / |
| 5 | IMH | 84 | Male | Chest pain for 2 hours | Chest CT*# | Abdominal CT*# |
| 6 | TBAD | 46 | Female | Abdominal pain for 6 days | Abdominal CT*# | / |
| 7 | IMH | 67 | Female | Pain in the chest and back for 7 days | Chest CT*# | / |
| 8 | TBAD | 93 | Male | Chest tightness for 2 hours | Chest CT*# | / |
| 9 | TBAD | 52 | Female | Chest and abdominal pain for 2 days | Abdominal CT*# | / |
| 10 | TBAD | 42 | Male | Pain in the chest and back for 7 hours | Chest CT*# | / |
| 11 | TBAD | 58 | Female | Back pain for 5 hours | Chest CT*# | Abdominal CT*# |
| 12 | TAAD | 87 | Female | Chest pain for 2 hours | Chest CT*# | / |
| 13 | TBAD | 88 | Male | Back pain for 1 day | Chest CT*# | / |
| 14 | TAAD | 53 | Male | Chest pain with chest tightness and shortness of breath for 2 hours | Chest CT*# | Pulmonary CT angiography*# |
| 15 | PAU | 54 | Male | Chest tightness and shortness of breath for 4 days | Chest CT | / |
| 16 | TBAD | 43 | Male | Abdominal pain for 3 days | Abdominal CT*# | / |

**Supplementary Table 8 |** Cases with initial false suspicion in the clinical evaluations of RW2 Cohort 3. The superscript * and # represents the cases that were successfully detected by original and upgraded model, respectively. Chest CT represents non-contrast chest CT; Abdominal CT represents non-contrast abdominal CT.

| Patient | Definitive diagnosis | Age | Gender | Chief complaint | Primary CT protocol | Rounds of investigations | Time from presentation to AAS triage (mins) Group A | Time from presentation to AAS triage (mins) Group B |
|---|---|---|---|---|---|---|---|---|
| T1 | TAAD | 54 | Female | Chest tightness for 2 hours | Chest CT | Pulmonary CT angiography | 223 | 43 |
| T2 | TBAD | 62 | Female | Abdominal pain and bloating for 9 hours | Abdominal CT | Chest CT | 325 | 67 |
| T3 | IMH | 71 | male | Abdominal pain for 7 hours | Abdominal CT | Chest CT | 273 | 56 |
| T4 | TAAD | 54 | male | Chest pain for 1 hour | Chest CT | / | 115 | 58 |
| T5 | IMH | 57 | male | Abdominal pain for 4 hours | Abdominal CT | / | 132 | 49 |
| T6 | IMH | 41 | male | Chest tightness and shortness of breath for 2 hours | Chest CT | / | 294 | 89 |
| T7 | TBAD | 57 | male | Upper abdominal pain with vomiting for 3 hours | Abdominal CT | / | 261 | 53 |
| T8 | TBAD | 38 | male | Upper abdominal pain for 13 hours | Abdominal CT | / | 235 | 66 |
| T9 | TAAD | 55 | male | Chest pain for 4 hours | Chest CT | / | 119 | 73 |
| F10 | PAU | 73 | male | Back pain for 2 days | Chest CT | / | 305 | 311 |
| F11 | PAU | 61 | Female | Upper abdominal pain for 21 hours | Abdominal CT | / | 259 | 241 |

**Supplementary Table 9 |** Cases with initial false suspicion in the prospective multi-center study (comparative study). Chest CT represents non-contrast chest CT; Abdominal CT represents non-contrast abdominal CT.

| Patient | Definitive diagnosis | Age | Gender | Chief complaint | Primary CT protocol | Time from presentation to diagnosis (mins) |
|---|---|---|---|---|---|---|
| TP1 | TAAD | 32 | male | Chest pain for 2 hours | Chest CT | 87 |
| TP2 | TAAD | 54 | female | Acid reflux for 1 month and chest pain for 4 hours | Esophageal CT | 92 |
| TP3 | TBAD | 43 | male | Right upper quadrant dull pain for 12 hours, worsening in the last 2 hours | Abdominal CT | 94 |
| TP4 | TBAD | 45 | male | Chest tightness for 11 hours | Chest CT | 96 |
| TP5 | TBAD | 54 | male | Nausea and vomiting for 2 hours | Abdominal CT | 101 |
| TP6 | TBAD | 71 | female | Upper abdominal pain for 5 hours | Abdominal CT | 88 |
| TP7 | IMH | 60 | male | Chest tightness for 6 hours | Chest CT | 107 |
| TP8 | IMH | 62 | male | Abdominal distension for 10 hours | Abdominal CT | 99 |
| TP9 | PAU | 63 | male | Chest pain for 1 day | Chest CT | 113 |
| TP10 | TAAD | 41 | female | Chest tightness and shortness of breath for 2 hours | CTPA | 121 |
| TP11 | TAAD | 62 | male | Chest pain for 3 hours | Chest CT | 105 |
| TP12 | TAAD | 68 | male | Chest tightness for 4 hours | Chest CT | 90 |
| TP13 | TBAD | 39 | male | Abdominal pain for 2 hours | Abdominal CT | 126 |
| TP14 | TBAD | 66 | female | Chest tightness for 10 hours | Chest CT | 75 |
| TP15 | TBAD | 78 | male | Upper abdominal pain for 6 hours | Abdominal CT | 133 |
| TP16 | TBAD | 32 | male | Back pain with weakness in the lower limbs for 8 hours | Lumbar CT | 111 |
| TP17 | TBAD | 40 | female | Chest pain for 2 hours | Chest CT | 108 |
| TP18 | IMH | 56 | male | Chest pain for 6 hours | Chest CT | 100 |
| TP19 | IMH | 59 | female | Upper left abdominal pain with vomiting for 12 hours | Abdominal CT | 95 |
| TP20 | IMH | 44 | male | Pain below the sternum for 3 hours | Chest CT | 87 |
| TP21 | PAU | 88 | male | Chest tightness for over 5 hours | Chest CT | 116 |
| FN1 | PAU | 76 | male | Chest pain for 4 days | Chest CT | 357 |

**Supplementary Table 10 |** Cases with initial false suspicion in the prospective multi-center study (pilot deployment study). Chest CT represents non-contrast chest CT; Abdominal CT represents non-contrast abdominal CT.